\documentclass[journal]{IEEEtran}

\usepackage[dvipsnames,table]{xcolor}  
\usepackage{booktabs}

\usepackage{cite}

\usepackage{multirow}
\usepackage{graphicx}
\usepackage{amssymb,amsmath,times,wasysym,amsthm}
\usepackage{pb-diagram}
\usepackage{psfrag,balance}
\usepackage{srcltx,verbatim} 

\usepackage{multicol}
\usepackage{algorithmic}
\usepackage{algorithm}
\usepackage{tikz}
\usepackage{pgfplots}
\usepackage[edges]{forest}
\usepackage{enumitem}
\usepackage{fancybox}
\usepackage{amsmath,bm}

\usepackage{adjustbox}
\newcommand{\rot}[1]{\adjustbox{angle=85,lap=1em}{\textbf{#1}}}

\usetikzlibrary{positioning, arrows.meta, fit}

 \definecolor{Lightgray}{RGB}{235,235,235}

\usepackage[para]{threeparttable}
\newcommand*\OK{\checkmark}

\usetikzlibrary{arrows}
\tikzset{decision/.style = {diamond, draw, fill=black!20, text width=4.5em, text badly centered, node distance=2.8cm, inner sep=0pt},
	block/.style    = {rectangle, draw, fill=blue!5, text width=5em, text centered, rounded corners, minimum height=2em},	
	block2/.style    = {rectangle, draw, dashed,  fill=blue!20, text width=5em, text centered, rounded corners, minimum height=3em},
	line/.style     = {draw, -latex', very thick},
	cloud/.style    = {draw, ellipse,fill=red!20, node distance=2.8cm, minimum height=2em}
} 

\def\checkmark{\tikz\fill[scale=0.35](0,.35) -- (.25,0) -- (.9,.7) -- (.25,.15) -- cycle;} 

\tikzset{
	basic/.style  = {draw, text width=2cm, drop shadow, font=\sffamily, rectangle},
	root/.style   = {basic, rounded corners, thin, align=center, fill=red!30},
	level-2/.style = {basic, rounded corners=6pt, thin,align=center, fill=blue!20, text width=3cm},
	level-3/.style = {basic, thin, align=center, fill=orange!20, text width=2.4cm}
}

\usetikzlibrary{trees,positioning,shapes,shadows,arrows}

\newcounter{appendx}

\newtheorem{rem}{Remark}

\renewcommand{\exp}[1]{\mathrm{e}^{#1}}

\renewcommand{\log}[2][]{\mathrm{log}_{#1}\left(#2\right)}

\title{{
{A Tutorial on} AI-Empowered Integrated \\ Sensing and Communications
} 
}

\author{Mojtaba Vaezi,~\IEEEmembership{Senior Member,~IEEE}, Gayan Aruma Baduge,~\IEEEmembership{Senior Member,~IEEE}, \\ Esa Ollila,~\IEEEmembership{Senior Member,~IEEE}, and Sergiy A. Vorobyov,~\IEEEmembership{Fellow,~IEEE}  \vspace{-0mm}
	\thanks{		
		Mojtaba Vaezi is with the ECE Department, Villanova University, Villanova, PA 19085, USA (email: mvaezi@villanova.edu). His work was supported by the U.S. National
Science Foundation (NSF) under Grant CCF-2326622. Gayan Aruma Baduge is with the School of ECBE, Southern Illinois University, Carbondale, IL 62918, USA (email: gayan.baduge@siu.edu). His work in part was supported by the U.S. NSF under Grant CCF-2326621. 

{Esa Ollila  and Sergiy A. Vorobyov are with the Department of Information and Communications Engineering, Aalto University, 02150 Espoo, Finland (e-mail: esa.ollila@aalto.fi; sergiy.vorobyov@aalto.fi). Esa Ollila's work was supported by the Research Council of Finland under the grant 359848.}
}     
}

\DeclareMathAlphabet\mathbfcal{OMS}{cmsy}{b}{n}

\begin{document}
\bstctlcite{IEEEexample:BSTcontrol}
\maketitle

  
\begin{abstract}

Integrating sensing and communication (ISAC) can help overcome the challenges of limited spectrum and expensive hardware, leading to improved energy and cost efficiency. While full cooperation between sensing and communication can result in significant performance gains, achieving optimal performance requires efficient designs of unified waveforms and beamformers for joint sensing and communication. Sophisticated statistical signal processing and multi-objective optimization techniques are necessary to balance the competing design requirements of joint sensing and communication tasks. 
{ As model-based approaches can be suboptimal or too complex, deep learning offers a powerful data-driven alternative, especially when optimal algorithms are unknown or impractical for real-time use.} Unified waveform and beamformer design problems for ISAC fall into this category, where fundamental design trade-offs exist between sensing and communication performance metrics, and the underlying models may be inadequate or incomplete. 
This {tutorial paper} explores the application of artificial intelligence (AI)  to enhance efficiency or reduce complexity in ISAC designs.
 We emphasize the  integration benefits through AI-driven ISAC designs, prioritizing the development of unified waveforms, constellations, and beamforming strategies for both sensing and communication. To illustrate the practical potential of AI-driven ISAC, we present  {three}  case studies on waveform, beamforming,  {and constellation} design, demonstrating how unsupervised learning and neural network–based optimization can effectively balance performance, complexity, and implementation constraints.

\end{abstract}



 
\section{Introduction}\label{sec:introduction}

With the goal of ``everything is sensed, everything is connected, and everything is intelligent", the sixth-generation (6G) wireless systems will incorporate intelligent sensing to enable innovative services such as localization, autonomous vehicles, area imaging,  activity detection,  identification, and  monitoring \cite{Zhang2022}. Historically,  radar sensing and communication have been designed and operated independently, thus using different frequency bands.  \textit{Integrated sensing and communications (ISAC)} enables spectrum and hardware sharing to reduce the cost and resource demands.   
According to the federal communications commission (FCC) \cite{FCC2022}, the coexistence of radar and wireless systems can be  observed  in the L-band (1-2\,GHz), S-band (2-4\,GHz), C-band (4-8\,GHz), and millimeter-wave  (mmWave)   band (30-300 GHz). 

ISAC exemplifies some of the key distinctions between 5G and 6G. 
In 6G, communication signals will also serve as a medium for sensing, allowing the network to simultaneously detect, track, and map its environment while transmitting data. For example, ISAC enables autonomous driving and unlocks new use cases such as  environmental awareness, precise localization, and context-aware services, pushing beyond the capabilities of 5G, where sensing remains a separate function. In addition,  ISAC leads to greater spectral and energy efficiency, reduces congestion, and prevents spectrum under-utilization
\cite{Liu2022a,Liu2022b,kumari2019adaptive}. However, the true potential of ISAC technology can be realized through efficient designs of unified waveforms and beamformers. Nonetheless, achieving optimal performance is challenging, in part because sensing and communication have contrasting objectives and performance metrics.  

\begin{figure}[htbp] 
	\centering
 \includegraphics[scale=.59]{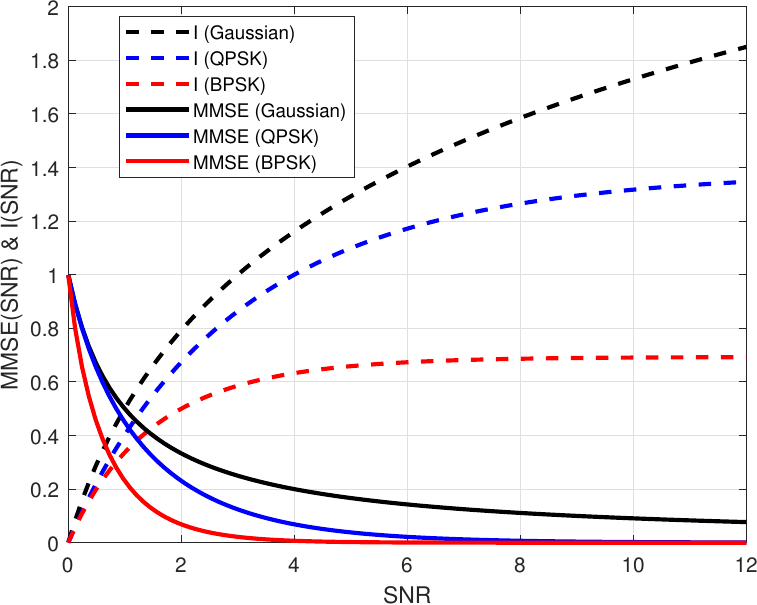}
\caption{Mutual information ($I$ in \textit{nats}/channel-use)  and MMSE versus SNR for an additive white Gaussian noise (AWGN) channel with three different input signals: Gaussian, QPSK, and BPSK. While Gaussian signaling achieves the maximum information rate, it also results in the highest (worst) MMSE.}

	\label{fig:IMMSE} 
\end{figure}

The performance of sensing tasks, such as estimation, detection, and tracking, is typically quantified  via mean squared error (MSE),  Cram\'er-Rao lower bound (CRLB), and receiver operating characteristic \cite{Liu2022a,Liu2022b,kumari2019adaptive}. On the other hand, the performance of wireless communications is mainly characterized by spectral and energy efficiency and reliability metrics. There are fundamental trade-offs between the performance metrics of sensing and communications. In particular, in a Gaussian channel,  the \textit{mutual information} (MI), the main metric for the transmission data rates, is related to the \textit{minimum mean square error} (MMSE) of sensing by
$\frac{\partial}{\partial \gamma} I(\gamma) = 0.5  \mathrm{MMSE}(\gamma),$
where \(\gamma\) represents the signal-to-noise ratio (SNR)\cite{Guo2005}. 
Thus, for a Gaussian channel, Gaussian inputs provide an upper bound on both the mutual information and MMSE.  
In other words, the Gaussian input is the most favorable for communication (as it maximizes $I(\gamma)$),  yet the least favorable for sensing (as it maximizes $\mathrm{MMSE}(\gamma)$).
This is demonstrated in Fig.~\ref{fig:IMMSE}, where the $\rm MMSE$ and  $\rm MI$ are also plotted for two practical constellations, namely binary phase-shift keying (BPSK) and quadrature phase-shift keying (QPSK). It is clear that QPSK is preferred if mutual information is the metric, whereas BPSK is preferred when MMSE is a concern.\footnote{In Section~\ref{sec:constellation}, we illustrate how novel constellations can be designed for ISAC to effectively balance the trade-off between sensing and communication.}

Given the fundamental trade-offs and the inherent multi-objective optimization requirements, advanced statistical signal processing and multi-objective optimization techniques are necessary to balance the competing design objectives of joint sensing and communication.
 Significant research efforts have been dedicated to achieving the goals of ISAC through classical model-based designs and theoretical analyses, leveraging communication and statistical signal processing theories \cite{Zhou2022,Zhang2022,Liu2022a,Zhang2021,Fortunati2020,Liu2020a}.
 However, designing unified constellations, waveforms and beamformers using model-based analytical approaches remains often prohibitively complex and challenging.
 
 { Artificial intelligence (AI) and its subsets, machine learning (ML) and deep learning,} have revolutionized wireless physical-layer design. \cite{zhang2019deep}.   
 Particularly, deep learning  (DL) is a powerful tool for extracting channel features and leveraging them in  various   tasks   through universal model approximations and solving non-linear multi-objective optimization problems.  The use of DL/ML offers significant advantages in this context by enabling performance optimization without relying on explicit models. 
 Instead, DL approaches leverage implicit learning to extract channel parameters directly from high-dimensional data,  making them well-suited for the dynamic and unpredictable conditions inherent to sensing and communication systems.
  {DL designs based on data-driven, model-driven, and their hybrids have been investigated for a variety of physical-layer communication and sensing problems  \cite{Shea2017,Zhang2021a,He2019,Jagannath2021,Zheng2021,Yu2022}.}

\subsection{Motivations {of AI-Based ISAC}}


{Classical model-based analytical approaches may be suboptimal due to model inadequacy or incompleteness, or become overly complex for ISAC problems because of the fundamental trade-offs between sensing and communication performance metrics, where optimal designs require sophisticated multi-objective optimization.
Motivated by these limitations} in developing efficient unified beamforming and waveform solutions, this tutorial  focuses o providing an overview of efficient AI-
empowered ISAC designs, introducing underlying learning-
based design foundations, presenting practical case studies,
evaluating their performance, and comparing them
against the classical model-based analytical counterparts.

A critical question is when and why learning-based approaches are needed in the context of ISAC. 
 {
Performance gains of DL-based designs over model-based analytical counterparts in physical-layer are prevalent   in cases where (i) the optimal algorithms are unknown, (ii) the known algorithms can acquire the optimal solution but are prohibitively  complexity for real-time implementation, and (iii) the   system models are either inadequate or incomplete \cite{Bjornson2020_AI}.  Unified waveform and beamformer design problems for ISAC systems fall into one or more of the aforementioned categories.} We elaborate on this in what follows.  


\subsubsection{{Model Inadequacy \& Hardware Impairments}}
Learning-based approaches become essential when traditional model-based methods fail to adequately account for imperfections and mismatches in models or hardware. These mismatches can stem from imperfect (e.g., estimated or quantized) channel state information (CSI), dynamic environments, where  CSI changes rapidly, hardware limitations, or the inherent complexity of ISAC. Learning-based techniques can adapt to these uncertainties by leveraging real-time data, enabling robust waveform and beamformer designs. This adaptability is crucial for optimizing ISAC performance in scenarios where precise analytical models are either insufficient or computationally intractable.
{For example, \cite{nguyen2024joint} shows that, compared to convex optimization methods, deep learning can significantly enhance the performance of both sensing and communication while reducing complexity. Similarly, \cite{mateos2022end} demonstrates that learning-based approaches in ISAC outperform traditional optimization techniques, particularly in terms of error rates under hardware impairments. Additional examples can be found in \cite{temiz2025deep}.

\begin{table*}[h] 
 	\caption {Existing ISAC surveys and tutorials and their primary focus}
 	\label{tab:related-survey} 
 	\centering
 	\begin{tabular}{@{} crl*{13}{c} }
 		\rowcolor{blue!20} \cellcolor{white}
 		& Year & Reference & Main Topic & \rot{Radar/Sensing} & \rot{Communications} & \rot{Positioning/Mapping} & \rot{Performance Metrics} & \rot{Fundamental Limits} & \rot{Constellation Design} & \rot{Waveform Design} & \rot{Beamforming Design} & \rot{Channel Estimation} & \rot{Deep Learning} & \rot{End-to-Tnd  Learning} & \rot{Security } \\
 		\cmidrule{2-16}

            &2026 & This paper & AI-empowered ISAC waveforms& \OK & \OK &  & \OK & & \OK  & \OK  &  \OK &  & \OK& \OK&   \\


 		\rowcolor{black!15} \cellcolor{white}
         &2025 &\cite{Liu2025sensing} & Sensing using communication signals  & \OK & \OK &  & \OK & & \OK& \OK&  &   & & &  \\
         
 		&2025 &\cite{Wen2024_survey} & Sensing, communication, and computation & \OK & \OK &   & & & & \OK & \OK&  & & & \\

          
 		\rowcolor{black!15} \cellcolor{white}
 		
 	
 		&2025 &\cite{Zhu2024_survey} & Secure ISAC in 6G & \OK & \OK && \OK & & & \OK  &  \OK & & & & \OK \\
        &2025 &\cite{Niu2024_survey} & Interference management for ISAC & \OK & \OK &  & &  & & \OK &\OK & & \OK &  &  \\
 		\rowcolor{black!15} \cellcolor{white}
 		&2024 &\cite{Gonzalez2024_survey}  & Sensing, positioning, and mapping in ISAC & \OK & \OK & \OK & & & & \OK &\OK &\OK &\OK & & \\
 		&2023 &\cite{Wei2023_survey} & ISAC signal design and optimization & \OK &\OK &  &\OK & & &\OK & &   & & & \\
 		\rowcolor{black!15} \cellcolor{white}
 		&2023 &\cite{cheng2023_survey} &  Multi-modal
sensing-communication & \OK & \OK&  & & & & \OK &\OK &\OK & \OK& &  \\
 		&2022 &\cite{Liu2022a} &  ISAC background and key applications    & \OK & \OK &  & \OK & \OK & & \OK&\OK & & & & \\
 		\rowcolor{black!15} \cellcolor{white}
 		&2022 &\cite{LiuAn2022} & Fundamental limits of ISAC   &  \OK & \OK &   & \OK & \OK &  &  \OK& \OK & & & &  \\
 		&2022 &\cite{Zhang2022} &  Integrating sensing to mobile networks & \OK & \OK &  & &  &  & \OK &\OK  & & & & \\

 		\cmidrule[1pt]{2-16}
 	\end{tabular}
\end{table*}

\subsubsection{Simplifying System Design}
 DL architectures, such as \textit{end-to-end learning} \cite{o2017introduction}, offer a powerful alternative to the block-by-block system design. End-to-end learning enables to jointly optimize  multiple blocks, e.g., encoding, modulation, and beamforming, thereby revolutionizing and simplifying system design compared to conventional methods. 
This approach is a significant  step toward AI-native air interface which is a long-term goal in AI-based communication system
design  \cite{hoydis2021toward}.}
The benefits of DL, and autoencoders in particular, stem  from their ability to learn low-dimensional structures embedded within high-dimensional data \cite{zhang2024deep}.  In ISAC systems, received radar echoes or pilot signals of the communication channel can be viewed as high-dimensional observations of sparse structures, making autoencoders suitable for joint optimization tasks such as channel estimation, target detection, beamforming, and constellation or waveform design in ISAC systems \cite{mateos2022end}.  DL models can also jointly optimize tasks such as range-Doppler estimation,
target classification, beam pattern design, and waveform adaptation \cite{temiz2025deep}. Further, when optimization algorithms are too complex or computationally intensive, DL methods such as plug-and-play \cite{shlezinger2023model} and algorithm unrolling \cite{monga2021algorithm,Jagannath2021,nguyen2024joint} reduce design complexity compared to traditional optimization solutions.

%

{\subsubsection{Seamless Integration to 6G Networks}
6G networks are expected to be AI-driven. In addition, sensing and localization capabilities will be integrated with traditional communication functions, creating intelligent networks that simultaneously map environments, track objects, and transmit data with higher efficiency \cite{Gonzalez2024_survey}.  Since ISAC is anticipated to play a key role in 6G systems, incorporating AI into ISAC designs is a natural step toward enabling its seamless adaptation to the evolving demands of 6G networks, making it intelligent by design.}
 
%

The availability of real-world datasets is another important factor facilitating and motivating AI-based ISAC. Many ISAC applications, such as vehicle-to-everything systems \cite{Sun2020, Mishra2019, Yuan2021}, naturally generate large datasets, which enable data-driven learning and model fine-tuning. This, in turn,  makes AI-based ISAC designs more practical and supports consistent benchmarking.
Examples of such datasets include \textit{DeepSense 6G} \cite{alkhateeb2023deepsense}, a large-scale real-world dataset featuring synchronized communication, radar, camera, and GPS data collected in both indoor/outdoor scenarios; and the \textit{ISAC UAV Dataset} \cite{mateos2022isac-uav}, an outdoor radar-based channel measurements along with ground-truth drone trajectories, ideal for radar-centric ISAC tasks.

 

\subsection{Related Tutorials and Paper Scope} 

{ We review related survey and tutorial papers,  outline the scope of our work, and highlight how it differs from existing surveys and tutorials.

\subsubsection{Existing Work}
Several papers listed in Table~\ref{tab:related-survey} have surveyed prior research efforts contributing to the advancement of ISAC technology.
The integration of radar/radio sensing into mobile networks is explored in \cite{Zhang2022}, which surveys radar-centric, communication-centric, and joint design approaches.
 The fundamental limits of ISAC have been studied in \cite{Liu2022a} and \cite{LiuAn2022}. In particular, \cite{LiuAn2022} provides an in-depth discussion of key performance metrics, such as the CRLB and MSE, along with relevant theoretical limits, whereas \cite{Liu2022a} explores rate-distortion tradeoff as well as detection and estimation bounds.  

All papers in Table~\ref{tab:related-survey} have topics related to signal/waveform  design.  However, \cite{Wei2023_survey} specifically focuses on this topic and surveys orthogonal frequency-division
multiplexing (OFDM) and orthogonal time frequency space (OTFS) modulation for ISAC.  OFDM-based sensing is also discussed in \cite{Gonzalez2024_survey} and \cite{LiuAn2022}. A recent tutorial on sensing with OFDM signals if offered in \cite{Liu2025sensing} which also includes a discussion on constellation design. 

 The scope of some of these papers extends beyond sensing and communication.
 Specifically, \cite{Gonzalez2024_survey} not only reviews network sensing and sensing-aided communication techniques for cellular systems, but also addresses positioning, localization, and mapping.
 On the other hand, \cite{Wen2024_survey} also considers the role of computing, discussing integration gains, system and signal design, and network resource management for joint sensing, communication, and computation.

Finally, the scope of some other papers is more narrowly focused. For example, \cite{Niu2024_survey} presents a survey of interference management techniques for ISAC. In \cite{cheng2023_survey}, intelligent multi-modal sensing–communication integration is explored using the concept of machine synesthesia. Meanwhile, \cite{Zhu2024_survey} surveys secure ISAC for intelligent connectivity, emphasizing the importance of dynamic adaptability and self-learning capabilities, and \cite{luo2025isac} is about ISAC layered structure and standardization.

%

%

\subsubsection{Paper Scope} This work presents a tutorial dedicated to AI-based ISAC, with a primary focus on how AI can enhance ISAC design through the  optimization of waveform, constellation and beamforming. We discuss various ML/DL–based optimization approaches for theses design tasks, emphasizing their potential advantages over traditional analytical and numerical methods typically used for multi-objective optimization.

Existing ISAC surveys are typically centered around signal processing and multi-objective optimization methods. Although some have explored the use of DL in the ISAC context (see Table~\ref{tab:related-survey}), they primarily apply learning techniques to channel estimation.
In contrast, this work aims to familiarize readers with a broader range of ML/DL-based approaches, with an emphasis on constellation, waveform, and beamforming design in ISAC systems.

 }
\subsection{Contribution}

This tutorial provides a comprehensive exploration of ISAC systems, with a particular focus on AI-driven advancements in the context of 6G networks. It emphasizes the limitations of traditional modeling approaches and showcases the potential of various AI-based techniques—such  unsupervised learning—to overcome these challenges. The growing role of AI in next-generation networks is discussed, leading to an in-depth examination of AI-based ISAC design.

We begin with a primer on ISAC, introducing key concepts in both communication and sensing, along with performance metrics relevant to each domain. Radar technology is presented as a core sensing component, and its integration with communication systems is discussed. We then frame ISAC problems as multi-objective optimization tasks, comparing conventional approaches—such as weighted sum, $\epsilon$-constraint, and Pareto optimization—with modern learning-based solutions, including multi-task learning and multi-objective learning.

A major contribution of this tutorial is its detailed examination of waveform, beamforming, and constellation design for ISAC and the role AI plays in enhancing it. We cover traditional beamforming techniques in communication, radar, and ISAC and explore learning-based approaches, including supervised, unsupervised, and  reinforcement learning, as well as model-based learning techniques like plug-and-play methods and algorithm unrolling.

To make AI-driven ISAC designs more tangible, we present {three} concrete  case studies on AI-powered waveform, beamforming, and {constellation} design.
\begin{itemize}
    \item 
 The first case study explores \textit{unsupervised learning}-based waveform design, and demonstrates how the sensing-communication trade-off can be captured via a custom loss function, with constraints enforced using a proper layer. Performance is evaluated and compared to existing approaches using achievable sum rate, probability of detection, and transmit beampatterns, while also analyzing computational complexity and design trade-offs. 
 \item The second case study presents a hybrid beamforming method using a learned optimizer, which  optimizes analog and digital precoders by reformulating the problem as constrained least-squares optimization, and solves it via neural network-based projection. Numerical results show that the proposed \textit{unfolded} hybrid beamforming method achieves spectral efficiency close to optimal digital beamforming, particularly with more receiving antennas.
 \item The third case study employs a combination of \textit{end-to-end learning} for communication and \textit{supervised learning} for sensing.
Specifically, an autoencoder-based architecture is designed and trained to learn a modulation constellation that achieves low bit error rates.
Besides,  a radar detector is trained using labeled data to identify the presence or absence of a target. The shared encoder enables a single transmitted signal to serve both functions, and allows balancing sensing and communication performance through joint optimization.

 \end{itemize}

The three case studies collectively highlight AI’s adaptability across key ISAC physical-layer designs. The first applies unsupervised learning for waveform optimization, embedding sensing–communication trade-offs into the loss function. The second  demonstrates a learned optimizer for hybrid beamforming, efficiently approximating constrained solutions. The third combines end-to-end and supervised learning for joint constellation design and radar detection, enabling unified signal generation. Together, they showcase AI’s potential for data-driven waveform, beamforming, and constellation design for ISAC in next-generation systems.

 {  Finally, we elaborate on the lessons learned and outline potential future research directions. 
The \textit{Lessons Learned} section includes detailed discussions on: \textit{(i)} when to use each AI model;
    \textit{(ii)} the benefits of AI-empowered ISAC systems; and
    \textit{(iii)}  the limitations of AI-based ISAC systems.
The \textit{Future Research} section covers promising topics such as} the application of transformer neural networks for unified waveform and beamformer design, and digital twin-aided ISAC architectures. By bridging the gap between traditional signal processing and modern AI-driven techniques, this tutorial serves as a valuable resource for researchers and practitioners at the forefront of ISAC technology.

\begin{table}[!tbp]
\caption{List of Key Abbreviations.}
\centering
\begin{threeparttable}
\begin{tabular}{|l|l|}
\hline
\textbf{Acronym} & \textbf{Description} \\ [0.5ex] 
\hline \hline
3GPP & 3rd Generation Partnership Project \\ \hline
6G & Sixth generation \\ \hline
AI & Artificial intelligence \\ \hline
AWGN & Additive white Gaussian noise \\ \hline
BER & Bit error rate \\ \hline
BS & Base station \\ \hline
CRLB & Cram\'er-Rao lower bound \\ \hline
CSI & Channel state information \\ \hline
CW & Continuous wave \\ \hline
DBF & 
Digital beamforming \\ \hline
DFRC & Dual-functional radar-communication \\ \hline
 DL & Deep learning \\ \hline
DNN & Deep neural network \\ \hline
EVM & Error vector magnitude \\ \hline
GNN & Graph neural network \\ \hline
IoT & Internet of things \\ \hline
ISAC & Integrated sensing and communications \\ \hline
LiDAR & Light detection and ranging  \\ \hline
LSTM & Long short-term memory \\ \hline
MI    & Mutual information \\ \hline
MIMO & Multiple-input multiple-output \\ \hline
ML & Machine learning \\ \hline
MMSE & Minimum mean squared error \\ \hline
MOL & Multi-objective learning \\ \hline
MSE & Mean squared error \\ \hline
MTL & Multi-task learning \\ \hline
MUI&  Multi-user
interference  \\ \hline
OFDM & Orthogonal frequency division multiplexing \\ \hline
OTFS & Orthogonal time frequency spacing \\ \hline
PSK & Phase-shift keying \\ \hline
PAPR & Peak-to-average power ratio  \\ \hline
PGA & Projected gradient ascent \\ \hline
PnP & Plug-and-play  \\ \hline
QPSK & Quadrature phase-shift keying \\ \hline
ReLU & Rectified linear unit \\ \hline
 RCS &  Radar cross section \\ \hline
 RF & Radio frequency \\ \hline
ROC & Receiver operating characteristic \\ \hline
SDR & Semi-definite relaxation \\ \hline
SER & Symbol error rate \\ \hline
SINR & Signal-to-interference-plus-noise ratio \\ \hline
SNR & Signal-to-noise ratio \\ \hline
SVD & Singular value decomposition \\ \hline
\end{tabular}
\end{threeparttable}
\label{tableABB}
\end{table}

\section{A Primer to ISAC}\label{sec:Waveform}

 In 6G, communication signals will not only facilitate data transmission but will also serve as a medium for sensing and radar applications. This dual functionality will enable next-generation networks to simultaneously detect, track, and map its surrounding environment while maintaining reliable communication. By integrating sensing capabilities directly into the communication infrastructure, 6G networks will support advanced use cases such as environmental monitoring, high-resolution imaging, and object detection.
 This convergence of communication, radar, and sensing technologies will pave the way to new possibilities for applications like autonomous driving, smart cities, and new IoT systems, where real-time situational awareness and efficient data exchange are critical.

 Sensing and radar are closely related concepts, particularly in the field of remote sensing and detection technologies. Below, we explain each of these topics, their applications, their interrelation, and their integration with communication technologies. We also discuss the typical performance metrics used in both communication and radar systems.

\begin{table*}[t] 
	\centering
	\caption{Common Performance Metrics for Communication Systems} 
	\label{table:comm:metric}
	\begin{tabular}{c c c}  
		\toprule
		\begin{minipage}{2.2cm}\centering\textbf{Metric}\end{minipage} &
		\begin{minipage}{10.70cm}\textbf{Definition and Description}\end{minipage} &
		\begin{minipage}{2cm}\centering\textbf{References}\end{minipage} \\
		\midrule
		
		\begin{minipage}{2.2cm}\centering BER/SER/BLER\end{minipage} &
		\begin{minipage}{10.70cm}Probability of \textit{bit} or \textit{symbol}, \textit{block} errors in received data. Lower values indicate more reliable communication. These metrics depend on ${E_b}/{N_0}$ (or SNR), modulation scheme, and coding.\end{minipage} &
		\begin{minipage}{2cm}\centering\cite{proakis2008digital}\end{minipage} \\
		
		\midrule
		
		\begin{minipage}{2.2cm}\centering Spectral efficiency\end{minipage} &
		\begin{minipage}{10.70cm}Bits transmitted per second per Hz of bandwidth. Higher spectral efficiency allows for more data transmission in limited bandwidth.\end{minipage} &
		\begin{minipage}{2cm}\centering\cite{proakis2008digital}\end{minipage} \\
		
		\midrule
		
		\begin{minipage}{2.2cm}\centering Capacity\end{minipage} &
		\begin{minipage}{10.70cm}Theoretical upper bound on data transmission rate. For a point-to-point channel, Shannon's formula \( \rm C = W \, \log[2]{1 + SNR} \)  gives the maximum possible data rate with vanishing errors.\end{minipage} &
		\begin{minipage}{2cm}\centering\cite{proakis2008digital}\end{minipage} \\
		
		\midrule
		
		\begin{minipage}{2.2cm}\centering Achievable rate\end{minipage} &
		\begin{minipage}{10.70cm}Data transmission rate considering practical constraints. Determined by modulation, coding, and BER requirements. It is lower than capacity because of implementation limitations.\end{minipage} &
		\begin{minipage}{2cm}\centering\cite{proakis2008digital}\end{minipage} \\
		
		\midrule
		
		\begin{minipage}{2.2cm}\centering Throughput\end{minipage} &
		\begin{minipage}{10.70cm}Actual data successfully transmitted and received. Affected by network congestion, re-transmissions, and protocol overhead. It is always lower than the achievable data rate.\end{minipage} &
		\begin{minipage}{2cm}\centering\cite{haenggi2009outage,xiao2002throughput,zou2025throughput}\end{minipage} \\
		
		\midrule
		
		\begin{minipage}{2.2cm}\centering Outage probability\end{minipage} &
		\begin{minipage}{10.70cm}Probability that the received signal falls below a threshold. Important in fading channels to assess communication reliability.\end{minipage} &
		\begin{minipage}{2cm}\centering\cite{hunter2006outage,haenggi2009outage}\end{minipage} \\
		
		\midrule
		
		\begin{minipage}{2.2cm}\centering Latency\end{minipage} &
		\begin{minipage}{10.70cm}Delay in data transmission and reception. Low latency is crucial for real-time applications like voice, video, gaming, and autonomous systems.\end{minipage} &
		\begin{minipage}{2cm}\centering\cite{vaezi2022cellular}\end{minipage} \\
		
		\midrule
		
		\begin{minipage}{2.2cm}\centering EVM\end{minipage} &
		\begin{minipage}{10.70cm}Deviation between actual and ideal signal constellation points. Lower EVM indicates better modulation accuracy and signal quality.\end{minipage} &
		\begin{minipage}{2cm}\centering\cite{shafik2006extended,schmogrow2011error,brunner2024bistatic}\end{minipage} \\
		
		\bottomrule
	\end{tabular}
\end{table*}

\subsection{Communications} \label{sec:subsec2}

Communication systems involve the transmission of signals, such as voice, video, and data from a sender to a receiver through various physical media, including cables, optical fibers, or wireless channels. These systems rely on a series of processes, such as source and channel encoding, modulation, multiplexing, beamforming and waveform design to prepare the information for efficient transmission over the selected medium. At the receiver end, reverse operations are applied to reconstruct the original signal.

The primary objective of a communication system is to ensure the fast and reliable exchange of information by maximizing the possible data transmission rate while minimizing the possible  error rate.
Achieving this involves addressing challenges like noise, interference, channel distortions, hardware imperfectness, and synchronization, which can degrade signal quality. Modern communication systems also aim to optimize factors such as bandwidth efficiency, power consumption, and latency to meet the demands of various applications, from high-speed internet and mobile communications to streaming and real-time conferencing.

5G networks are designed to support diverse applications by providing enhanced performance metrics.
Building upon the advancements of 5G,  6G technology aims to significantly enhance key performance indicators. Projected targets for 6G include a peak data rate of 1 terabit per second, an experienced data rate of 1 gigabit per second, and a peak spectral efficiency of 60 bits/s/Hz \cite{vaezi2022cellular}. Additionally, 6G aspires to achieve user plane latency as low as 100 microseconds (\(\mu\)s) and reliability levels reaching \(10^{-7}\). These ambitious goals are designed to support applications such as immersive communications, hyper-reliable low-latency services, autonomous driving, and the extensive connectivity required for the IoT.

\subsubsection{Communications Performance Metrics} \label{sec:subsec2}

The main figure of merit in digital communication is the \textit{energy per bit to noise power spectral density ratio}, \({E_b}/{N_0}\), where $E_b$ denotes the energy per bit and $N_0$   corresponds to the noise power spectral density. It is related to the SNR by 
${E_b}/{N_0} = \text{SNR} \cdot {W}/{R_b}$,
where \(W\) denotes the system bandwidth, and \(R_b\) is the bit rate \cite{sklar2021digital}. The ratio \({R_b}/{W}\) is known as bandwidth efficiency, or \textit{spectral efficiency}, and is measured in \textit{bits/s/Hz}. 
With ideal pulse shaping, for  $M$-ary digital modulation schemes like QPSK this efficiency is  $\mathrm{log}_2 M$  \cite{proakis2008digital}.
 {Other common  metrics include the bit error rate (BER), symbol error rate (SER), and block error rate (BLER). BER represents the probability of bit errors in the received data, SER quantifies the probability of symbol misdetection, and BLER measures the probability that an entire message block is received incorrectly. These metrics are typically plotted as functions of  \( E_b/N_0 \) or SNR.
}

The \textit{achievable data rate} refers to the data transmission rate in a channel, considering practical constraints.  
The \textit{capacity} of a channel is defined as the highest achievable data rate when the BER goes to zero. 
For an AWGN channel with bandwidth $W$, Shannon’s formula  \( \rm C = W\, \log[2]{1 + {SNR}}\) defines the capacity in \textit{bits/s}. 
There are other important metrics such as throughput,  outage probability,
error vector magnitude (EVM), and latency, all defined in Table~\ref{table:comm:metric}.

\subsection{Sensing} \label{sec:sensing}

Sensing refers to the process of detecting or measuring physical properties, objects, or conditions in the environment using various sensors.
Sensors convert physical phenomena into measurable signals, which can be processed and analyzed.
Sensing can be performed using different types of sensors, such as optical sensors, thermal sensors, acoustic sensors, and electromagnetic sensors.

Sensing has a central role in enabling data-driven decision-making, enhancing efficiency, improving safety, and advancing technological capabilities across various industries and everyday life. Application areas of sensing include, but are not limited to, the following: 
\begin{itemize}[leftmargin=*]
    \item 
 \textit{IoT:} Enabling interconnected devices to collect and share data for smart home automation, agriculture, transportation, and more. 
 \item 
\textit{Autonomous Systems:} Enabling  vehicles, drones, and robots to perceive and interact with their surroundings using sensors such as radar, light detection and ranging (LiDAR), cameras, and ultrasonic sensors.

\item \textit{Energy Management:} Optimizing energy use and efficiency in smart grids, renewable energy systems, and smart homes.
\item  \textit{Natural Disaster Monitoring:} Early detection and warning systems for earthquakes, tsunamis, floods, and wildfires using sensors to mitigate risks and improve disaster response.
\item  \textit{Biomedical Sensing:} Monitoring and diagnosing medical conditions through sensors that detect biological signals, analyze biomarkers, and deliver targeted therapies.
\end{itemize}

As communication standards is evolving to 6G, the integration of advanced sensing technologies is expected to drive innovations in communication and computing.

\subsubsection{Sensing Performance Metrics}

The estimation-theoretic metrics are typically used to characterize the sensing performance. 
In parameter estimation, the performance of the estimator is commonly evaluated using the MSE metric as $\mathrm{MSE}_{\boldsymbol{\theta}} = \mathbb E   [\|\boldsymbol{\theta} - \hat {\boldsymbol{\theta} } \|^2]$,     where $\boldsymbol{\theta} $ is a  vector of deterministic (true) parameters  and $\hat{\boldsymbol{\theta}} $ is its estimate. The  estimator that  minimizes the MSE is termed an MMSE estimator, and it may not be  realizable for many  practical scenarios.  From a practical perspective, we constrain the estimator's bias to be zero to construct an estimator that minimizes the variance. Such an estimator is termed a minimum variance unbiased estimator and its MSE is just the variance.  The variance of any unbiased estimator is lower-bounded by the CRLB; $\mathrm{CRB} = \mathbf I^{-1} (\boldsymbol{\theta})$, where $\mathbf I(\boldsymbol{\theta})$ is the Fisher's information matrix. The CRLB is widely used as a sensing performance metric.
The notion of CRLB can also be extended to random parameters with a known a-priori distribution. Hence, one can define a posterior CRLB for Bayesian estimators as it serves as a lower bound for Bayesian MSE. Moreover,  Weiss-Weinstein and Ziv-Zakai bounds are two tighter Bayesian lower bounds that can be derived when the prior distribution of the parameter vector is known.

\subsection{Radar} \label{sec:radar}

Radio detection and ranging, or \textit{radar}, is a  type of sensing technology that uses radio waves to detect and determine the distance, speed, and other characteristics of objects. By transmitting waves towards an object and receiving the reflected signals, radar operates in an active sensing manner. 
The time delay, frequency shifts (Doppler effect), and the angle of arrival (AoA) of the received signals provide information about the object’s properties.
Radar's applications include air traffic control, automotive safety systems, weather forecasting, military surveillance, motion detection and perimeter security, fall detection and healthcare, as well as general security and surveillance.

Radars are typically classified as either active or passive. \textit{Active radar} systems emit their own electromagnetic waves towards a target and detect the reflections that bounce back from objects.
In contrast, \textit{passive radar} systems do not emit their own waves but rely on external sources of illumination for detection. They detect reflections of signals sent from broadcast stations, communication satellites, other radars, etc. While this method allows for covert operation and reduces the risk of detection by potential adversaries, passive radars face challenges related to signal availability, processing complexity, and operational limitations compared to active radar systems.

\begin{table*}[!th] 
	\centering
	\caption{Radar/Sensing Performance Metrics}  
	\label{table:radar:metric}	
	\begin{tabular}{c c c}  
		\toprule
		\begin{minipage}{2.35cm}\centering\textbf{Metric}\end{minipage} &
		\begin{minipage}{11.5cm}\textbf{Definition and Description}\end{minipage} &
		\begin{minipage}{1.2cm}\centering\textbf{References}\end{minipage} \\
		\midrule
		
		\begin{minipage}{2.35cm}\centering Range resolution\end{minipage} &
		\begin{minipage}{11.5cm}Ability to distinguish closely spaced targets. Higher resolution allows for better target discrimination.\end{minipage} &
		\begin{minipage}{1.2cm}\centering\cite{levanon2004radar,sturm2011waveform,moreira2013tutorial}\end{minipage} \\
		
		\midrule
		
		\begin{minipage}{2.35cm}\centering Doppler resolution\end{minipage} &
		\begin{minipage}{11.5cm}Ability to distinguish targets by velocity. Important for detecting moving targets.\end{minipage} &
		\begin{minipage}{1.2cm}\centering\cite{levanon2004radar,sturm2011waveform}\end{minipage} \\
		
		\midrule
		
		\begin{minipage}{2.35cm}\centering Ambiguity function\end{minipage} &
		\begin{minipage}{11.5cm}The resolution and ambiguity in range and Doppler. Helps analyze the trade-off between range and Doppler resolution.\end{minipage} &
		\begin{minipage}{1.2cm}\centering\cite{levanon2004radar}\end{minipage} \\
		
		\midrule
		
		\begin{minipage}{2.35cm}\centering \( P_D \)\end{minipage} &
		\begin{minipage}{11.5cm}Likelihood of correctly detecting a target. Higher \( P_D \) is desirable for reliable detection.\end{minipage} &
		\begin{minipage}{1.2cm}\centering\cite{chang2010multiparameter}\end{minipage} \\
		
		\midrule
		
		\begin{minipage}{2.35cm}\centering \( P_{\mathrm{FA}} \)\end{minipage} &
		\begin{minipage}{11.5cm}Likelihood of false target detection. Lower \( P_{\mathrm{FA}} \) is desirable to reduce false alarms.\end{minipage} &
		\begin{minipage}{1.2cm}\centering\cite{chang2010multiparameter}\end{minipage} \\
		
		\midrule
		
		\begin{minipage}{2.35cm}\centering Receiver operating characteristic (ROC)\end{minipage} &
		\begin{minipage}{11.5cm}\( P_D \) versus \( P_{\mathrm{FA}} \) for a given decision threshold. The ROC curve evaluates a system's ability to detect the presence of a target versus making a false alarm.\end{minipage} &
		\begin{minipage}{1.2cm}\centering\cite{chang2010multiparameter}\end{minipage} \\
		
		\midrule
		
		\begin{minipage}{2.35cm}\centering Clutter rejection\end{minipage} &
		\begin{minipage}{11.5cm}Ability to distinguish targets from clutter. Effective clutter rejection improves target detection.\end{minipage} &
		\begin{minipage}{1.2cm}\centering\cite{levanon2004radar}\end{minipage} \\
		
		\midrule
		
		\begin{minipage}{2.35cm}\centering Azimuth/elevation resolution\end{minipage} &
		\begin{minipage}{11.5cm}Ability to distinguish targets by angular position. Higher angular resolution allows for better spatial discrimination.\end{minipage} &
		\begin{minipage}{1.2cm}\centering\cite{moreira2013tutorial}\end{minipage} \\
		
		\midrule
		
		\begin{minipage}{2.35cm}\centering CRLB\end{minipage} &
		\begin{minipage}{11.5cm}Theoretical lower bound on the variance of unbiased parameter estimators. Provides a benchmark for the best achievable estimation accuracy (e.g., range, velocity).\end{minipage} &
		\begin{minipage}{1.2cm}\centering\cite{LiuAn2022}\end{minipage} \\
		
		\midrule
		
		\begin{minipage}{2.35cm}\centering MSE\end{minipage} &
		\begin{minipage}{11.5cm}Average mean squared error between estimated and true parameter values. Lower MSE indicates better estimation performance (e.g., in range or velocity).\end{minipage} &
		\begin{minipage}{1.2cm}\centering\cite{LiuAn2022}\end{minipage} \\
		
		\bottomrule
	\end{tabular}
\end{table*}

\textit{Pulsed radar} works by sending short bursts of radio waves and detecting the echoes from targets. It transmits briefly and listens during the intervals between pulses, thus not listening while transmitting.
 The time delay between transmitted and received signals determines target distance, and the frequency shift due to the Doppler effect gives the relative speed. 
Pulsed radars are classified into two main types: moving target indicator (MTI) radar and Doppler radar.
MTI radar is focused on filtering out stationary clutter to detect moving targets, but it does not measure target velocities as precisely.
Doppler radar is designed to measure the velocity of moving targets with high accuracy. 
\textit{Impulse radar} is a variant of pulsed radar. While both use electromagnetic pulses for detection, impulse radar operates with shorter pulses, making it ideal for high-resolution and precise distance or profile measurements.

A \textit{continuous wave (CW) radar}, in contrast to  pulsed radars,  transmits continuously, meaning that it listens for target returns simultaneously with transmitting. 
CW radars can beted or unmodulated. Two prominent examples of modulated CW radars are frequency-modulated continuous wave (FMCW) and phase-modulated continuous wave (PMCW) radars.
CW radar  offers a longer range and higher resolution than pulsed radar but faces challenges like requiring separate transmitters and receivers, difficulty measuring target distance, and susceptibility to interference and jamming.

\subsubsection{Radar Performance Metrics}


 { 
   Radar's main functionalities are detecting, tracking, imaging, and classifying targets. The key metrics associated with these radar and  sensing tasks are summarized in Table~\ref{table:radar:metric}.} 
In radar detection, the probability of detection ($P_D$), probability of false alarm ($P_{\rm FA}$), and resolution are the widely used performance metrics. 
The range resolution of radar systems is defined as $\Delta R = c/2W$, where $c$ is the speed of light and $W$ is the bandwidth. Similarly, the velocity resolution is defined as $\Delta v = \lambda/(2NT) $, where $\lambda$ is the wavelength, $N$ is the number of coherently processed pulses, and  $T$ is the pulse repetition interval.  Finally, the half-power angular resolution is defined as $\Delta \theta = 0.886\lambda/D_A$, where $D_A$ is the size of the array's aperture.
{For tracking, range, Doppler, and AoA are needed to determine target position and motion. For classification, additional discriminative features such as micro-Doppler signatures become critical for distinguishing between different target types or activities.}

\subsection{Integrated Sensing and Communication} \label{sec:subsec2}

In its broad sense, ISAC refers to the integration of various sensing technologies, including radar, with communication systems.
The goal is to leverage shared resources, infrastructure, or information exchange between sensing (like environmental monitoring, surveillance) and communication functions (like data transmission, networking). It often focuses on optimizing sensor data utilization within communication networks or enhancing sensing capabilities through communication systems.

However, ISAC typically denotes the specific integration of radar systems with communication systems. This involves sharing hardware components, signal processing resources, or network infrastructure between radar functions (such as target detection, tracking) and communication functions (data transmission, networking). The emphasis is on synergizing radar functionalities with communication technologies to improve efficiency and effectiveness in both domains.
In this paper, we focus on ISAC within this specific domain.

Integrating radar and communication systems poses several significant challenges. Achieving the diverse operational requirements and technical specifications of radar and communication systems is one main challenge. Radar systems typically prioritize high-resolution target detection and tracking capabilities. On the other hand, communication systems focus on efficient data transmission, network scalability, and reliability. Integrating these systems requires addressing conflicting requirements, optimizing resource allocation, and ensuring seamless interoperability without compromising performance in either domain.

\subsubsection{ISAC Performance Metrics}

The performance of ISAC designs can be evaluated by characterizing fundamental trade-offs between sensing and communication metrics \cite{kumari2019adaptive,Chiriyath2016, An2023}. Generally, an ISAC metric combines communication and radar matrices from Tables~\ref{table:comm:metric} and~\ref{table:radar:metric}, along with newly defined notions.

The concept of radar estimation information rate, introduced in \cite{Chiriyath2016, Kumari2017}, parallels the data information rate in communication systems. It treats radar illumination on a target as a form of non-cooperative communication, where parameters such as delay, Doppler, and spatial directions are transmitted. Consequently, the radar channel can be viewed as a non-cooperative communication channel. This information rate is defined as the mutual information between radar and target, derived from the entropy of random estimation parameters and the entropy of estimation uncertainty \cite{Chiriyath2016}. 
Using the data processing inequality, the mutual information between the observation \( Y \) and the true parameter \( \theta \) is lower-bounded as \cite{LiuAn2022}
$
    I(Y;\theta) \geq \frac{1}{2} \log {\frac{P}{D}},
$
where \( \theta \) is Gaussian-distributed with variance \( P \), and \( D \) is the MSE distortion of estimating \( \theta \). This bound links the equivalent estimation information rate to the MSE distortion.

Alternatively, an equivalent communication information rate was introduced in \cite{Kumari2017}, leveraging rate-distortion theory to relate MSE distortion \( D \) to the information rate \( R \) as
$D = 2^{-R}$
assuming a Gaussian channel. The trade-off between sensing distortion and communication capacity can also be characterized via the capacity-distortion function \cite{LiuAn2022}. Furthermore, fundamental trade-offs among achievable sum rate, detection probability, and CRLBs can be quantified to assess ISAC performance \cite{An2023}.

\section{Multi-Objective Optimization: Traditional versus  Learning-Based Approaches}\label{sec:ML}
ISAC problems are inherently multi-objective optimization problems. Such a problem involves simultaneously optimizing two or more conflicting objectives which can be formulated as \cite{marler2004survey}

 \begin{center}
\cornersize{.2}
\setlength{\fboxsep}{5pt}
\ovalbox{
\begin{minipage}{0.435\textwidth}
\begin{eqnarray} \label{eq:MO}
&\underset{\bm{x}}{\text{Minimize}}&  \quad \bm{f}(\bm{x}) = [f_1(\bm{x}), f_2(\bm{x}), \ldots, f_i(\bm{x})]  \nonumber \\
&{\text{Subject to}}& \quad  {g}_j(\bm{x}) = 0, \quad  j = 1, 2, \ldots, m
\nonumber \\
& & \quad  {h}_k(\bm{x}) \leq 0, \quad   k = 1, 2, \ldots, n \qquad \qquad
\end{eqnarray} 
\end{minipage}
}
\end{center}

\noindent in which \( f_i(\bm{x}) \) are the objective functions, \( g_j(\bm{x}) \) and \( h_k(\bm{x}) \) are the equality and inequality constraints, respectively, and \( \bm{x} \) is the decision variable vector.

{Unlike in single-objective problems, multi-objective optimization typically has no single global optimum. Instead, the goal is to identify a set of solutions that satisfy a chosen optimality criterion. The most common is Pareto optimality. A  solution is Pareto-optimal if no other solution can improve one objective without degrading at least one other objective. There are several established approaches to solving multi-objective optimization problems \cite{marler2004survey,miettinen1999nonlinear}. In this section, we discuss a few common methods, grouped into two categories: \textit{(i)} traditional optimization methods, and \textit{(ii)} learning-based techniques.}

\subsection{Traditional Optimization Methods}

{For a given problem, the Pareto optimal set may contain infinitely many points. It is therefore important to distinguish between methods that generate the entire set (or a subset of it) and those that aim to obtain a single final solution.}
Using the \textit{scalarization method} \cite{miettinen1999nonlinear}, we can transform a multi-objective optimization problem into a single-objective problem and find a single solution. This can be done in various ways \cite{marler2004survey}. We discuss weighted sum method and $\epsilon$-constraint method, as two prominent examples of vectorization. {Other scalarization methods, such as the \textit{min–max} method, \textit{lexicographic} method, \textit{weighted product} method, and \textit{goal programming} exist and can be found in \cite{marler2004survey,cho2017survey}.
}
 
\subsubsection{Weighted Sum Method} It is also known as linear scalarization. The  design is equivalent to optimization of a joint cost function that balances the trade-off between communication and radar sensing objectives (optimizing the corresponding performance metrics). It can be described as the following optimization problem.
 
 \begin{center}
\cornersize{.2}
\setlength{\fboxsep}{5pt}
\ovalbox{
\begin{minipage}{0.4\textwidth}
\begin{eqnarray} \label{eq:weighted}
&\underset{}{\text{Minimize}}&  \quad \eta \,f_{\rm sens}(\cdot)    +  \bar \eta \, f_{\rm comm} (\cdot)   \nonumber \\
&{\text{Subject to}}& \quad  {g}(\cdot) = 0, \quad 
\nonumber \\
& & \quad  {h}(\cdot) \leq 0, \quad 
\label{eq:weightedsum}
\end{eqnarray} 
\end{minipage}
}
\end{center}

 \noindent
where $\eta$ ($ 0 \leq \eta \leq 1 $) and $\bar \eta = 1 -\eta $ are the normalization and weighting factors that assign priorities for the sensing and communication tasks, respectively, and $f_{\rm sens} ({\rm sensing \; metric})$ and $f_{\rm comm} ({\rm communication \; metric})$ are some functions of the sensing and communication metrics, respectively. The weights can be adaptively adjusted with respect to the requirements imposed by different scenarios. 
Various sensing and communication metrics can then be used.

{\textbf{Example:} The weighted-sum objective approach is widely used in ISAC designs, e.g., in an ISAC-enabled base station (BS)  tracking user mobility while maintaining high-rate communication \cite{zhang2023toward}.
 Prior works have also applied this method to minimize interference and beampattern mismatch under power constraints \cite{Liu2018}, maximize combined sensing and communication mutual information with spatio-temporal power allocation \cite{Yuan2021_example,He2024_example,Bazzi2025}, design precoders for MIMO systems to optimize achievable rates \cite{Wang2025}, and jointly tune phase-shift matrices and precoders for metrics such as SNR, SINR, rates, and CRLB \cite{Li2023,Wei2024}.
}

{
\begin{rem}
    The weighted sum fails on the non-convex Pareto front,\footnote{Pareto front represents the set of solutions where no objective can be improved without worsening at least one other objective.} missing some optimal points. Thus, only some simplified forms of \eqref{eq:weighted} under some simplifying assumptions, which may not hold in practice, can be solved using this method. However, weighted sum is often simpler and faster to compute, especially for linear or convex problems.
\end{rem}     
}

\subsubsection{$\epsilon$-Constraint Method} {Introduced in \cite{haimes1971bicriterion},} this method optimizes one objective while treating the remaining objectives as constraints with acceptable bounds ($\epsilon$).  By adjusting the $\epsilon$-values, different Pareto-optimal solutions can be generated \cite{marler2004survey,miettinen1999nonlinear}. This method is particularly useful when there is a priority among objectives, as it prioritize one objective while constraining the others.

Mathematically, the optimization can be expressed as
\begin{center}
\cornersize{.2}
\setlength{\fboxsep}{5pt}
\ovalbox{
\begin{minipage}{0.4\textwidth}
\begin{align} \label{eq:eps}
    \text{Minimize} \quad  f_i(\cdot) &  \nonumber\\
    \text{Subject to} \quad  f_j(\cdot)  &\leq \epsilon_j, \quad  j \in  \{{\rm sens}, {\rm comm}\} \backslash \{i\}  \nonumber \\
       {g}(\cdot) &= 0, \quad 
\nonumber \\
  \quad  {h}(\cdot) &\leq 0, \quad 
\end{align}
\end{minipage}
}
\end{center}

where
\begin{itemize}
    \item $f_i(\cdot)$ is the primary objective function to minimize; 
    \item $f_j(\cdot)$ are the other objective functions, constrained by the thresholds $\epsilon_j$; and,
    \item $\epsilon_j$ are user-defined constraint bounds.
\end{itemize}

{ \textbf{Example:} 
Many $\epsilon$-constraint optimization problems appear in the ISAC literature. Examples include probabilistic constellation shaping for ISAC that maximizes mutual information under a sensing constraint \cite{du2024pcs}, and conversely, \cite{liu2025probabilistic} which maximizes the sensing estimation rate under a communication constraint; SNR/CRB-constrained joint beamforming and reflection design \cite{liu2023snr}; and communication SINR-constrained CRLB minimization beamforming \cite{zhao2024joint}, among others.

}



{ 
\begin{rem} 

The $\epsilon$-constraint method offers an advantage over weighted-sum scalarization, as it can capture Pareto-optimal solutions in non-convex regions of the Pareto front~\cite{miettinen1999nonlinear}. Moreover, it does not require the selection of objective weights,  reducing potential bias. However, this benefit comes at the cost of selecting and systematically varying constraint bounds ($\epsilon$-values) and potentially increased computational effort.

\end{rem}
}

\subsubsection{Pareto Optimization} This approach, 
instead of finding a single optimal solution, aims to find the Pareto front. This can be achieved using techniques like evolutionary algorithms and Pareto-based methods. {Note that both the weighted-sum and $\epsilon$-constraint methods can generate Pareto-optimal solutions \cite{miettinen1999nonlinear}. However, the weighted-sum approach can obtain only those on the convex regions of the Pareto front, whereas the $\epsilon$-constraint method is capable of identifying all Pareto-optimal solutions, including those in non-convex regions.}

{\textbf{Example:} 
Several ISAC studies have employed this method for optimization \cite{gao2023cooperative,gao2010cooperative,sen2013papr,Wang2024_Pareto,Liu2025_Pareto}. The  motivation is to avoid reducing inherently multi-objective ISAC problems to single-objective formulations. Scalarization techniques can be problematic because the optimal solution is often subjective, being sensitive to the predefined scalar weights  or constraint bounds. Also, in nonlinear and non-convex problems, such approaches may fail to capture all Pareto-optimal solutions.

}

\subsection{Learning-Based Techniques }\label{sec:ML2}
 
ML-based approaches can be divided into supervised, unsupervised, and reinforcement learning, in general. Each approach can be implemented in various ways, such as multi-objective versus single-objective learning, multi-task versus single-task learning, and end-to-end learning, or a combination of them. In this section, we first elaborate on multi-objective and multi-task learning, highlighting their differences. We then cover end-to-end learning and hybrid methods.

\begin{figure*}[htbp] 
	\centering
  \scalebox{0.99}{
  \begin{tikzpicture}

    \node at (-5, 3.5) {{\textcolor{blue}{Single-task Learning}}};
    
    \node[] (st1) at (-2.75, 2) {{Task 1}};
    \node[draw, rectangle,minimum height=1cm, fill=blue!30,thick] (st1_layer1) at (-5, 2) {};
    \node[draw, rectangle,minimum height=1cm, fill=blue!30,thick] (st1_layer2) at (-6, 2) {};
    \node (st1_layer2b) at (-4.5, 2) {{\large $\ldots$}};
     \node[draw, rectangle,minimum height=1cm, fill=blue!30,thick] (st1_layer3) at (-3.5, 2) {};
    \draw[->,thick] (st1_layer2) -- (st1_layer1);
     \draw[->,thick] (st1_layer2b) -- (st1_layer3);
    
    \node[] (st2) at (-2.75, 0.5) {{Task 2}};
    \node[draw, rectangle,minimum height=1cm, fill=blue!30,thick] (st2_layer1) at (-5, 0.5) {};
    \node[draw, rectangle,minimum height=1cm, fill=blue!30,thick] (st2_layer2) at (-6, 0.5) {};
    \node (st2_layer2b) at (-4.5, 0.5) {{\large $\ldots$}};
     \node[draw, rectangle,minimum height=1cm, fill=blue!30,thick] (st2_layer3) at (-3.5, 0.5) {};
    \draw[->,thick] (st2_layer2) -- (st2_layer1);
     \draw[->,thick] (st2_layer2b) -- (st2_layer3);
   
    \node[] (st3) at (-2.75, -1) {{Task 3}};
    \node[draw, rectangle,minimum height=1cm, fill=blue!30,thick] (st3_layer1) at (-5, -1) {};
    \node[draw, rectangle,minimum height=1cm, fill=blue!30,thick] (st3_layer2) at (-6, -1) {};
    \node (st3_layer2b) at (-4.5, -1) {{\large $\ldots$}};
     \node[draw, rectangle,minimum height=1cm, fill=blue!30,thick] (st3_layer3) at (-3.5, -1) {};
    \draw[->,thick] (st3_layer2) -- (st3_layer1);
     \draw[->,thick] (st3_layer2b) -- (st3_layer3);
   
    \node at (-4.7, 2.85) {Task-specific Layers};

    \node at (3, 3.5) {{\textcolor{blue}{Multi-task Learning}}};

    \hspace{1cm}

    \node[draw, rectangle,rectangle,minimum height=1cm, fill=blue!30,thick] (shared1) at (-1.25, 0.5) {};
    \node[draw, rectangle,minimum height=1cm, fill=blue!30,thick] (shared2) at (-0.5, 0.5) {};
    \node (shared2b) at (0, 0.5) {{\large $\ldots$}};
    \node[draw, rectangle,minimum height=1cm, fill=blue!30,thick] (shared3) at (1.0, 0.5) {};
    \draw[->,thick] (shared1) -- (shared2);
    \draw[->,thick] (shared2b) -- (shared3);
    
    \node at (-0.25, 1.4) {Shared Layers};
    
    \node[] (mt1) at (5.25, 2) {Task 1};
    \node[draw, rectangle,minimum height=1cm, fill=blue!30,thick,thick] (mt1_layer1) at (2, 2) {};
    \node[draw, rectangle,minimum height=1cm, fill=blue!30,thick,thick] (mt1_layer2) at (3, 2) {};
    \node (mt1_layer2b) at (3.5, 2) {{\large $\ldots$}};
     \node[draw, rectangle,minimum height=1cm, fill=blue!30,thick,thick] (mt1_layer3) at (4.5, 2) {};
    \draw[->,thick] (mt1_layer1) -- (mt1_layer2);
     \draw[->,thick] (mt1_layer2b) -- (mt1_layer3);
   
   \node[] (mt2) at (5.25, 0.5) {Task 2};
    \node[draw, rectangle,minimum height=1cm, fill=blue!30,thick] (mt2_layer1) at (2, 0.5) {};
    \node[draw, rectangle,minimum height=1cm, fill=blue!30,thick] (mt2_layer2) at (3, 0.5) {};
    \node (mt2_layer2b) at (3.5, 0.5) {{\large $\ldots$}};
     \node[draw, rectangle,minimum height=1cm, fill=blue!30,thick] (mt2_layer3) at (4.5, 0.5) {};
    \draw[->,thick] (mt2_layer1) -- (mt2_layer2);
     \draw[->,thick] (mt2_layer2b) -- (mt2_layer3);
   
   \node[] (mt3) at (5.25, -1.0) {Task 3};
    \node[draw, rectangle,minimum height=1cm, fill=blue!30,thick] (mt3_layer1) at (2, -1) {};
    \node[draw, rectangle,minimum height=1cm, fill=blue!30,thick] (mt3_layer2) at (3, -1) {};
    \node (mt3_layer2b) at (3.5, -1) {{\large $\ldots$}};
     \node[draw, rectangle,minimum height=1cm, fill=blue!30,thick] (mt3_layer3) at (4.5, -1) {};
    \draw[->,thick] (mt3_layer1) -- (mt3_layer2);
     \draw[->,thick] (mt3_layer2b) -- (mt3_layer3);
    
    \node at (3.3, 2.85) {Task-specific Layers};
    
    \draw[->,thick] (shared3) -- (mt1_layer1);
    \draw[->,thick] (shared3) -- (mt2_layer1);
    \draw[->,thick] (shared3) -- (mt3_layer1);

\end{tikzpicture}
}
\caption{Comparison of single-task and multi-task learning for ISAC systems. {\it Left}: single-task learning trains separate models for each task independently. {\it Right}: Multi-task learning leverages shared layers to learn common features before branching into task-specific layers.}
	\label{fig:multitask} 
\end{figure*}

\subsubsection{Multi-Task Learning (MTL)}
MTL  is an ML paradigm where a
model learns to perform multiple tasks simultaneously.
Each task has its own set of outputs, and the model learns to predict those outputs jointly. The main idea is that learning shared representations across tasks can improve generalization and performance on each individual task. In contrast, single-task learning focuses on training a model for one specific task at a time, without considering interactions or dependencies with other tasks. Single-task learning is suitable when the goal is to solve a single, well-defined problem without considering the interactions or dependencies with other tasks. It is commonly used when the tasks are distinct and unrelated, or when there is insufficient data or computational resources to train a multi-task model. In multi-task learning, the goal is to train a model to perform multiple tasks simultaneously.  Figure~\ref{fig:multitask} illustrates the difference between these two types of learning. 

The model combines these task-specific losses into a single joint loss function, usually by applying weighted sums
\begin{equation}
    \mathcal{L}_{\text{MTL}} = \sum_{i=1}^{T} \lambda_i \mathcal{L}_i
\end{equation}

\noindent where \( \mathcal{L}_i \) is the loss for task \( i \), \( \lambda_i, \; i=1, \cdots , T \) are weights balancing the importance of the tasks, and \( T \) is the number of tasks. The overall objective is to minimize the joint loss across all tasks.

\textbf{Example:} 
Consider a self-driving car neural network model that has two tasks: detecting pedestrians (object detection) and  predicting the road's curvature (regression). These are two distinct tasks, and the model  combines the losses from both tasks to optimize neural network's performance, ensuring that it effectively detects pedestrians while accurately predicting road curvature
\begin{equation}
    \mathcal{L}_{\text{MTL}} = \lambda_1 \cdot \mathcal{L}_{\text{classification}} + \lambda_2 \cdot \mathcal{L}_{\text{regression}} .
\end{equation}
The goal is to optimize performance on multiple tasks using shared information or features.

The core idea of multi-task learning is that learning related tasks together improve performance by allowing the model to share information and transfer knowledge across tasks. This is especially effective when tasks have common features or underlying similarities.

Multi-task learning is implemented using either hard or soft parameter sharing approaches \cite{sener2018multi}, explained below. 

\begin{itemize}
    \item In \textit{hard parameter sharing}, a subset of parameters is shared across all tasks, while the remaining parameters are task-specific. This approach reduces the risk of overfitting by utilizing shared information, making it particularly effective when tasks are closely related.
\item In \textit{soft parameter sharing}, 
  each task maintains its own task-specific parameters, but these are jointly regularized to encourage  knowledge transfer between tasks. This is commonly achieved through techniques like Bayesian priors or a shared dictionary, allowing the model to learn task relationships without enforcing direct parameter sharing. 
\end{itemize}

\subsubsection{Multi-Objective Learning (MOL)}
Similar to multi-objective optimization, MOL  aims to optimize multiple, possibly conflicting, objective functions simultaneously. These objectives represent different criteria that need to be balanced within the same task. The goal is to find a \textit{balance} between these objectives, often resulting in a set of solutions known as the Pareto front.

In multi-objective learning, the model optimizes multiple objectives within a single task. Each objective has its own loss function, but instead of combining them into a single joint loss, the goal is to balance or trade off between the competing objectives. 
A common approach is to use weighted sum scalarization to combine the objectives into a single scalar loss function
\begin{equation}
    \mathcal{L}_{\text{MOL}} = \sum_{i=1}^{O} \alpha_i \mathcal{L}_i
\end{equation}
\noindent 
where \( \mathcal{L}_i \) is the loss for objective \( i \),  \(  i=1,\dots, O \), $\alpha_i$  are weights, and \( O \) is the number of objectives. Alternatively, other methods like the $\epsilon$-constraint method or Pareto optimization can be used  to solve the multi-objective optimization problem without explicitly combining the losses into one scalar function.

\textbf{Example:} Consider a single-task car model trained solely to detect pedestrians. This could still have multiple objectives, e.g., one objective might be to maximize accuracy, while another is to minimize inference time. These are two \textit{conflicting objectives} for the same task, requiring the model to find an optimal trade-off between them, e.g., by
 \begin{equation}
    \mathcal{L}_{\text{MOL}} = \alpha_1 \cdot \mathcal{L}_{\text{accuracy}} + \alpha_2 \cdot \mathcal{L}_{\text{latency}} .
\end{equation}
%
The goal is to optimize multiple `conflicting' objectives simultaneously. The model does not simply combine the losses but tries to balance them, often leading to Pareto optimal solutions where improving one objective worsens another.

\begin{rem}
The key difference between a task and an objective is that a task refers to what the model is designed to do, such as detecting objects, predicting the road's curvature, or removing interference. An objective, on the other hand, defines how the model's performance is measured and optimized, such as minimizing error, maximizing accuracy, or minimizing latency. In multi-task learning, the model is trained to solve multiple tasks simultaneously, whereas in multi-objective learning, the focus is on optimizing multiple objectives for a single task.
\end{rem}

In summary, multi-task learning is about performing multiple tasks simultaneously and using \textit{shared representations} to improve performance. Multi-objective learning is about \textit{balancing multiple conflicting objectives} during the optimization process. {
The boundaries between multi-task learning and multi-objective learning  in ISAC are scenario-dependent, as explained below.  
}

\begin{rem} \label{rem:MOL-MTL}
The distinction between multi-task learning and multi-objective learning in ISAC is often scenario-dependent and conceptually subtle. 
When sensing and communication are treated as{distinct tasks} producing separate outputs (e.g., channel estimation and target detection), the formulation follows the MTL paradigm. 
Conversely, when the focus lies on {balancing conflicting objectives} within a single system (e.g., optimizing rate–sensing trade-offs), the problem aligns more closely with MOL. 
In practice, the boundary between the two is blurred: ISAC systems can be modeled as MTL problems while still being optimized via multi-objective formulations \cite{sener2018multi}. 
Indeed, several recent works treat sensing and communication as distinct objectives within a multi-objective optimization framework \cite{wang2024multi,kumari2021adaptive,Liu2022a}. 
Therefore, the distinction ultimately depends on whether the emphasis is placed on the MTL or the MOL.
\end{rem}

\section{Waveform and Beamforming Design}\label{sec:Waveform}
In this section, we cover the basic definitions, required criteria, and common types and examples for waveforms and beamforming in radar, communication, and ISAC.

\subsection{Waveform Basics} \label{sec:Waveform}

A waveform refers to the shape or structure of the electromagnetic signal that is transmitted over a medium (channel) for communication, radar, ISAC, or other purposes. It describes how the signal varies over time, typically in terms of its amplitude, frequency, or phase. 
While the core definition of a waveform is consistent across radar and communication systems, the objectives and purposes behind transmitting the waveform differ largely between the two domains, as do their performance criteria.

\subsubsection{Waveform in Radar} 
In radar systems, the waveform is designed primarily for sensing and detecting objects. 
A well-designed waveform can greatly enhance system performance by improving the signal-to-interference ratio, optimizing spectrum usage, aligning with desired beampatterns, and improving the accuracy of target parameter estimation, among other factors \cite{he2012waveform,baker2004radar,rohling2008radar}.

Waveform design in radar involves selecting specific parameters such as pulse duration, pulse repetition frequency, carrier frequency, modulation scheme, and waveform shape (e.g., pulsed, continuous wave, frequency-modulated) to achieve desired performance objectives. These objectives may include maximizing detection range, improving target discrimination, 
and minimizing the radar's detectability.
The choice of waveform can significantly impact radar system performance in terms of range, accuracy, and sensitivity to various types of targets and interference. For example, short-duration pulses may be used for high-resolution imaging radar, while longer pulses or continuous waveforms might be employed for detecting moving targets or tracking objects over long ranges. Common radar waveforms are \cite{patole2017automotive}
\begin{itemize}
    \item \textit{Pulsed Waveforms:} The radar sends out short bursts of high-power energy. 
 \item \textit{Chirp Waveforms}: A linear frequency-modulated waveform, where the frequency of the signal changes linearly over time. Chirps are widely used because they provide high resolution in range and are robust to noise.
 \item \textit{Continuous Waves}: They involve continuous transmission without interruptions and can be unmodulated or modulated. The former is used for velocity measurement via the Doppler effect, while modulated CW radars—such as FMCW and PMCW—enable range estimation.
\item  \textit{OFDM Waveforms}: Originally developed for communication systems, OFDM waveforms have been adapted for radar because of their flexibility in spectrum shaping and potential for joint radar-communication applications \cite{barneto2019full}.

\end{itemize}

\subsubsection{Waveform in Communication} 

In communication systems, a waveform serves as the carrier of information from a transmitter to a receiver, encoding data into a format that can be efficiently transmitted over a communication channel and decoded at the receiver. Waveform design in communication involves selecting parameters such as modulation scheme, symbol rate, carrier frequency, bandwidth, and pulse shape to optimize efficiency and reliability of the transmission for the intended application and environmental conditions.

OFDM has been the primary waveform in modern wireless systems, including WiFi and cellular networks since 4G. One main drawback of OFDM is the high peak-to-average power ratio (PAPR), the ratio between the peak power and the average power of a signal. A waveform with high PAPR requires highly linear power amplifiers to avoid signal distortion. Such amplifiers are typically less power-efficient, leading to increased hardware costs and higher energy consumption—particularly challenging for battery-powered devices.

In full-duplex systems, the BS transmits and receives simultaneously over the same frequency band, introducing additional waveform design constraints due to self-interference, spectral coexistence, and hardware limitations.  
High-PAPR waveforms  impose stringent linearity and dynamic-range requirements on full-duplex transmitters. When self-interference suppression is limited, the residual self-interference after analog cancellation is often dominated by power amplifier-induced nonlinear distortion, which scales with PAPR. Consequently, waveforms with lower PAPR, or employing effective PAPR reduction techniques, can indirectly enhance self-interference suppression performance by reducing nonlinear residual components. Conversely, higher self-interference suppression capability relaxes power amplifier back-off constraints and improves the robustness of high-PAPR waveforms such as OFDM and OTFS \cite{bharadia2013full,korpi2014full,zhang2016full}.

\subsubsection{Waveform in ISAC} 

In ISAC, a single unified waveform is used to accomplish both sensing and communication tasks. 
Communication waveforms can be directly employed for sensing. The multicarrier waveforms based on OFDM have shown good performance in both sensing  and communication without requiring substantial alterations to existing communication system designs \cite{keskin2021mimo,sturm2011waveform}. 

In OFDM-based systems like 5G, unused subcarriers or sync blocks can be repurposed as radar subcarriers to enhance sensing performance, though this may reduce SNR for communication users due to power reallocation to radar subcarriers. It can also increase PAPR due to the use of more subcarriers. Optimizing these radar subcarriers can minimize estimation error (e.g., via CRLB) and reduce PAPR, improving overall system efficiency \cite{barneto2021full}.
In \cite{varshney2023low}, an iterative algorithm based on the majorization-minimization technique is proposed to minimize PAPR under the constraint that the integrated periodic autocorrelation sidelobe level is zero.

Other waveforms such as OTFS \cite{hadani2018otfs} and orthogonal delay-Doppler division multiplexing  \cite{lin2022orthogonal}  are shown to better suit to time-varying channels, especially in high-mobility  scenarios. For example, OTFS shows greater robustness to Doppler shifts compared to OFDM in communication scenarios \cite{raviteja2018practical}, while both waveforms exhibit comparable performance for sensing applications \cite{gaudio2020effectiveness}.
The difficulty with such communication-centric waveforms is that their PAPR is greater than one, indicating that the envelope of the transmitted time-domain signal does not have a constant amplitude. For example, the expected PAPR of OFDM increases with the number of subcarriers, while the expected PAPR pf OTFS increases with the number of subsymbols transmitted over slow-time in one radar frame. As a result, a non-uniform envelope requires using highly linear power amplifiers at the transmitter to avoid signal distortion that affects both radar and communication performance. It conflicts radar requirements, since high power is always beneficial for sensing. Indeed, the attenuation of a signal power for both way propagation in sensing channel is bigger than that for one way propagation in communications channel. OTFS suffers from this drawback in ISAC, just like OFDM. To deal with such difficulties, AI-based solutions have shown their importance within this approach based on the use of communication-centric waveforms \cite{Jiang2024Jamming, Visa2024MBOL}.  

A robust waveform design method for multiple-input and multiple-output (MIMO) dual-functional radar-communication (DFRC) system in the presence of transmit hardware impairments has been proposed in \cite{Guo2024hardware}. In \cite{Wang2024Power}, the authors devised a power-domain non-orthogonal ISAC waveform design with OFDM modulation, eliminating the need for complex operations while providing flexible control over the trade-offs between sensing and communication performances.
In \cite{Tang2024ISI}, to address the problem of short sensing range achieved by radar sensing in cyclic prefix OFDM based ISAC system,  a novel reference signal design and corresponding signal processing scheme were developed for OFDM long-range radar sensing. Relevant to waveforms design other ISAC functionalities are sidelobe information embedding in radar systems \cite{hassanien2015dual}, coexistence through array partitioning and waveform optimization \cite{liu2018mu}, or index modulation through exploitation of the different degrees of freedom in radar systems \cite{xu2022hybrid}.

{
Full duplex communication can increase spectral efficiency. It, however, introduces additional waveform design constraints due to self-interference, spectral coexistence, and hardware limitations, all of which can significantly impact the PAPR of transmitted signals, particularly in ISAC systems \cite{barneto2021full,barneto2019full}. Particularly, for self-interference suppression, waveforms must be designed to minimize leakage into the receiver path \cite{ahmed2015all}. This may require power shaping or signal clipping, which can increase PAPR due to waveform distortion.	

}

\begin{figure}[!t]
\centering
 \includegraphics[width=\columnwidth]{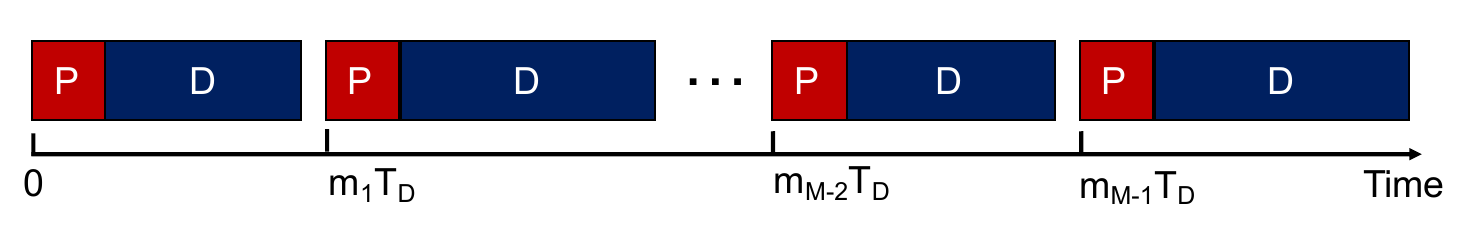}
 \caption{A coherent processing interval of $T$ seconds duration with $M$ ISAC frames. Each frame contains a fixed-length preamble of $P$ symbols and a varying length data segment. The length of each frame is an integer multiple, $q_m$, of the Nyquist sampling interval in the Doppler domain, $T_{\rm D}$.}
\label{fig:GeneralFrame}
\end{figure}

A generic waveform structure that allocates a fraction of symbols to communication (used exclusively for data transmission) and the remaining fraction to preamble symbols (dedicated entirely to sensing) within a coherent processing interval of \( T \) seconds spanning \( Q \) frames can mitigate the major limitations associated with communication-centric waveforms.
 Such generic waveform structure has been introduced for ISAC, especially for systems with high requirements to accuracy of velocity estimation, in \cite{kumari2019adaptive, kumari2017ieee}. 
 The IEEE 802.11ad standard can support such a multi-frame approach, where, as illustrated in Fig.~\ref{fig:GeneralFrame}, each frame consists of a fixed preamble duration \( P \) and a variable data payload, resulting in a varying overall frame length. The preambles are used as radar waveforms (e.g., pulsed waveforms), which are distinct from the communication waveforms (e.g., OFDM-based waveforms) employed for data transmission.

Using the frame structure in Fig.~\ref{fig:GeneralFrame}, the general ISAC optimization problem that aims to achieve the best possibly sensing performance by using the least possible resources (number of preambles in Fig.~\ref{fig:GeneralFrame}) can be formulated as
\begin{eqnarray} \label{eq:weightedWave}
&\underset{Q,\{ m_q \}_{q=1}^{Q-1}}{\text{Minimize}}& \eta \,f_{\rm sens}(Q,\{ m_q \}_{q}, \cdot)    +  \bar \eta \, f_{\rm comm} (Q,\{ m_q \}_{q}, \cdot)  \nonumber \\
&{\text{Subject to}}& \{ T, K \} = \mathrm{constants,} \nonumber\\
&&  0 < m_1 < \cdots < m_{Q-1} < T/T_{\rm D},
\end{eqnarray}
where $\eta$, $\bar \eta$, $f_{\rm sens}(Q,\{ m_q \}_{q=1}^{Q-1}, \cdot)$ and $f_{\rm comm} (Q,\{ m_q \}_{q=1}^{Q-1}, \cdot)$ have been introduced after \eqref{eq:weighted}, and $K$ is the number of targets to be sensed. 
Alternatively, problem \eqref{eq:weighted} can be modified as a minimization of one of the objectives with the second one written as an $\epsilon$-constraint (see \eqref{eq:eps}) that would guarantee an acceptable performance for one of the tasks. 
Such problem simplifies whenever a specific sparse pulse configuration/preamble positions, such as coprime pulses, is assumed. However, in general, the problem of optimal preamble location finding is hards (similar to a sphere packing problem), and there is no efficient optimization-based solution to it. It may however be addressed using AI methods. Thus, despite not being fully explored yet, AI can serve as an enabling approach for ISAC waveform design, when exact mathematical or optimization-based approaches fail.

\subsection{Beamforming Design} \label{sec:subsec1}

Beamforming, or precoding, the control of signal phase and amplitude, is a key signal processing technique used in both wireless communications and radar.
While both applications share the same fundamental principles, their goals and implementations differ.

\begin{itemize}
\item In wireless communications, beamforming directs signals toward intended users, minimizes interference, enhances spectral efficiency, extends coverage, and improves overall system performance. Its focus is on optimizing data transmission.

\item In radar, beamforming is primarily used for target detection, localization, and resolution. It emphasizes directional sensitivity and interference rejection to improve detection accuracy and target tracking.
\end{itemize}

A review of 25 years of beamforming progress—from convex and non-convex optimization to modern machine learning approaches, with applications in radar, communications, and beyond—can be found in\cite{elbir2023twenty}.
 In the following, we describe different beamforming methods used in communication, radar, and ISAC.

\subsubsection{Beamforming in Communication} 
In  wireless communications, beamforming is a critical technology enabling higher data rates, improved connectivity, and enhanced spectral efficiency. 
Beamforming can be implemented in digital, analog, or hybrid forms, each with its own advantages and trade-offs.

\begin{table*}[h]
	\caption{Comparison of Digital Beamforming (DBF) Techniques in Communication}
	\label{tab:beamforming}
	\centering
	\renewcommand{\arraystretch}{1.1}
	\begin{tabular}{c c c c c}  
		\toprule
		\begin{minipage}{1.3cm}\centering\textbf{Technique}\end{minipage} &
		\begin{minipage}{3.6cm}\centering\textbf{Description}\end{minipage} &
		\begin{minipage}{6.4cm}\centering\textbf{Advantages \& Disadvantages}\end{minipage} &
		\begin{minipage}{3.5cm}\centering\textbf{Beamforming Formula}\end{minipage} &
		\begin{minipage}{0.4cm}\centering\textbf{Ref.}\end{minipage} \\
		\midrule
		
		\begin{minipage}{1.3cm}\centering SVD\end{minipage} &
		\begin{minipage}{3.6cm}Uses SVD to decompose the correlated channel $\mathbf{H}$ into a set of orthogonal channels.\end{minipage} &
		\begin{minipage}{6.4cm}
			\begin{itemize}[leftmargin=*, itemsep=2pt, parsep=0pt]
				\item[\textbf{+}] Eliminates interference between antennas; optimal for single-user MIMO.
				\item[\textbf{--}] Requires perfect CSI; computationally expensive.
			\end{itemize}
		\end{minipage} &
		\begin{minipage}{3.4cm}
			$\text{Channel:} \; \mathbf{H} = \mathbf{U} \mathbf{\Sigma} \mathbf{V}^H,$ \\
			$\mathbf{W}_{\text{SVD}} = \mathbf{V}$
		\end{minipage} &
		\begin{minipage}{0.5cm}\centering\cite{cover1999elements, Tse2005}\end{minipage} \\
		\midrule
		
		\begin{minipage}{1.3cm}\centering EVD\end{minipage} &
		\begin{minipage}{3.6cm}Uses eigenvalue decomposition on the signal covariance matrix $\mathbf{Q}$ to extract dominant spatial directions for beamforming.\end{minipage} &
		\begin{minipage}{6.4cm}
			\begin{itemize}[leftmargin=*, itemsep=2pt, parsep=0pt]
				\item[\textbf{+}] Effective for spatial filtering and directional beamforming.
				\item[\textbf{--}] Requires statistical knowledge of the channel; high computational cost.
			\end{itemize}
		\end{minipage} &
		\begin{minipage}{3.4cm}
			$\text{Covariance:} \; \mathbf{Q} = \mathbf{V} \mathbf{\Lambda} \mathbf{V}^H,$ \\
			$\mathbf{W}_{\text{EVD}} = \mathbf{V}$
		\end{minipage} &
		\begin{minipage}{0.5cm}\centering\cite{zhang2020rotation}\end{minipage} \\
		\midrule
		
		\begin{minipage}{1.3cm}\centering ZF\end{minipage} &
		\begin{minipage}{3.6cm}Eliminates multi-user interference by inverting the channel matrix.\end{minipage} &
		\begin{minipage}{6.4cm}
			\begin{itemize}[leftmargin=*, itemsep=2pt, parsep=0pt]
				\item[\textbf{+}] Simple implementation, effective at high SNR.
				\item[\textbf{--}] Amplifies noise at low SNR; performance degrades in noisy environments.
			\end{itemize}
		\end{minipage} &
		\begin{minipage}{3.4cm}$\mathbf{W}_{\text{ZF}} = \mathbf{H}^H (\mathbf{H} \mathbf{H}^H)^{-1}$\end{minipage} &
		\begin{minipage}{0.5cm}\centering[3]\end{minipage} \\
		\midrule
		
		\begin{minipage}{1.3cm}\centering MMSE\end{minipage} &
		\begin{minipage}{3.6cm}Balances interference suppression and noise amplification by utilizing a regularization factor $\alpha$ in the channel inversion.\end{minipage} &
		\begin{minipage}{6.4cm}
			\begin{itemize}[leftmargin=*, itemsep=2pt, parsep=0pt]
				\item[\textbf{+}] Performs well at both low and high SNR, more robust than ZF.
				\item[\textbf{--}] Higher computational complexity than ZF.
			\end{itemize}
		\end{minipage} &
		\begin{minipage}{3.4cm}$\mathbf{W}_{\text{MMSE}} = \mathbf{H}^H (\mathbf{H} \mathbf{H}^H + \alpha \mathbf{I})^{-1}$\end{minipage} &
		\begin{minipage}{0.5cm}\centering[4]\end{minipage} \\
		\midrule
		
		\begin{minipage}{1.3cm}\centering MRT/MRC\end{minipage} &
		\begin{minipage}{3.6cm}Maximizes SNR by transmitting/receiving in the direction of the strongest channel gain.\end{minipage} &
		\begin{minipage}{6.4cm}
			\begin{itemize}[leftmargin=*, itemsep=2pt, parsep=0pt]
				\item[\textbf{+}] Maximizes received power, simple to implement.
				\item[\textbf{--}] Does not cancel interference; not suitable for multi-user scenarios.
			\end{itemize}
		\end{minipage} &
		\begin{minipage}{3.4cm}$\mathbf{w}_{\text{MRT}} = \frac{\mathbf{h}^\dagger}{\|\mathbf{h}\|}$\end{minipage} &
		\begin{minipage}{0.5cm}\centering\cite{Lo1999}\end{minipage} \\
		\midrule
		
		\begin{minipage}{1.3cm}\centering Capon (MVDR)\end{minipage} &
		\begin{minipage}{3.6cm}Minimize interference from other directions while keeping an undistorted response in the look direction.\end{minipage} &
		\begin{minipage}{6.4cm}
			\begin{itemize}[leftmargin=*, itemsep=2pt, parsep=0pt]
				\item[\textbf{+}] Strong interference suppression; adaptive to varying conditions.
				\item[\textbf{--}] Requires accurate covariance matrix estimation; sensitive to environmental changes.
			\end{itemize}
		\end{minipage} &
		\begin{minipage}{3.4cm}
		$\mathbf{w}_{\scalebox{0.7}{$\mathrm{MVDR}$}}
		 = \frac{\mathbf{Q}^{-1} \mathbf{a}(\theta_0)}{\mathbf{a}(\theta_0)^H \mathbf{Q}^{-1} \mathbf{a}(\theta_0)} $ \\
			$\mathbf{a}(\theta_0)$: steering vector for direction $\theta_0$, \\
			$\mathbf{Q}$: covariance matrix of the received signal
		\end{minipage} &
		\begin{minipage}{0.5cm}\centering\cite{capon1969high,li2003robust,elbir2023twenty}\end{minipage} \\
		\midrule
		
		\begin{minipage}{1.3cm}\centering ML-based\end{minipage} &
		\begin{minipage}{3.6cm}Uses deep learning to optimize beamforming in dynamic and complex environments.\end{minipage} &
		\begin{minipage}{6.4cm}
			\begin{itemize}[leftmargin=*, itemsep=2pt, parsep=0pt]
				\item[\textbf{+}] Adapts to non-stationary channels; reduces computational overhead.
				\item[\textbf{--}] Needs large datasets for training.
			\end{itemize}
		\end{minipage} &
		\begin{minipage}{3.4cm}AI-based optimization, no explicit formula\end{minipage} &
		\begin{minipage}{0.5cm}\centering\cite{huang2019fast, zhang2021multi, elbir2023twenty}\end{minipage} \\
		\bottomrule
	\end{tabular}
\end{table*}

\paragraph{Digital Beamforming} Digital beamforming (DBF) operates entirely in the baseband domain, leveraging digital signal processing techniques to manipulate signals amplitude/phase  before transmission. Each antenna element is connected to an independent RF chain, allowing precise beam steering and adaptive beamforming capabilities. This flexibility enables multi-user communication and dynamic interference mitigation, making digital beamforming a key block of advanced MIMO systems. However, its implementation requires high-speed analog-to-digital and digital-to-analog  converters, which lead to increased power consumption and hardware complexity. Despite these challenges, DBF continues to be the most adaptable and widely utilized approach for multi-user communication where fine-grained control over multiple simultaneous beams is needed.

DBF in communication uses various techniques depending on system goals, complexity, and  constraints. Singular value decomposition (SVD) and eigenvalue decomposition (EVD) provide optimal performance in single-user MIMO, while zero-forcing (ZF) and MMSE beamforming are widely used in multi-user MIMO to handle interference. Maximum ratio transmission (MRT) and  maximum ratio combining (MRC) are simple techniques, but they lack interference mitigation, whereas Capon beamforming, also known as  minimum variance distortionless response (MVDR) beamforming, is effective for adaptive spatial filtering. With emerging ML–based sensing and beamforming, future beamforming techniques are expected to become increasingly environment-aware, adaptive, and intelligent, paving the way for ISAC and 6G systems.
 Common DBF techniques are summarized in Table~\ref{tab:beamforming}.

\paragraph{Analog Beamforming} 
Analog beamforming, in contrast, controls the phase and amplitude of signals directly in the RF domain using phase shifters before converting them to baseband. This approach forms a single beam per RF chain, making it highly power-efficient and well-suited for high-frequency  mmWave communications. Its lower hardware complexity makes it an attractive choice for cost-sensitive deployments, especially in systems where power consumption is the main concern. However, analog beamforming lacks the ability to generate multiple independent beams simultaneously, limiting its flexibility in supporting multi-user MIMO scenarios. The reduced control over beamforming patterns also poses challenges in dynamically adapting to varying channel conditions, making it less effective in highly dynamic environments.

\paragraph{Hybrid Beamforming} 
Hybrid beamforming compromises between digital and analog techniques, balancing flexibility and efficiency by reducing the number of RF chains \cite{elbir2023twenty} while maintaining some level of digital control. This approach is particularly advantageous in massive MIMO and mmWave systems, where the cost and power consumption of fully digital beamforming become prohibitive. By integrating digital processing at the subarray level and analog processing at the RF front-end, hybrid beamforming allows for multiple beam generation while optimizing energy efficiency. However, its implementation requires sophisticated optimization algorithms to effectively manage the trade-offs between beamforming precision and hardware constraints. The performance of hybrid beamforming depends heavily on antenna architecture and system design, but it represents a promising solution for next-generation wireless networks, including 5G and beyond.

\subsubsection{Beamforming in Radar} 
 Like in communication systems, beamforming in radar is used to direct and focus transmitted energy in specific directions while enhancing target detection and reducing interference.
 It improves performance by shaping the radiation pattern of an antenna array \cite{friedlander2012transmit,Fortunati2020,lipor2014fourier,pfeffer2013fmcw,hassanien2008robust}. However, radar beamforming differs from communication beamforming in objectives, processing techniques, and system constraints. In radar, the primary goal is target detection, tracking, and imaging, while in communication, beamforming aims to enhance data transmission fidelity and spectral efficiency. 
  
 \begin{itemize} 
\item  \textit{Conventional Beamforming:} A fixed-direction technique that uses a static antenna pattern, achieved via mechanical steering (e.g., rotating parabolic dishes) or phased arrays, to direct the beam across a region.

\item  \textit{Adaptive Beamforming (Digital Beamforming):}  Dynamically adjusts antenna weights based on received signal characteristics, optimizing the beampattern in real time for clutter suppression, interference rejection, and multi-beam processing. It is often implemented in phased array radars, provides improved resolution, tracking accuracy, and flexibility compared to fixed beamforming.

\item \textit{Analog Beamforming:}
Similar to communication,  analog beamforming in  radar relies on RF phase shifters to control signal direction. Unlike digital beamforming, which processes signals in the digital domain, analog beamforming operates directly on the analog signals.

\item  \textit{Hybrid Beamforming:} Combines analog and digital techniques for cost-performance balance.

\item  \textit{MIMO Beamforming:} Employs multiple transmit/receive antennas for improved angular resolution and target separation. It excels in low SNR conditions and is particularly useful for multi-target tracking.

\item \textit{SAR and ISAR Beamforming:}
Synthetic aperture radar (SAR)
 and inverse SAR (ISAR) beamforming synthesize a large aperture using motion to achieve high-resolution imaging.  SAR is commonly used in earth observation and remote sensing, while ISAR is applied to moving target imaging, such as ships and aircraft.

 \end{itemize}

\subsubsection{Beamforming in ISAC} 
An ISAC beamformer  must optimize sensing and communication accounting for trade-offs.
In \cite{Peng2024Integrated}, a joint active beamforming design for BS and passive beamforming design for intelligent reflecting surfaces  was proposed to enhance both communication and localization performance in the downlink scenario. Similarly, in \cite{Li2024Mutual}, the authors studied ISAC beamforming design by maximizing the MI between the target response matrix of a point radar target and the echo signals, while ensuring the data rate requirements of the communication users.
Three types of echo interference were considered: no interference, point interference, and extended interference. 

Interference management has also been explored in other studies.  In \cite{Wang2023QoS}, the authors investigated the precoder optimization problem to maximize radar sensing signal-to-interference-plus-noise ratio (SINR),  considering both signal-independent interference and signal-dependent clutter. The proposed approach ensured a minimum SINR at multiple communication users while maintaining per-antenna power constraints.  Additionally, in \cite{Nguyen2023Multiuser}, a communication-centric subcarrier allocation strategy was developed for wideband ISAC systems, where a subset of subcarriers was tailored for radar sensing while the entire bandwidth was used for communications. An efficient hybrid beamforming design was proposed on the basis of this allocation scheme.

 To improve spectral efficiency in ISAC, researchers have explored {ISAC systems with full-duplex communication.\footnote{Note that sensing is inherently full-duplex  as a sensing BS transmits and receives radar signals simultaneously. However, communication can operate in downlink, uplink, or both directions. The latter corresponds to full-duplex communication, while the former two represent half-duplex communication.}
For example, beamforming design for an ISAC system, where the BS performs target detection while simultaneously communicating with two different groups of users in the uplink and downlink, is studied in \cite{He2023FD}.} An  energy-efficient robust beamforming design for {ISAC systems with full-duplex communication} under norm-bounded CSI errors was introduced in \cite{Allu2024Robust}.

 {Robustness under imperfect CSI was studied in \cite{Bazzi2023outage}, which proposed a robust  beamforming for DFRC based on outage SINR probability.}
In \cite{xu2024integrated}, a multi-beam design scheme was proposed where one beam is used to communicate with a single-antenna user in the downlink while one or multiple dedicated sensing beams are used to sense the unknown and random angle parameter of a target. Similarly, \cite{liu2024beam} introduced a beam pattern modulation-embedded hybrid transceiver design for mmWave ISAC systems, where the ISAC transmitter provides multi beams for single-user communication and scanning beams for sensing.

In \cite{Meng2024Multi}, transmit beamforming was studied for a multi-target, multi-user MIMO-ISAC system, where a general form and tight upper bound of sensing MI were derived, along with a multi-objective optimization framework based on user SINR and sensing MI.
 In \cite{Diluka2024NFISAC}, an iterative beamforming algorithm is developed to mitigateinterference between target echo signals while capturing near-field channel characteristics in an ISAC system.
In \cite{He2024MSE}, the authors exploited both the training and transmission signals for sensing, and proposed two MSE-based schemes for training and transmission design.
In \cite{di2025reconfigurable}  holographic-surface–based ultra-massive MIMO and its application to ISAC is investigated.

\subsection{Joint Beamforming and Waveform Design} \label{sec:subsec2}

Joint beamforming and waveform design can be approached  with various objectives in mind. Maximizing the SNR is crucial for radar applications, while minimizing different types of interference (maximizing SINR) is essential for communications quality and transmission rate. 
 In \cite{Mikko2024HBF}, the authors designed transmit precoders and receive combiners to minimize inter-user, intra-user, and radar-communications interference for ISAC systems.
In \cite{Liu2022b}, considering point and extended target scenarios, the authors proposed minimizing the CRLB of radar sensing while ensuring a predefined level of SINR for each communication user.

Constellation design is another related topic.   The waveform and constellation are closely related as constellation serves as the ``alphabet" of a modulation scheme, defining the set of possible symbols, while the waveform is the ``spoken sentence" that delivers those symbols over the air. In OFDM systems, high-rate communication prefers constellations with varying amplitudes—such as QAM—resulting in non-constant amplitude waveforms with higher PAPR but improved spectral efficiency and error performance. In contrast, constant-modulus constellations like PSK yield waveforms with lower PAPR and better auto-correlation properties, making them more suitable for sensing tasks such as radar. Therefore, ISAC systems, waveform and constellation design must strike a balance between the randomness required for communication and the determinism favored by sensing \cite{yang2024constellation}. Moreover, constellation structure plays a key role in enabling effective interference cancellation.

\section{ML-based Waveform and Beamforming Design for ISAC}

 We describe different ways to optimize \eqref{eq:MO}, or its variants in \eqref{eq:weighted} and \eqref{eq:eps},  using learning-based techniques in what follows.

\subsection{Single-Task  Learning-Based Techniques}

  Many of the existing works on ISAC fall in this category.  
 
 Single-task  learning aims to train a network to optimize  $\eta \, f_{\rm sens}(\cdot)    +  \bar \eta \, f_{\rm comm} (\cdot)$  as one task where sensing and communication objectives are optimized concurrently. In this approach,  $\eta$ could be an input parameter in addition to other inputs such as channel coefficients. The weight could also be fixed if there is a preference for one of the objectives. This itself could be performed using supervised, unsupervised, and reinforcement learning. 
 Many of the existing works on ISAC fall in this category.

\subsubsection{Supervised Learning}

In \cite{wu2024efficient}, a graph neural network (GNN)-based binary classifier was designed to determine irrelevant nodes, thereby reducing computational complexity of branch and bound algorithm tailored to the general multi-user transmit beamforming design problem for ISAC systems.
In \cite{patel2024harnessing}, a  framework for sensor-aided beam training in a multi-user mmWave MIMO system is proposed. It designs an ML–based multimodal fusion network trained on data from camera, LiDAR, and position sensors, along with the beamspace representations of the  channel.
In \cite{Hu2024ISAC},  a  model-based ML approach was proposed for ISAC receiver design using a simplified sliding transformer.
Li \textit{et al.} \cite{li2023graph}  proposed a GNN-based approach using heterogeneous graph representation to jointly optimize service mode selection and target association in THz vehicular networks, aiming to maximize communication data rates while meeting sensing requirements.
In \cite{li2024machine}, a model-driven regression network combining a multi-layer neural network and a single-layer perceptron is proposed to enhance near-field emitter localization accuracy.

\subsubsection{Unsupervised Learning} 


Several studies have leveraged deep learning to enhance predictive beamforming. To enhance hybrid beamforming, \cite{nguyen2023joint} proposed a modified projected gradient ascent method and an interpretable deep unfolded projected gradient ascent framework.
In \cite{Liu2022Learning}  a convolutional long short-term memory (LSTM) network is proposed that uses historical channels for predictive beamforming in ISAC vehicular networks, while \cite{Jang2024Neural} uses a neural network for statistical joint optimization of beamforming and localization based on environmental data from near-field channel estimation.

In \cite{Qi2024Deep}, a joint design of sensing  waveform and communication receive beamforming was proposed to mitigate mutual interference in a non-orthogonal uplink ISAC system,  with an unsupervised deep neural network (DNN) structure.  Using deep learning to learn key parameters of the weighted MMSE, \cite{Jin2023Model} proposed manifold optimization for hybrid precoding in mmWave multiuser MIMO systems.
In \cite{liang2024data}, the authors addressed Gaussian channel uncertainties by developing a modified optimal beamforming structure and a specialized GNN for bipartite message-passing inference.

In \cite{Lavi2023Learn}, a deep unrolling approach was introduced to design downlink hybrid MIMO precoders with speed, robustness, and interpretability. Similarly, \cite{Zhu2023Learning} proposed a beamforming learning architecture by unfolding a parallel gradient projection algorithm. Moreover, generative adversarial networks (GANs) have been applied to ISAC tasks; for example, \cite{Jiang2024Jamming} utilized GANs to adaptively generate jamming waveforms based on specific jamming effect requirements. In \cite{Ddassanayake2025unsupervised} an ISAC waveform design based on unsupervised learning is proposed to address the high complexity of classical optimization, using a custom loss function that balances sensing and communication trade-offs. It demonstrats competitive performance with reduced complexity which is  especially suitable for large antenna arrays.

\subsubsection{Reinforcement Learning}

In \cite{Visa2024MBOL}, the authors introduced a model-based reinforcement learning framework for waveform optimization in multi-carrier ISAC systems operating under dynamic environments, offering enhanced explainability and sample efficiency.
In \cite{yang2024doubly}, a constrained deep reinforcement learning approach was proposed for real-time precoding in ISAC systems, enabling simultaneous target tracking and multi-user communication. To improve training efficiency, the method integrates a primal-dual deep deterministic policy gradient algorithm with the Wolpertinger architecture. Most reinforcement learning-based ISAC schemes focus on simple point-to-point channels, and their extension to multi-user and multi-cell \cite{vaezi2023deep} scenarios are unexplored. 


\subsection{Multi-Task Learning} \label{sec:multi}

While ISAC problems can be addressed using either single-task or multi-task learning, the latter is generally more suitable. 
As an alternative to single-task learning, {we can train a single neural network with shared layers and some task-specific layers, as depicted in Fig.~\ref{fig:multitask},} to handle multiple tasks simultaneously.
 {
Despite it potential, multi-task learning-based ISAC has received considerably less attention in the literature compared to single-task learning ISAC. Representative examples can be found in \cite{liu2024multi}, which jointly optimizes sensing, communication, and computation through a multi-objective formulation, solved using a multi-task learning model with a multiple-gradient descent algorithm. The goal is to simultaneously boost communication rate, enhance sensing precision, and reduce computing power consumption.

A multi-task learning architecture for channel estimation and position estimation in visible light integrated positioning and communication systems is proposed in \cite{wei2022visible}. The architecture consists of a shared network and two task-oriented sub-networks.
Similarly, \cite{yang2025wirelessgpt} investigates multi-task learning in communication and sensing, addressing tasks such as channel estimation, channel prediction, and human activity recognition. Their approach employs a Transformer-based architecture with self-supervised pretraining, followed by task-specific fine-tuning.

}

{As an alternative to the multi-task learning structure in Fig.~\ref{fig:multitask}, which includes shared and task-specific layers, one may use only shared layers and distinguish tasks in a different way, e.g., by providing a binary vector, like in \cite{zhang2021multi}, as an additional input to indicate the task. In this approach, we can train the network to handle multiple tasks without explicit task-specific branches. Specifically, we may train a network} to generate a waveform for $f_{\rm sens}(\cdot)$ and then retrain it for $f_{\rm comm}(\cdot)$.
Even the radar component itself can involve multiple tasks. Traditional radar processing is modular, consisting of separate stages such as signal detection, target tracking, and classification. More specifically, radar processing typically involves sequential stages:

\begin{itemize} \item \textit{Signal Processing}: Tasks such as filtering, detection, and parameter estimation (e.g., range and velocity). \item \textit{Tracking}: Using algorithms like Kalman filtering to track detected targets. \item \textit{Classification}: Identifying target types based on additional signal features. \end{itemize}

Each stage is independently designed and optimized, which can lead to suboptimal overall performance. However, deep learning—particularly multi-task learning—integrates all these stages into a single model that is optimized jointly, from raw data acquisition to final decision-making, for the desired outcomes.
Similarly, the communication component itself may involve multiple tasks.

\subsection{Model-Based Learning}
In both sensing and communications, we often have access to system or process models. However, even when these models are adopted in corresponding optimization problems, acceptable complexity of optimization-based solutions is not guaranteed. In addition, optimization-based algorithms have to be typically manually designed and allow for few or no tuning parameters. Then the model-based learning or learning-to-optimize approach is preferable. Using the model, the learned optimizer is trained over a set of similar optimizees and as a result is able to solve unseen optimizees from the same distribution \cite{chen2022learning}. The following two approaches are typically used to design such learned optimizers. 

\subsubsection{Plug-and-Play} {
When an optimization algorithm is too complex or computationally intensive, the plug-and-play (PnP) approach offers a practical alternative. In this framework, the algorithm is decomposed into modular update steps, and the most computationally demanding components, typically proximal operators or denoisers, are replaced with neural networks  \cite{shlezinger2023model}. These networks are trained on a distribution of similar problems (optimizees), as previously discussed. This method is particularly effective when the structure of the original algorithm supports such modularity, allowing learning-based components to replace specific steps without changing the overall optimization logic.

To date, PnP methods have been primarily applied in the field of image reconstruction and have seen limited adoption in wireless communications or ISAC. Main reason may be that many PnP priors are designed for image processing and may not generalize well to  communication tasks, where the statistical properties and physical constraints differ significantly. One notable exception is   \cite{wan2024deep}, where a deep PnP prior framework is used for multitask channel reconstruction in massive MIMO systems. In this approach, a shared regularization term acts as a learned wireless channel prior. However, the potential of PnP methods for ISAC applications remains largely unexplored and presents an open direction for future research. { Unlike in imaging tasks,  suitable PnP priors are still largely undeveloped. Therefore, while PnP may become useful in future ISAC scenarios, its applicability today remains limited due to the lack of domain-aware priors}


\subsubsection{Algorithm Unrolling} 
Algorithm unrolling is an alternative approach for model-based learning. It consists of three steps \cite{monga2021algorithm}. First, an iterative algorithm is picked/designed for solving an optimization problem. It is preferred that such iterative algorithm is simple, because in the second step, it is unrolled into a neural network. Unrolling stands for representation of computations in one iteration of an algorithm as a layer of a neural network, while the neural network is then build out of several such layers that correspond to the algorithm iterations. Third, a set of neural network parameters to be learned is selected. These may be, for example, step size parameters over the iterations of the corresponding optimization algorithm. 
There are many examples of algorithm unrolling applied to various communications and ISAC problems \cite{nguyen2024joint}. Such problems are typically non-convex, and this is the case when algorithm unrolling shows its power since unrolled (from an algorithm) and trained AI model is less affected by the existence of local minima, escapes them, and as a result finds a better suboptimal solution than an algorithm from which such AI model is unrolled. It also requires fewer layers in an unrolled network architecture than the number of iterations for an algorithm. However, the computations at each layer of an unrolled network are just equivalent to the computations at each iteration of an algorithm. Therefore, the computation saving for the pre-trained network come only from the reduced number of layers compared to the number of iteration of an algorithm, which is a factor reduction only. An illustrative example of algorithm unrolling for hybrid beamforming design problem, which have in detail discussions of the main points discussed above, can be found in Section~\ref{sec:caseII}.

\subsection{End-to-End Learning} \label{sec:e2e}

Using deep autoencoders for end-to-end communication is a novel approach with strong potential to improve BER in both single- and multi-antenna systems \cite{o2017introduction,zhang2021svd}. Unlike traditional systems that separately design modulation, coding, and decoding to minimize BER and maximize spectral efficiency, autoencoder-based methods treat the entire communication chain as a single neural network. This enables joint optimization from transmitter to receiver by learning an end-to-end mapping that minimizes reconstruction error.

Autoencoders have shown promising performance for various communication scenarios \cite{jiang2019turbo, bourtsoulatze2019deep, jang2019deep, Jafarkhani2024noma, nartasilpa2018communications, felix2018ofdm}. They have  also been applied to radar interference mitigation \cite{fuchs2020automotive, waldschmidt2021automotive, brown2024aircraft, chen2021dnn}.
In these designs images are used to train the model.
Autoencoders have also been applied to target detection \cite{wagner2023small}, focusing on designing and training networks that model background statistics and reconstruct input data while suppressing background-induced peaks. Assuming a known background, this approach emphasizes change detection for dynamic targets, such as people and drones. The dataset used consists of simulated high-range resolution profiles for both background and targets. These profiles represent amplitude variations of the reflected radar signal as a function of range, where peaks correspond to objects or features with significant reflectivity. This approach is conceptually similar to anomaly detection, a common application of autoencoders \cite{vaezi2022cellular}, as the model effectively clusters normal (background) and abnormal (background with target) profiles separately.

In the context of ISAC, several studies are referred to as end-to-end ISAC \cite{mateos2022end, mateos2024semi, zheng2024end}. However, these approaches typically apply end-to-end learning only to the communication component, while relying on separate methods—often supervised learning—for the radar component.
Specifically, in \cite{mateos2022end}, the radar task is detecting the presence of a target and  estimating its angle of arrival. The other two papers extend the above approach to the multi-target \cite{mateos2024semi} or multi-user \cite{zheng2024end}  cases.  
In \cite{wang2023deep}, an attention-based LSTM network is proposed to directly map reflected signal samples to beamforming vectors, capturing temporal correlations to maximize achievable rates. Using an unsupervised learning framework, this approach eliminates intermediate estimation errors, outperforming  methods like extended Kalman filtering and deep learning-based two-step designs, while achieving performance close to that of perfect CSI-based beamforming.

In \cite{Wang2023LSTM}, an attention-based LSTM network was proposed for end-to-end predictive beamforming design in ISAC. This scheme  captures the temporal correlation in reflected signal samples and directly determines the beamformer, offering a streamlined  solution.
To address high computational complexity caused by the need to continuously track vehicle motion parameters in vehicular ISAC, the authors in \cite{Zhang2024Predictive} designed an end-to-end predictive beamforming framework based on joint CRLBs. They evaluated sensing performance while ensuring an acceptable downlink communication sum-rate.
Similarly, in \cite{Wang2024Learning}, to adapt precoding policies to dynamic environments, the authors proposed an end-to-end learning method to directly learn downlink multi-user analog and digital hybrid precoders from the received uplink sounding reference signals. In addition, they developed a parallel proactive optimization network to simultaneously learn hybrid precoding.

\section{Representative Examples and Performance Evaluation}\label{sec:performance_analysis}

This section presents {three} learning-based ISAC waveform,  beamformer and constellation designs. We describe the problem setups, explain the learning-based methods, and analyze their effectiveness through the following case studies:

\begin{itemize}
    \item \textbf{Case Study I} employs unsupervised learning to design an ISAC waveform based on a customized loss function. 
    \item {\textbf{Case Study II} utilizes a modified deep unrolling method to design hybrid analog/digital beamformers.}
    \item {\textbf{Case Study III} leverages autoencoders to perform end-to-end constellation design for ISAC.}
\end{itemize} 

\subsection{Case Study I: Unsupervised Learning-Based ISAC  Waveform Design} 
We consider a dual-functional MIMO   system, which enables simultaneous downlink communication with multiple users while performing radar sensing. The BS transmits a unified ISAC waveform for both communication and sensing functionalities.  The BS communicates with $K$ single-antenna users and detects radar targets concurrently. The BS is also equipped with a uniform linear array (ULA) of $M$ antennas.

\subsubsection{The Communication Channel Model}
We denote  the channel between the  BS and   the $k$th user 
by  $\mathbf{h}_k \in \mathbb{C}^{M}$. We use the Rician distribution to model  $\mathbf{h}_k$  such that deterministic line-of-sight (LoS) and random non-LoS channel components can be captured as   
	\begin{eqnarray}\label{eqn:Channel_gk}
		\mathbf{h}_k &=&  \sqrt{ \frac{ K_{h_k} \eta_{h_k} }{K_{h_k} + 1}} \bar{\mathbf{h}}_k +   \sqrt{\frac{ \eta_{h_k} }{K_{h_k} + 1}} \tilde{\mathbf{h}}_k
	\end{eqnarray}
	where $K_{h_k}$ is the Rician factor, which is the power ratio between the LoS and non-LoS channel components. The large-scale fading coefficient, including the path-loss and log-normal shadowing, is represented by $\eta_{h_k}$. 
        In (\ref{eqn:Channel_gk}),  $\bar{\mathbf{h}}_k$ captures the LoS channel component, and it can be modeled as 
\begin{eqnarray}\label{eqn:deterministic channel}
		\bar{\mathbf{h}}_{k} = [1, e^{j 2 \pi \Delta \sin{\theta_k}}, \cdots,  e^{j 2 \pi (M-1)\Delta \sin{\theta_k} }] 
	\end{eqnarray}
	\vspace{-5mm}
	
	\noindent
	where  $\theta_{k}$  is the \textit{angle-of-departure} for user $k$ at the BS, and   $\Delta$ is  the  normalized  spacing (by the wavelength)  between adjacent antennas. 
    In \eqref{eqn:Channel_gk}, $\tilde{\mathbf{h}}_k$  represents the non-LoS channel components, and its elements are modeled via a complex Gaussian distribution  with zero mean and unit variance as 
    $\tilde{\mathbf{h}}_k  \sim\mathcal {CN}\left(\boldsymbol{0}_{{M} \times 1}, \mathbf{I}_{M} \right) $. 
  Using (\ref{eqn:Channel_gk}), the communication channel matrix between the BS antenna array (of $M$ elements) and $K$ single-antenna users  can be aggregated  as 
    \begin{eqnarray}
        \mathbf H = [\mathbf{h}_1, \cdots, \mathbf{h}_k, \cdots, \mathbf{h}_K]^T \in \mathbb{C}^{K \times M}.
    \end{eqnarray}

\subsubsection{The Sensing Channel}\label{sec:sen_cha}

	We model the sensing channel between the  BS and the target by assuming LoS propagation through the   steering vector of the BS antenna array as 
	\begin{eqnarray}\label{eqn:sensing_channel}
		\mathbf{v}(\theta_t) = [1, e^{j 2 \pi \Delta \sin{\theta_t}}, \cdots,  e^{j 2 \pi (M-1)\Delta \sin{\theta_t} }] , 
	\end{eqnarray}
    where $\theta_{t}$ is the  \textit{angle-of-arrival} of the  target with respect to the receive array of the BS.

\subsubsection{ISAC Signal Model}

We denote the   signal matrix  transmitted by the ULA of  BS by  $\mathbf{X} = [\mathbf{x}_1, \cdots , \mathbf{x}_{\tau_d}] \in \mathbb{C}^{M \times \tau_d}$, where $\tau_d$ represents the  length of communication frame. Hence, the signal matrix $\mathbf{X}$ serves as a dual-functional transmit ISAC waveform.  It should be noted that the BS precoder is also embedded in $\mathbf{X}$. 
We assume that the BS perfectly estimates $\mathbf{H}$ by using uplink pilots. Then, we can write the received signal matrix  at the communication users in the downlink  as 
\begin{equation} \label{eqn:received_signal}
    \mathbf{Y} = \mathbf{H}\mathbf{X} + \mathbf{N}
\end{equation}
where $\mathbf{N} = [\mathbf{n}_1, \cdots , \mathbf{n}_{\tau_d}] \in \mathbb{C}^{K \times \tau_d}$ is an Gaussian  noise matrix having entries modeled as  $\mathbf{n}_i \sim \mathcal{CN}(\mathbf{0}, \sigma^2 \mathbf{I}_K), \forall i$. 
Next, we denote the desired received signal matrix for $K$  users by $\mathbf{D}$. Then, we can rewrite  \eqref{eqn:received_signal} as \cite{Liu2018}
\begin{equation} \label{eqn:modified_received_signal}
    \mathbf{Y} = \mathbf{D} + \underbrace{(\mathbf{H}\mathbf{X} - \mathbf{D})}_{\text{MUI}} + \mathbf{N}
\end{equation}
where $\mathbf{D}$  is the aggregated data matrix for $K$ communication users, and its entries  can be drawn from a constellation of any modulation scheme. The second term of (\ref{eqn:modified_received_signal}) represents   multi-user interference (MUI). The total power  of the MUI term in (\ref{eqn:modified_received_signal}) can be written as
\begin{equation} \label{eqn:MUI_energy}
    P_{\text{MUI}} = \|\mathbf{H}\mathbf{X} - \mathbf{D}\|^2_F,
\end{equation}
where \(\|\cdot\|_F \) denotes the Frobenius norm.


The optimization of radar beam patterns is fundamentally related to the design of the correlation matrix of the transmit waveform \cite{Li2007,Stoica2007}. Hence, the correlation matrix of the ISAC  waveform $\mathbf{X}$ can be defined as
\begin{equation} \label{eqn:covariance}
    \mathbf{\Sigma}_X = \frac{1}{\tau_d} \mathbf{X}\mathbf{X}^H.
\end{equation}
Then, the transmit beam pattern of the ISAC waveform $\mathbf{X}$   can be defined  through the correlation matrix 
(\ref{eqn:covariance}) of the ISAC waveform  as \cite{Liu2018}
\begin{equation} \label{eqn:tx_beam_pattern}
    P(\theta) = \mathbf{v}^H (\theta) \mathbf{\Sigma}_X \mathbf{v}(\theta),
\end{equation}
 where $\theta$ denotes the detection angle of the target, and 
$\mathbf{v}(\theta)$ is the array steering vector defined in (\ref{eqn:sensing_channel}).
 
\subsubsection{ISAC Waveform Optimization}

The ISAC waveform $\mathbf{X}$ can be optimized to prioritize communication, sensing, or both functionalities subject to fundamental constraints \cite{Liu2018}. Hence, we review the corresponding optimization problem formulations and their optimization-based solutions. Then, we present an alternative ISAC waveform design technique based on unsupervised learning.

 \begin{itemize}[noitemsep]

     \item \textbf{$\epsilon$-Constraint Method:} The optimization problem can be solved in two different ways such that it prioritizes either communication or sensing functionalities. 
     
     First, we consider the case in which communication functionality is   prioritized. Then, the $\epsilon$-constraint optimization can be formulated as     
\begin{subequations}   
    \begin{eqnarray} \label{eqn:epsilon1_optimization}
       \underset{\mathbf{X}}{\text{minimize}} &&   \|\mathbf{H}\mathbf{X} - \mathbf{D}\|^2_F  \label{eqn:obj_fcn_3_eps1} \\
        \text{subject to} && \|\mathbf{X} - \mathbf{X}_0\|^2_F - \epsilon_{\rm sens} \le 0, \\
         && \frac{1}{\tau_d} \|\mathbf{X}\|^2_F = P_T,
        \label{eqn:cons_3_1_eps1}
    \end{eqnarray}
\end{subequations}
where \( P_T \) denotes the ISAC waveform power per frame, and \( \mathbf{X}_0 \) is a reference radar waveform with desirable auto- and cross-correlation properties for target sensing \cite{Li2018_good_Correlation,Liu2018}. Specifically, \( \mathbf{X}_0 \) can be designed based on \cite[Eq.~(9)]{Liu2018} for omnidirectional beampatterns and \cite[Eq.~(15)]{Liu2018} for directional ones.

Alternatively, by prioritizing the sensing functionality, the optimization problem can be formulated as   
\begin{subequations}   
    \begin{eqnarray} \label{eqn:epsilon2_optimization}
       \underset{\mathbf{X}}{\text{minimize}} && \|\mathbf{X} - \mathbf{X}_0\|^2_F    \label{eqn:obj_fcn_3_eps2} \\
        \text{subject to} && \|\mathbf{H}\mathbf{X} - \mathbf{D}\|^2_F 
 - \epsilon_{\rm comm} \le 0, \\
         && \frac{1}{\tau_d} \|\mathbf{X}\|^2_F = P_T
        \label{eqn:cons_3_1_eps2}
    \end{eqnarray}
\end{subequations}
where the goal is to design 
$\mathbf{X}$ to resemble reference radar waveform $\mathbf{X}_0$. 

The above two optimization problems in (\ref{eqn:epsilon1_optimization}) 
and (\ref{eqn:epsilon2_optimization}) 
can be solved optimally using classical convex  optimization theory \cite{Liu2018,Liu2020,Zhou2022}.  

 
\item \textbf{Weighted Sum Method:} 
The waveforms obtained by solving the above two optimization problems may be skewed towards the communications or sensing, respectively.
In order to  circumvent such circumstances, the ISAC waveforms can be designed by jointly optimizing an objective function computed through a  weighted sum of sensing and communication performance metrics by using multi-objective optimization techniques discussed in Section \ref{sec:ML}. This joint design technique focuses on optimizing the  fundamental   sensing-communication trade-offs. 
For example, a joint objective can combine the weighted sum of MUI power and the mismatch between the designed ISAC waveform $\mathbf X$ and  reference waveforms $\mathbf{X}_0$.
This joint objective function can be then minimized subject to a transmit power constraint as  
     \begin{subequations}   
    \begin{eqnarray} \label{eqn:scalar_optimization}
 \!\!\!    \!\!\!\!\!\!  \!\!\!\!\!\!  \underset{\mathbf{X}}{\text{minimize}} \!\!\! \!\!\!&& \eta \|\mathbf{H}\mathbf{X} - \mathbf{D}\|^2_F + (1-\eta) \|\mathbf{X} - \mathbf{X}_0\|^2_F \label{eqn:obj_fcn_3} \\
    \!\!\! \!\!\!\!\!\! \!\!\!\!\!\!    \text{subject to}\!\!\!\!\!\! && \frac{1}{\tau_d} \|\mathbf{X}\|^2_F = P_T, \label{eqn:P_const}\label{eqn:cons_3_1}
    \end{eqnarray}
\end{subequations}

\noindent where $0 \leq \eta \leq 1$ is a weighting factor   that controls the trade-off between radar sensing and communication performance.  

The optimization problem \eqref{eqn:obj_fcn_3}-\eqref{eqn:cons_3_1} can be equivalently  reformulated as a \textit{quadratically constrained quadratic program}. It can 
  also be   transformed into a \textit{semi-definite program}  by using the \textit{semi-definite relaxation} (SDR) techniques, and then solved using standard optimization solvers as in  \cite{Liu2018}.
 \end{itemize}

The computational complexity of such classical convex optimization-based    procedures can be prohibitively high when system scales in terms of the numbers of BS antennas, users, and targets.  
 For instance, the computational complexity of  the SDR-based optimal solution  proposed in \cite{Luo2010}  for the joint optimization problem  in \eqref{eqn:scalar_optimization}-\eqref{eqn:cons_3_1}  is $\mathcal{O}((M\tau_d)^{3.5})$.

Efficient learning-based techniques can alternatively be used  to provide low-complexity yet sub-optimal    ISAC waveform designs with comparable performance  \cite{Ddassanayake2025unsupervised}.    
Both supervised and unsupervised learning-based techniques  can be applied to solve   optimization problems. However, unsupervised learning eliminates the need for labeled data-sets and, thus, is preferable. Specifically, in ISAC systems,  collecting labeled 
data-sets can be both costly and time consuming. Hence, the unsupervised learning-based techniques are more appealing for solveing optimization problems that appear in ISAC waveform designs.

\subsubsection{A Novel Unsupervised Learning-based Technique}

In learning-based techniques, the design trade-offs between the sensing and communication functionalities can be captured through a custom loss function, which is  formulated as a sum of two weighted performance metrics for sensing and communications \cite{Ddassanayake2025unsupervised}.  For example,  minimization of this  objective function can be translated into the problem of jointly minimizing the power of MUI for communication users, while retaining the desired correlation properties of the ISAC waveform to detect targets \cite{Liu2018}.  
This joint objective function can be minimized subject to a set of constraints,   enforced through multiple Lambda layers, which are custom layers implemented on Pytorch \cite{Lin2020,Qi2024,Ddassanayake2025unsupervised}.  

\noindent\textbf{Lambda layers in DNNs:}
A Lambda layer in DNNs is a customizable layer. It can be used to embed arbitrary mathematical expressions  when constructing    sequential functional application programming interfaces (APIs) designed for  learning models \cite{Lin2020,Qi2024,Ddassanayake2025unsupervised}.
The Lambda layers are particularly useful when the intended learning model cannot be implemented by only using standard layers in APIs. They are   useful for applying custom feature engineering at the input and enforcing model-specific constraints at the output. Both Pytorch and TensorFlow allow Lambda layers to incorporate arbitrary, user-defined function models into a learning architecture, specifically when standard layers do not provide the required functionality.  For instance, in Keras, a Lambda layer can be used to  wrap arbitrary expressions or functions as a custom Keras layer,  thus enabling execution of   specific computations within a custom learning  model architecture. 
In TensorFlow, a Lambda layer can be implemented by using \textit{tf.keras.layers.Lambda} that wraps a custom function.  Pytorch also provides Lambda-like behavior when the \textit{nn.Module} class is used with a custom \textit{forward($\cdot$)} method or through inline lambda functions in functional APIs. Lambda layers are useful when implementing and solving  constraint optimization problems   using deep learning techniques.

 A Lambda layer can   be used to apply a custom activation function, perform a specific mathematical operation, or implement a custom loss function in designing AI-assisted waveforms for ISAC systems \cite{Ddassanayake2025unsupervised}. 
Through unsupervised learning techniques, an unlabeled data-set can be used to train a DNN by iteratively minimizing the custom loss function subject to constraints introduced by the Lambda layers.



%

  \begin{figure}[!t]\centering  
 \includegraphics[width=0.85\columnwidth]{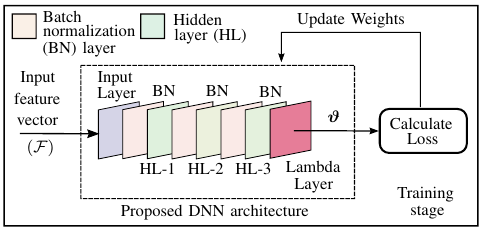}
 	\caption{An unsupervised learning-based DNN for ISAC waveforms \cite{Ddassanayake2025unsupervised}. } \label{fig:NN_archi}
 \end{figure}

 \begin{table}[]
	\centering \vspace{-0mm}
	\caption{The DNN architecture}\label{tab:ANN} \vspace{-0mm}
	\label{tab:DNN structure}
	\begin{tabular}{|c|c|c|}
		\hline
		Layer      & Dimension & Activation function \\ \hline
		Input Layer  & $2N$      & -                \\ \hline
		Hidden Layer 1 & $20N$          & ReLU             \\ \hline
		Hidden Layer 2 & $10N$          & ReLU             \\ \hline
		Hidden Layer 3 & $2M\tau_d$          & Tanh             \\ \hline
		Output/Lambda Layer  & $2M\tau_d$  & -          \\ \hline
	\end{tabular} \vspace{-0mm}
\end{table}

Here, we employ an unsupervised learning-based, single-task fully connected  DNN  to solve the optimization problem \eqref{eqn:obj_fcn_3}---a special form of \eqref{eq:weighted}---and design the waveform. The DNN's   architecture is depicted in Fig. \ref{fig:NN_archi}, and its structure is furnished in Table \ref{tab:ANN}. 


  The ISAC waveform design in (\ref{eqn:obj_fcn_3})-(\ref{eqn:cons_3_1}) depends on the   desired symbol matrix $\mathbf{D}$, channel matrix $\mathbf{H}$, and the reference waveform $ \mathbf{X}_0$. Therefore, the  input to the  DNN consists of $\mathbf{H}$, $\mathbf{D}$, and $\mathbf{X}_0$. 
  The input feature vector $\mathcal{F}  \in \mathbb{R}^{2N\times 1}$, where $N   =  K(M+\tau_d) +  M\tau_d $,   
  is formed by stacking the real and imaginary parts of the entries  of  the complex-valued matrices $\mathbf{H}$, $\mathbf{D}$, and $\mathbf{X}_0$ in the corresponding long column vectors. The input feature size is thus depends on the number of BS antennas ($M$), number of communication users ($K$), and  the communication frame length ($\tau_d$).
  
  The output layer   consists of a column vector stacked by the real and imaginary parts of the learned ISAC waveform.  
    Specifically, we implement this output layer through a custom Lambda layer  to enforce  the constraints of the optimization problem \eqref{eqn:cons_3_1} as
\begin{eqnarray}\label{eqn:constraint_handling}
	\boldsymbol{\vartheta}=
	\begin{cases}
		\boldsymbol{\vartheta},& \|\boldsymbol{\vartheta}\|^2_2 \leq \tau_d P_T\\
		\tau_dP_T \boldsymbol{\vartheta}/ \|\boldsymbol{\vartheta}\|_2, & \|\boldsymbol{\vartheta}\|^2_2 > \tau_d P_T.
	\end{cases}
\end{eqnarray}
  We translate the objective \eqref{eqn:obj_fcn_3} into the unsupervised learning framework by introducing  a custom loss function   to handle the sensing-communication trade-off    as 
\begin{eqnarray} \label{eqn:loss}
  	\mathrm{Loss} \!=\!   \frac{1}{N_s}\!  \sum_{n=1}^{N_s}  \eta\|\mathbf{H}_n\mathbf{X}_n\!-\!\mathbf{D}_n\|_F^2 \!+\! (1\!-\!\eta)\|\mathbf{X}_n\!\!-\!   {\mathbf{X_0}}_n\|_F^2,
\end{eqnarray}
 where $N_s$ is the mini-batch size. For the $n$th training  sample, $\mathbf{H}_n$ is the channel matrix, $\mathbf{D}_n$ is the desired symbol matrix, 
${\mathbf{X_0}}_n$ is the reference waveform, and 
 $\mathbf{X}_n$ is the optimized output waveform.  
 
During the offline training phase, we train the DNN   using  an unlabeled data-set containing input feature vector samples $\mathcal{F}$. We generated this dateset consisting of $\mathcal{F}$ by following the procedure below. 
The  channel matrices $\mathbf{H}$  are generated  according to the channel model in (\ref{eqn:Channel_gk}) by using Monte-Carlo simulations. The  user symbol matrices  $\mathbf{D}$ are generated  by using a unit-power QPSK  modulation scheme. 
We generate the desired sensing waveform $\mathbf{X}_0$ in (\ref{eqn:obj_fcn_3}) and (\ref{eqn:loss})  by  adhering to ideal omnidirectional/directional waveforms by following \cite{Liu2020}. Then, the input feature vector $\mathcal{F}$ 
can be  obtained by stacking the real and imaginary pars of $\mathbf{H}$, $\mathbf{D}$, and $ \mathbf{X}_0$ for a particular realization. The corresponding  dataset can then be generated by stacking multiple samples of input feature vectors $\mathcal{F}$.
	The mini-batch training is utilized. The DNN model weights can be updated after each batch by applying  Adam optimizer. It minimizes  the custom loss function  \eqref{eqn:loss} via an iterative weight update procedure depicted in Fig.~\ref{fig:NN_archi} in Pytorch. 	
	First,  the DNN model is trained and tested by using our dataset. Then, the pre-trained model is used for online prediction of the ISAC waveforms based on new input feature vectors consisting of $\mathbf{H}$, $\mathbf{D}$, and $ {\mathbf{X}}_0$. 
 

\subsubsection{Communication Performance Metrics}
We use the downlink achievable sum rate to evaluate the performance of the designed ISAC waveform.
 The custom cost function aims to minimize the  MUI power (\ref{eqn:MUI_energy}), as the achievable  rate is a decreasing function of MUI power.  The  achievable sum rate   can be defined as  
\begin{equation} \label{eqn:user_rate}
	\mathcal R_k  = \sum^K_{k=1} \log[2]{1 + \gamma_k},
\end{equation}
where $\gamma_k$ is the SINR per symbol for the $k$th user in the downlink, and it is given as \cite{Liu2018}
\begin{equation} \label{eqn:SINR}
	\gamma_k  =  \frac{\mathbb{E} \left[ |D_{kq}|^2 \right]}{  \mathbb{E} \left[ \left| \mathbf{h}^T_k \mathbf{x}_q - D_{kq} \right|^2 \right]  + \sigma^2} .
\end{equation}

\noindent Here $D_{kq}$ is the $(k,q)$th symbol of $\mathbf{D} $, and the expectation is evaluated  over symbols  \cite{Liu2018}. The term  $  \mathbb{E}  [ \left| \mathbf{h}^T_k \mathbf{x}_q - D_{kq} \right|^2  ] $ captures the MUI. 
Since $\mathbb{E} \left[ |D_{kq}|^2 \right]$ is fixed for a given constellation with constant energy, the achievable sum rate can be maximized by minimizing the MUI.
Hence, the user SINR (\ref{eqn:SINR}) and achievable sum rate (\ref{eqn:user_rate}) serve as communication performance metrics of the ISAC waveform designs. 

\subsubsection{Sensing Performance Metrics}
Good correlation properties of ISAC waveforms can enhance sensing success. The correlation matrix of the ISAC waveform is given by (\ref{eqn:covariance}). The transmit  beampattern of the ISAC waveform for a given target direction ($\theta$), computed using (\ref{eqn:tx_beam_pattern}), can be used to compare ISAC waveforms. 
The probability of detection, probability of false alarm, and ROC can also be used to evaluate the sensing performance of the designe ISAC waveform. The sensing detector can be designed based on the generalized likelihood ratio test for a binary hypothesis test \cite{Khawar2015}.

{ \subsubsection{Benchmarks for Performance Comparisons}\label{sec:benchmarks}

The performance of the learning-based ISAC waveform designs is compared against the analytical benchmarks, which  have been obtained through classical optimization theory. Our analytical benchmarks are elaborated below. 

\begin{enumerate}
    \item [I.] \textit{Omni-directional sensing-centric waveforms:} 
A typical sensing-centric   waveform   can be designed to  minimize the MUI subject to a predefined beampattern  (or equivalently transmit correlation matrix)  as  \cite{Liu2018}
 \begin{subequations}
	\begin{eqnarray}\label{eqn:sensing_centric}
	\underset{\mathbf{X}}{\text{min}} &&  || \mathbf{H}\mathbf{X} - \mathbf{D} ||^2_F, \label{eqn:sensing_centric_obj} \\
	\text{subject to}  && \frac{1}{\tau_d} \mathbf{X}^H \mathbf{X} =  \mathbf C_{\rm D}, \label{eqn:sensing_centric_const}
	\end{eqnarray}
\end{subequations}

\noindent
where $\mathbf C_{\rm D}$ is the desired transmit correlation matrix of the waveform. An omnidirectional beampattern   is required for  initial radar probing. For this case,  we can set 
$\mathbf C_{\rm D}=  P_T \mathbf{I}/M$ in (\ref{eqn:sensing_centric_const}), where $P_T$ is the total transmit power. 
For an omnidirectional  beampattern, the optimization problem in (\ref{eqn:sensing_centric_obj})-(\ref{eqn:sensing_centric_const}) can be solved optimally to obtain an analytical solution  through SVD \cite{Liu2018}. To this end, the ISAC transmit waveform $(\mathbf X)$ satisfying (\ref{eqn:sensing_centric_obj})-(\ref{eqn:sensing_centric_const}) for an omnidirectional  beampattern is  computed in closed-form as reported in   \cite[Eq. (9)]{Liu2018}.

\item  [II.] \textit{Directional sensing-centric waveforms:}
A directional waveform is required when the radar is in its target tracking mode.
When a directional beampattern is needed, a desired   positive  definite covariance matrix can be assigned to  $\mathbf C_{\rm D}$ in (\ref{eqn:sensing_centric_const}) by minimizing  the mismatch between the desired beampattern and the obtained beampattern from \eqref{eqn:tx_beam_pattern} as reported in \cite{Liu2020}.
An analytical solution for an optional  directional beampattern is obtained  by solving the optimization problem in (\ref{eqn:sensing_centric_obj})-(\ref{eqn:sensing_centric_const}) by using SVD techniques   as shown  in  \cite[Eq. (15)]{Liu2018}.

\item  [III.] \textit{Sensing-communication trade-off based waveforms:}
Owing  to the strict equality constraint (\ref{eqn:sensing_centric_const}),   the sensing-centric    waveform design  in  (\ref{eqn:sensing_centric}) may skew towards optimizing the sensing performance metrics. This design approach may  hinder the achievable rate performance for the communication users. To mitigate such rate losses, in our numerical result comparisons,
the fundamental   sensing-communication trade-off is optimized to design the ISAC waveforms by solving the optimization problem in \eqref{eqn:obj_fcn_3}-\eqref{eqn:cons_3_1}. To this end, the MUI for the communication users and the mismatch between the designed   ($\mathbf X$) and   reference  ($   \mathbf X_0$) ISAC waveforms have been  minimized jointly to obtain an analytical solution for comparison purposes. 
We designed  $\tilde{\mathbf{X}}$ for    omnidirectional and directional beampatterns by following \cite[Eq. (9)]{Liu2018} and \cite[Eq. (15)]{Liu2018}, respectively.  
 Next, an analytical solution for  the trade-off based ISAC waveform is obtained by first reformulating the optimization problem in \eqref{eqn:obj_fcn_3}-\eqref{eqn:cons_3_1} as a  quadratically constrained quadratic program problem  and then  solving it through SDR techniques.  

\item  [IV.] \textit{Ideal communication-centric waveforms:}
A genie-aided zero-MUI waveform is also used to obtain a communication performance upper bound. Specifically, the achievable   rate at the $k$th user by using this zero-MUI is given by $\gamma_k  =    \mathbb{E} \left[ |D_{kq}|^2 \right]/    \sigma^2 $.
 
\end{enumerate}

}

\subsubsection{Numerical Results and Performance Comparisons}

The numerical results and performance comparisons are presented below. The solutions obtained through convex optimization techniques are used as baselines for comparison purposes. 

In our example, the carrier frequency is set to 3.2\,GHz. The 3GPP urban microcell model \cite[Table 5.1]{3GPP2022} is used to model the path loss, and the Rician factors are set to $K\in [1, 3]$. The user-to-BS distance is set to be in $[50, 200]$~m range. Unless otherwise specified, $P_T = 1$\,W, $M= \{16,32\}$, $K=4$, and $\tau_d = 10$ are used. Moreover, a unit power QPSK modulation is adopted to generate the desired symbols in $\mathbf D$. 
 
DNN implemented in Pytorch is used. We generate $10^5$ samples for the training, testing, and prediction stages. With $N   =  K(M+\tau_d) +  M\tau_d$, for $M=16$  the input layer has $2N = 528$ neurons,   and the output layer has $M\tau_d = 160$ neurons. About  60\,\% and 20\,\% of the total number samples are used during the offline training and testing phases, respectively. The  remaining  20\,\%  of generated samples are used to predict the learning-based ISAC waveforms during the online deployment phase. 
For training, Adam optimizer with the maximum epoch size of 500 and mini-batch size of 1000 is used. The initial learning rate for the optimizer is set to $0.001$, and 
the overfitting is circumvented by using early-stopping with patience 20.


\begin{figure}[t!]\centering\vspace{0mm}
	\includegraphics[width=0.43\textwidth]{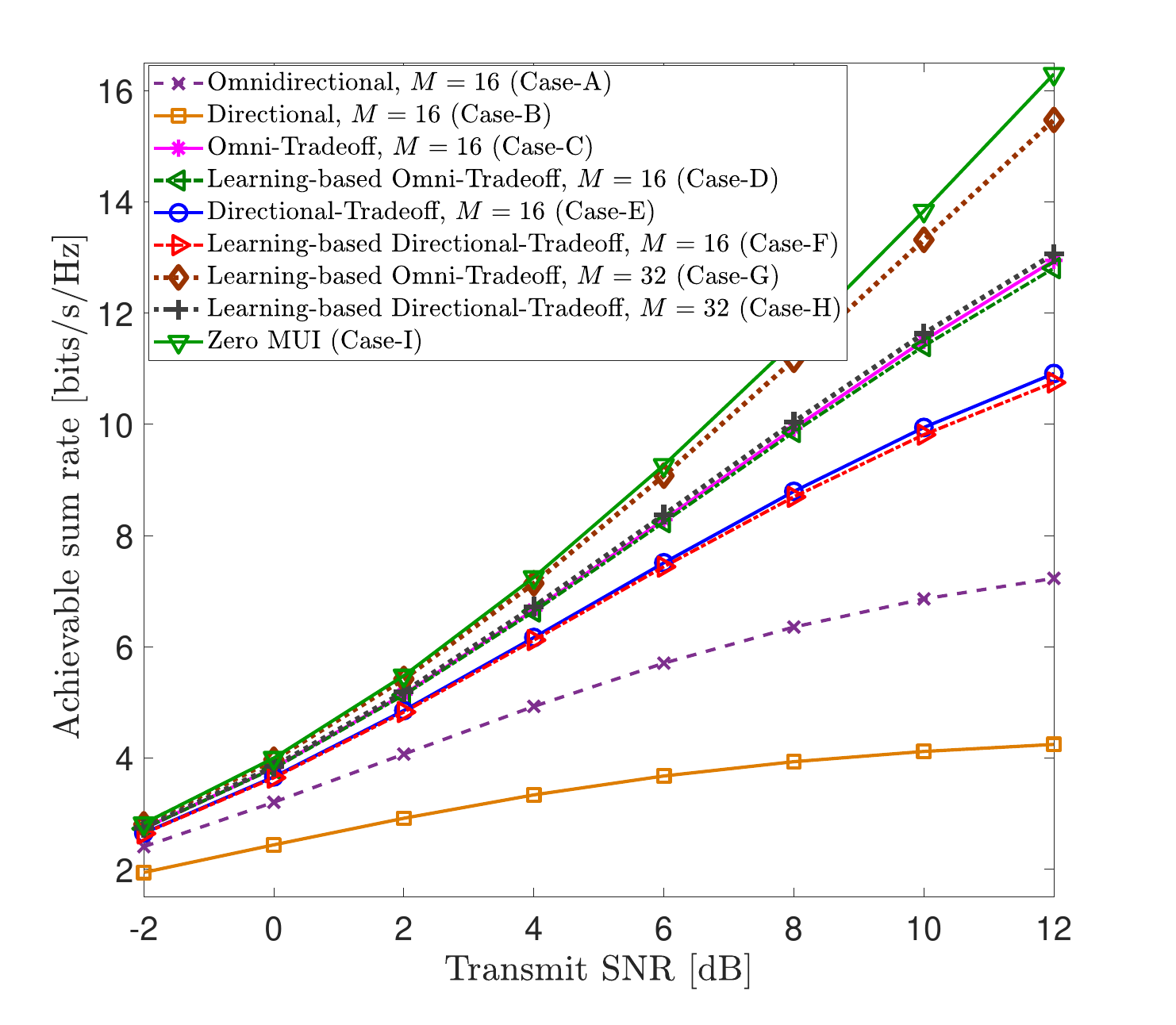}\vspace{-2mm}
	\caption{Comparison of sum rate between analytical and learning-based ISAC waveform designs for $M=\{16,32\}$, $K=4$, $\tau_d = 10$, and $\eta = 0.2$.} \vspace{-0mm}
	\label{fig:Sum_rates}\vspace{-2mm}
\end{figure}
{ 

    \textit{9.1) Comparison  of Communication Performance:}
In Fig.~\ref{fig:Sum_rates}, a  performance comparison in terms of a communication metric  is presented between the  learning-based ISAC waveform designs and classical optimization based analytical waveform designs. Specifically, the achievable sum rate of the ISAC waveforms designed by the proposed unsupervised learning-based approach is compared to that of four analytical waveform benchmarks  described in Section \ref{sec:benchmarks}. 
The channel model (\ref{eqn:Channel_gk}) is used, with Rician factors of 1.5, 2.7, 1.2, and 2.5 for the four users. The BS has $M=16$ or $M=32$   antennas. 
The  genie-aided zero-MUI case (Case-I) is included in Fig.~\ref{fig:Sum_rates} to serve as a sum rate upper bound achieved using an ideal communication-centric ISAC waveform. 
Case-D and Case-F shown in Fig.~\ref{fig:Sum_rates}, we present the sum rate  achievable by the proposed unsupervised learning-based technique for omnidirectional and directional desired beampatterns, respectively. 

It can be unveiled from Fig. \ref{fig:Sum_rates} that despite its sub-optimality, the learning-based ISAC waveform provides a comparable sum rate achievable by  more  complicated optimization-based counterparts.   
To be specific, by comparing Case-C with Case-E and Case-D with Case-F at a transmit SNR of 10\,dB, the sum rate gaps between the learning-based and analytical baseline waveforms are as small as  0.88\,\% and 1.31\,\%, respectively. Nevertheless, the classical optimization-based analytical waveform designs achieve this slight sum rate gain at the expense of significant computational complexity cost as elaborated in Section  \ref{sec:com_complexity}.

We also  plot the sum rates achievable by the learning-based technique for both omnidirectional and directional ISAC waveforms (Case-G and Case-H) by increasing the number of BS antennas  from  16 to 32 to demonstrate the sum rate gains and the scalability of the learning-based ISAC waveform design with respect to the number of BS antennas.

 }
 \begin{figure}[t!]\centering\vspace{0mm}
 	\includegraphics[width=0.45\textwidth]{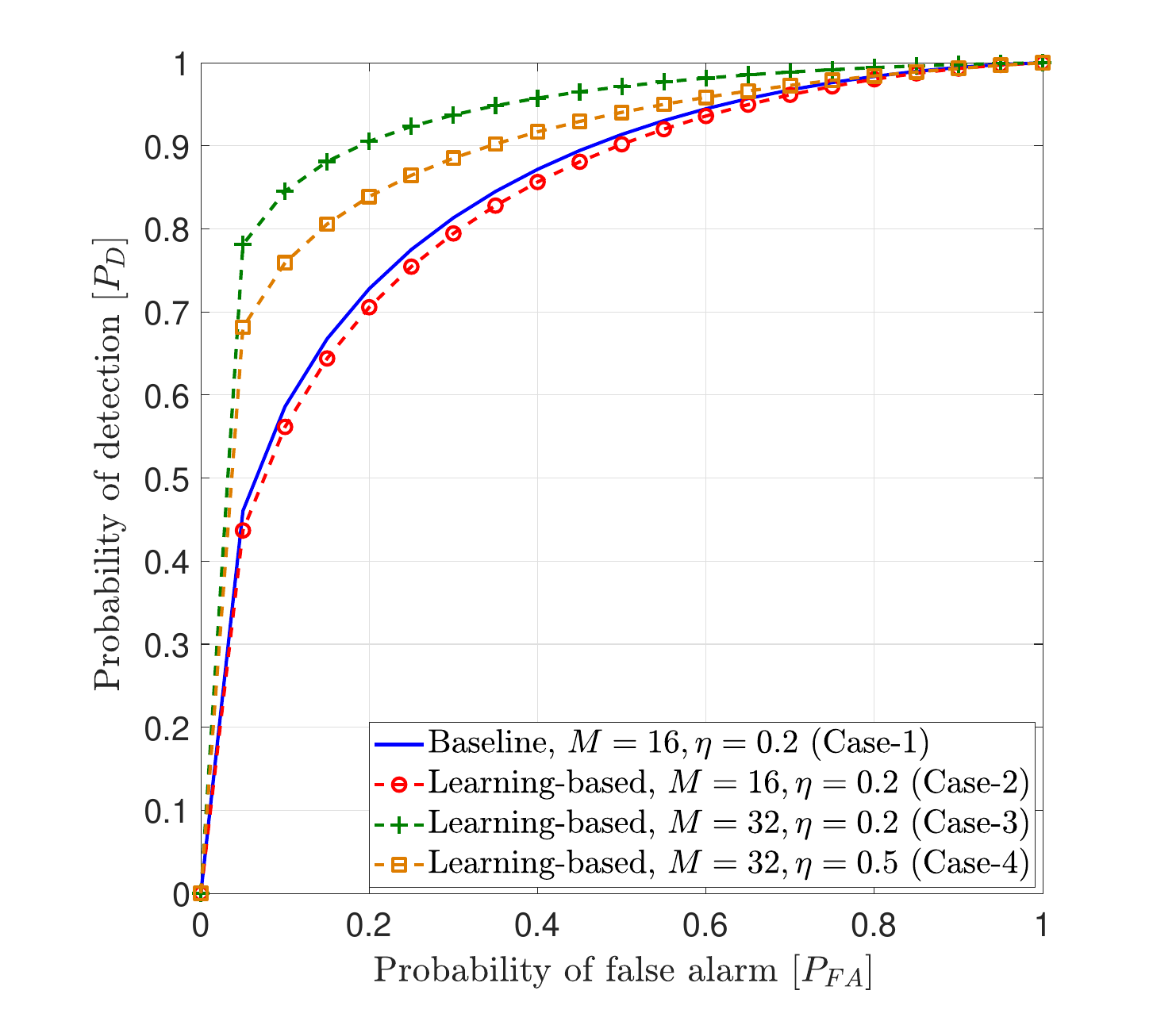}\vspace{-2mm}
 	\caption{Comparison of target detection ROC performance between learning-based and analytical ISAC waveform designs for $M=\{16,32\}$, $K=4$, $\tau_d = 10$, and $\eta = \{0.2, 0.5\}$. } \vspace{-0mm}
 	\label{fig:P_D vs P_FA}\vspace{-2mm}
 \end{figure}

 { 

  \textit{9.2) Comparison  of Sensing Performance:}
In Fig.~\ref{fig:P_D vs P_FA}, a comparison of performance in terms of sensing metrics is given between the ISAC waveforms designed through the proposed learning-based and classical optimization-based techniques. 
Specifically,  Fig.~\ref{fig:P_D vs P_FA} shows the ROC curves of detection probability (\(P_D\)) versus false alarm probability (\(P_{\rm FA}\)) for target detection.
As a benchmark comparison, in Case-1, the detector ROC curves with an ISAC waveform obtained by analytically solving the sensing-communication trade-off based design (\ref{eqn:obj_fcn_3})-(\ref{eqn:cons_3_1}) elaborated in Section \ref{sec:benchmarks} are also plotted.  
Moreover, the target detector proposed in   \cite[Eq. (69)]{Khawar2015} is used to compare the ROC performance. The ROC curves for   Cases-2, 3 and 4 are plotted for the trade-off-based ISAC waveform designs obtained through the proposed unsupervised learning-based approach.
 Through our benchmark comparison, Fig.~\ref{fig:P_D vs P_FA} unveils that the performance gap in terms of sensing metrics between the proposed learning-based and analytical-based ISAC waveform designs is again very small.  For example, at   $P_{\rm FA}= 0.5$, the gap between Case-1 (analytical-based) and Case-2 (learning-based) in terms of $P_{D}$ is about 0.0119. 
   
Figure~\ref{fig:P_D vs P_FA}  also shows how $P_D$ can be improved for a given $P_{\rm FA}$ by prioritizing sensing over the achievable rate by suitably varying the weight factor ($\eta$) in the custom loss function (\ref{eqn:loss}).  For   example, by referring to Case-3 and Case-4  at  $P_{\rm FA} = 0.2$,  we observe that  $P_D$ can be improved by about 7.91\,\%  by decreasing $\eta$ from 0.5 to 0.2. We also show in Fig.~\ref{fig:P_D vs P_FA}  that $P_D$ can be improved by using a larger array at the BS, compared in Case-2  (with $M=16$ BS antennas) with Case-3  (with $M=32$ BS antennas). For example,  at  $P_{\rm FA} = 0.2$, $P_D$ can be  increased by  about 18.88\,\% by doubling the number of BS antennas from 16 to 32.

}

\begin{figure}[t!]\centering\vspace{2mm}
	\includegraphics[width=0.42\textwidth]{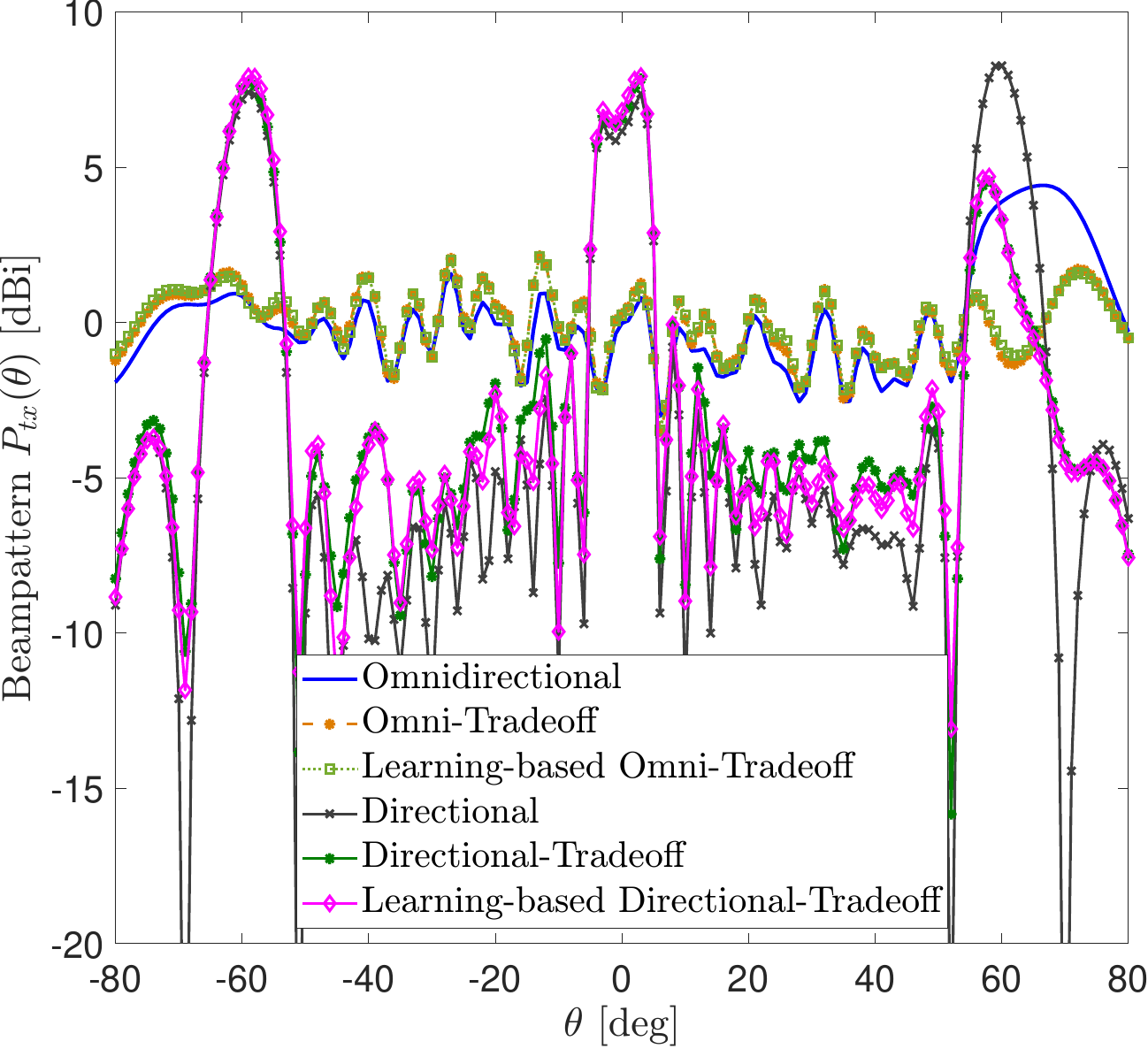}\vspace{-2mm}
	\caption{Comparison of beampatterns between analytical and learning-based ISAC waveform designs for  $M=32$, $K=4$, $\tau_d = 10$, and $\eta = 0.2$.} \vspace{-0mm}
	\label{fig:Beampattern}\vspace{-2mm}
\end{figure}

 \textit{9.3) Comparison  of Unified Beampatterns for ISAC:}
In Fig.~\ref{fig:Beampattern}, the unified ISAC transmit  beampatterns (\ref{eqn:tx_beam_pattern}) of the learning-based ISAC waveform  are compared with those of  the optimization-based analyical benchmarks presented in Section \ref{sec:benchmarks}. We consider three target directions for the  directional beampattern design as  $-\pi/3$, 0, and $\pi/3$. 
Fig.~\ref{fig:Beampattern} reveals that the unified ISAC beampatterns provided by the proposed unsupervised learning-based technique is closely aligned with the optimal analytical beampatterns derived through SDR optimization-based technique for (\ref{eqn:obj_fcn_3})-(\ref{eqn:cons_3_1}).
For both analytical and learning-based techniques,  the maximum gains of the beampatterms appear in the chosen target directions ($-\pi/3$, 0, and $\pi/3$). As further elaborated in Section \ref{sec:com_complexity}, the analytical-based ISAC waveform designs possess a considerable computational complexity over the learning-based alternatives. Hence, the observations in  Fig.~\ref{fig:Beampattern} conclude  that    low-complexity learning-based approaches can be beneficial in circumventing the high computational complexity of optimization-based ISAC waveform designs while achieving near-optimal unified ISAC beampatterns.

 \subsubsection{Computational Complexity Comparison} \label{sec:com_complexity}

 In this subsection, a computational complexity comparison is presented between the proposed learning-based  and analytical optimization-based ISAC waveform design approaches.  
 The underlying  computational complexity  can be compared in terms of floating-point operations. 
 First, the computational complexity of analytical ISAC waveform designs, which have been presented as benchmarks in Section \ref{sec:benchmarks},
 is presented as follows.
     For the optimization-based analytical ISAC waveform design with a sensing-centric omnidirectional beampattern, the complexity scales as \( \mathcal{O}(MK\tau_d + M\tau_d^2) \), due to two matrix multiplications and one SVD operation \cite{Liu2018}. Since the underlying optimization problem can  be cast as an orthogonal procrustes problem (OPP) \cite{Viklands2008},  it converges to a  globally optimal solution with a linear convergence rate.    
 The complexity of the optimization-based ISAC waveform design with sensing-centric directional beampattern scales with $\mathcal{O}(M{\tau_d}^2 + M^2 \tau_d + MK\tau_d + M^3 + M^2K )$ as it has one Cholesky decomposition, one SVD operation, and four matrix multiplications. Again, this optimization procedure relies on an OPP, and hence, it converges to a globally optimal solution with a linear convergence rate.
 The computational  complexity of the trade-off based joint ISAC waveform design in  \eqref{eqn:obj_fcn_3}-\eqref{eqn:cons_3_1} scales as  $\mathcal{O}((M\tau_d)^{3.5})$ when it is solved by leveraging SDR optimization-based technique in \cite{Luo2010}.  This SDR is tight, and its solution is rank-1,  yielding a global optimizer.  The corresponding low-complexity   algorithm uses the Golden-section search, having a  linear convergence rate \cite{Kiefer1953}.
  
Next, we evalute the computational complexity of the proposed 
learning-based ISAC waveform.  
The computational complexity of a fully-connected layer scales as  $\mathcal{O}((2D_I - 1)D_O)$, where $D_I$ and $D_O$ are the input and output dimensions, respectively \cite{Lin2019}. In the worst-case scenario,  they depend on the dimension of the input space $\mathcal{F} \in \mathbb{R}^{2N\times 1}$, where  $N = K(M+\tau_d) +  M\tau_d $. During the ISAC waveform prediction stage,  the overall computational complexity of the   unsupervised learning-based technique  scales as $\mathcal{O}(KM^2\tau_d + KM\tau^2_d + M^2\tau^2_d)$.

The above  analysis reveals that the unsupervised learning-based ISAC waveform design approached proposed to solve (\ref{eqn:obj_fcn_3})-(\ref{eqn:cons_3_1}) has a much lower computational complexity than that of the corresponding optimization-based solutions. Although suboptimal, the  learning-based ISAC waveform designs can be useful as a computationally efficient alternative to more complicated optimization-based ISAC waveform designs, particularly when the system scales in terms of the numbers of BS antennas and the communication users. 

 \subsubsection{Generalization Limitations} \label{sec:gen_limit}

We analyze performance degradation in the proposed unsupervised learning-based unified ISAC waveform design due to generalization limitations. Particularly, two such limitations,  channel changes/aging due to user mobility and system topology changes due to increments of the total number of users, are investigated as follows:

\begin{enumerate}

    \item [I.] \textit{Channel Aging:} The ISAC waveform optimized for an aged channel   $\mathbf{h}_k[n-1]$ is used to  transmit data for a case with user mobility. In this case,  the  true channel $\mathbf{h}_k[n]$ differs from the aged channel $\mathbf{h}_k[n-1]$, and their relationship 
can be modeled as follows:
		\begin{eqnarray}\label{eqn:Channel_hk_at_n}
			\!\!\!\!\!\!\!\!\!\mathbf{h}_k[n] &=&  \sqrt{\frac{ K_{h_k} \eta_{h_k} }{K_{h_k} + 1}} \bar{\mathbf{h}}_k[n] + \sqrt{\frac{ \zeta_{h_k} }{K_{h_k} + 1}} \tilde{\mathbf{h}}_k[n], 
		\end{eqnarray}
where $\mathbf{h}_k[n]$ is a sampled version of (\ref{eqn:Channel_gk}). Under  user mobility, the LoS component $\bar{\mathbf{h}}_k[n]$ is written as    \cite{Demir2020}
		 \begin{eqnarray}\label{eqn:LoS_change}
		  \bar{\mathbf{h}}_k[n]  = \exp{j\theta'_k} \bar{\mathbf{h}}_k[n-1],
		 \end{eqnarray}
 where  $\bar{\mathbf{h}}_k[n-1]$ is the LoS component at time  $(n-1)$. Moreover,   $\theta'_k \in [-\pi, \pi]$ is  a phase shift incurred due to user mobility, and it is assumed to be  common to all antenna elements when  far-field propagation is considered. Next, the effects of channel aging in the non-LoS component can be modeled as \cite{Zheng2021_CA}
		 \begin{eqnarray}\label{eqn:nLoS_change}
		 	\tilde{\mathbf{h}}_k[n] = \chi \tilde{\mathbf{h}}_k[n-1] + \sqrt{1 - \chi^2}\mathbf{e}[n],
		 \end{eqnarray}
		 where $\chi$ is the temporal correlation coefficient, and $\mathbf{e}[n] \sim \mathcal{CN} (\boldsymbol{0}, \mathbf{I}_M)$ is the innovation component at the time instance $n$. Here, the temporal correlation coefficient corresponds to the autocorrelation function of Clarke-Gans/Jakes model. It can be defined as $\chi = \mathrm{J}_0(2 \pi f_D T_s)$, where $\mathrm{J}_0(\cdot)$ is the zeroth-order Bessel function of the first kind, $f_D = \nu f_c / c$ is the maximum Doppler shift,  $T_s$ is the channel sampling duration,   $f_c$ is the carrier frequency, and $c$ is the velocity of light.

\item [II.] \textit{Topology Variations:} Two system topologies are considered by changing the number of users as $K=4$ and $K=8$. 
First, the ISAC waveform,  which  is optimized through our  learning model  for four users, is used for data transmission when there actually exist eight users.
    Second, the optimal waveform for eight users is also obtained by retraining the model and used it for data transmission towards those eight users.

\end{enumerate}
The  performance will be degraded when the learning-based ISAC waveforms are deployed in real-time system operations due to their generalization limitations in dynamic propagation environments. This is because the inferred ISAC waveforms from the ML models that are trained for the aged channels/topologies will no longer be optimal  for dynamic systems with   user mobility and volatility in  concurrent connections.

	 \begin{figure}[t!]\centering\vspace{0mm}
		 	\includegraphics[width=0.43\textwidth]{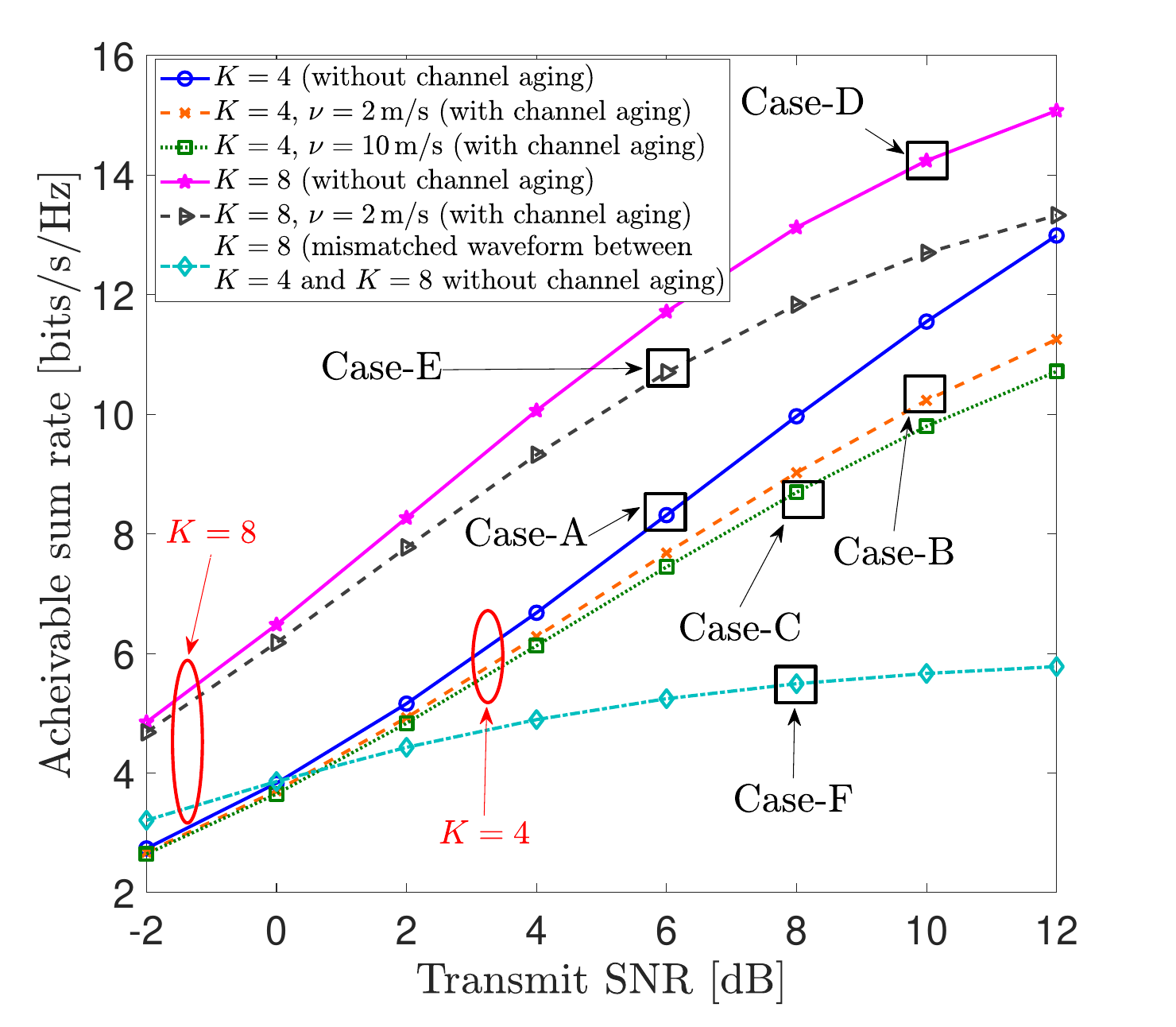}\vspace{-2mm}
		 	\caption{Performance degradations  of the learning-based ISAC waveforms; channel aging and topology changes. System parameters are set to  $M=16$, $K=\{4,8\}$, $\tau_d = 10$, and $\eta = 0.2$.} \vspace{-1mm}
		 	\label{fig:Sum_rates}\vspace{-2mm}
		 \end{figure}

         In Fig. \ref{fig:Sum_rates}, the performance degradations resulting from generalization limitations of the proposed  unsupervised learning-based ISAC waveform $(\mathbf{X})$ design due to channel aging and system topology changes are investigated.  The achievable sum rate curves are plotted for four cases. 
In Case-A, the system consists of  $K=4$ users, and $\mathbf{X}$ is optimized  via the proposed unsupervised learning framework. 
In Case-B, the impact of channel aging is studied with user mobility with a velocity of $\nu  = 2$\,m/s. To this end, the ISAC waveform is optimized for the aged channels  $(\mathbf{h}_k[n-1])$ and used it when  the true channels are given by  $\mathbf{h}_k[n]$. The  relationship between $\mathbf{h}_k[n-1]$ and $\mathbf{h}_k[n]$  is captured through our channel model under user mobility in  \eqref{eqn:Channel_hk_at_n}-\eqref{eqn:nLoS_change} with user mobility velocity of $\nu=2$\,m/s. Fig.   \ref{fig:Sum_rates} clearly unveils that the generalization limitation of the learned ISAC waveform under channel aging degrades the achievable sum rate. For example, when Case-B is  compared to Case-A, at $\rho_d =6$\,dB, a percentage sum rate degradation of 7.62\,\% can be observed. Moreover, in Case-C,  the user mobility velocity is  increased to $\nu=10$\,m/s, and the ISAC waveform is learned for the aged channels. When  Case-C is compared with Case-A  at  $\rho_d=6$\,dB, a higher  percentage sum rate loss of  10.42\,\% can be observed, mainly due to increased user mobility  velocity of Case-C over Case-B.    

Figure \ref{fig:Sum_rates} also shows the performance degradation arising from system topology changes in terms of the number of users $K\in\{4,8\}$ in Cases D, and E, and F.  In Case-D, the ISAC waveform is optimized to $K=8$ users   via our learning-based approach.  In Case-F, the number of users increases to  $K=8$ while the ISAC waveform optimized for four users is still in use at the BS.  
A comparison between Case-D and Case-F reveals that the sum rate degrades by 55.24\,\% at $\rho_d=6$\,dB when the BS adopts the ISAC waveform optimized for four users, while the system topology changes to eight users. In Case-E, the BS uses an ISAC waveform that is optimized to the aged channels under a user mobility with $\nu=2$\,m/s with $K=8$.
When Case-E is compared with Case-D, the sum rate is degraded by 8.66\,\%  at $\rho_d = 6$\,dB when the BS uses an ISAC waveform optimized to aged channels while the true channels have changed due to mobility scenario  with $K=8$. 
The above   sum rate degradations are primarily caused by the mismatched ISAC waveforms that are no longer optimal for suppressing inter-user interference  under extreme user mobility and volatility in concurrent connections.
Hence, the observations in Fig.~\ref{fig:Sum_rates}  reveal that the learning-based ISAC waveform designs are  vulnerable to generalization limitations due to channel aging and system topology changes.


\subsection{Case Study II: Hybrid Beamforming Using Learned \\Optimizer} 
\label{sec:caseII}

{ 
In this case study, we illustrate the algorithm unrolling (also known as algorithm unfolding \cite{Jagannath2021}) approach for designing neural network architectures. To keep the discussion simple and illustrative, we consider an example of hybrid precoding/beamforming design for wireless communications. The extensions to the case of ISAC can be found in \cite{nguyen2024joint}. Further extension to {\it robust} hybrid precoding/beamforming design for ISAC, where channels are assumed to be inaccurately known and channel mismatches are modeled as additive deterministic norm bounded, can be found in \cite{Wang2025Robust}. 

\begin{figure*}[htbp]
	\centering
	\includegraphics[width=.995\linewidth]{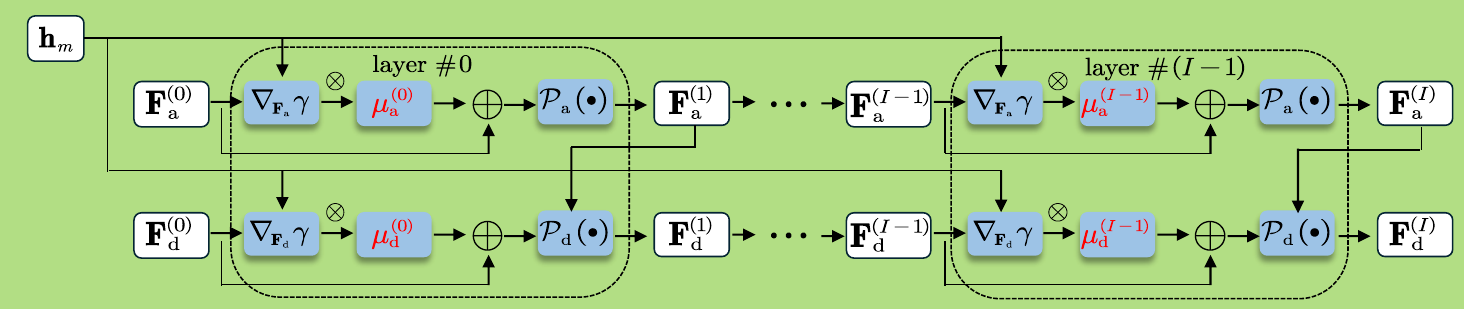}
	\caption{{ Unrolled PGA block illustration. The learnable parameters are marked in red color.}}
	\label{UPGA_architecture}
	
\end{figure*}

\subsubsection{System Model}
Consider a downlink multi-user MIMO system consisting of an $N$-antenna BS and $K$ single-antenna users. The hybrid beamforming architecture composed of analog beamformer $\mathbf{F}\in\mathbb{C}^{N\times L}$ and digital beamformer $\mathbf{W}=[\mathbf{w}_1,\ldots,\mathbf{w}_K]\in\mathbb{C}^{L\times K}$ is employed for data transmission. Here, $L$ denotes the number of RF chains. The communication data symbol is denoted as $\mathbf{s}=[s_1,\ldots,s_K]\in\mathbb{C}^{K\times 1}$ satisfying $\mathbf{s}\sim \mathcal{CN}(\mathbf{0},\mathbf{I}_K)$. The received signal at the $k$-th user side is expressed as
\begin{align}\label{eq_signal_model}
y_k = \mathbf{h}_k^{\mathsf{H}} \mathbf{F} \mathbf{w}_k s_k + \mathbf{h}_k^{\mathsf{H}} \sum_{j \neq k}^{K} \mathbf{F} \mathbf{w}_j s_j + n_k,
\end{align}
where $\mathbf{h}_k\in\mathbb{C}^{N \times 1}$ denotes the channel between BS and the $k$-th user, and $n_k$ is additive white Gaussian noise satisfying $n_k\sim \mathcal{CN}(0,\sigma^2_n)$ with $\sigma^2_n$ being the power of the noise. 

From \eqref{eq_signal_model}, the achievable sum rate   is given by
\begin{align}
R = \sum_{k=1}^{K} \mathrm{log} \left( 1 + \frac{|\mathbf{h}_k^{\mathrm{H}} \mathbf{F} \mathbf{w}_k|^2}{\sum_{j \neq k}^{K} |\mathbf{h}_k^{\mathrm{H}} \mathbf{F} \mathbf{w}_j|^2 + \sigma_n^2} \right).
\end{align}

Our goal is to jointly optimize the hybrid beamforming matrices $\mathbf{F}$ and $\mathbf{W}$ to maximize the achievable sum rate $R$. The corresponding optimization problem is formulated as
\begin{subequations}\label{eq_opt}
\begin{align}
\underset{\mathbf{F}, \mathbf{W}}{\text{maximize}} \quad & R \label{eq_sub_a} \\
\text{subject to} \quad & |\left[\mathbf{F}\right]_{i,j}| = 1, \quad \forall i,j, \label{eq_sub_b} \\
&\|\mathbf{F}\mathbf{W}\|_{\text{F}}^2 = P_t, \label{eq_sub_c}
\end{align}
\end{subequations}
where $|\left[\mathbf{F}\right]_{i,j}|$ denotes the magnitude of the $(i,j)$-th element of matrix $\mathbf{F}$, and $P_t$ is the power budget of the BS. The problem \eqref{eq_opt} is a non-convex problem due to the constant-modulus constraint \eqref{eq_sub_b} and unit-norm constraint \eqref{eq_sub_c}. We resort to projected gradient ascent (PGA) and the corresponding unrolled PGA algorithms to solve it, as will be elaborated in the next section.

\subsubsection{Projected Gradient Ascent Algorithm}

We implement PGA in an alternating-optimization manner. Specifically, keeping $\mathbf{W}$ fixed, $\mathbf{F}$ is updated at the $i$-th iteration via
\begin{align}\label{eq_grad_F}
\bar{\mathbf{F}}^{(i+1)} &= \mathbf{F}^{(i)} + \mu^{(i)}_{\text{F}} \nabla_{\mathbf{F}} R \big|_{\mathbf{F} = \mathbf{F}^{(i)}}, \\
\left[\mathbf{F}^{(i+1)}\right]_{i,j} &= \frac{\left[\bar{\mathbf{F}}^{(i+1)}\right]_{i,j}}{\left|\left[\bar{\mathbf{F}}^{(i+1)}\right]_{i,j}\right|}, \quad \forall i,j,
\end{align}
where $\mu^{(i)}_{\text{F}}$ is the step size for updating $\mathbf{F}$. 
Similarly, freezing $\mathbf{F}$, $\mathbf{W}$ is updated at the $i$-th iteration via 
\begin{align}\label{eq_grad_W}
\bar{\mathbf{W}}^{(i+1)} &= \mathbf{W}^{(i)} + \lambda^{(i)}_{\mathrm{W}} \nabla_{\mathbf{W}} R \big|_{\mathbf{W}=\mathbf{W}^{(i)}}, \\
\mathbf{W}^{(i+1)} &=  \frac{\sqrt{P_t}\bar{\mathbf{W}}^{(i+1)}}{\left\|\mathbf{F}^{(i+1)} \bar{\mathbf{W}}^{(i+1)}\right\|_{\text{F}}},
\end{align}
where $\mu^{(i)}_{\text{W}}$ is the step size for updating $\mathbf{W}$. The gradients $\nabla_{\mathbf{F}} R$ and $\nabla_{\mathbf{W}} R$ in \eqref{eq_grad_F} and \eqref{eq_grad_W} are expressed as~\cite{Nguyen2023MIMO,Nguyen2024ISAC}
\begin{align}
\nabla_{\mathbf{F}} R &=\frac{1}{\ln 2} \sum_{k=1}^{K}\left( \frac{\tilde{\mathbf{H}}_k \mathbf{FV}}{ \text{tr}\left(\mathbf{Z} \tilde{\mathbf{H}}_k\right) + \sigma_n^2 } -  \frac{\tilde{\mathbf{H}}_k\mathbf{F} \mathbf{V}_{\bar{k}}}{ \text{tr}\left(\mathbf{Z}_{\bar{k}} \tilde{\mathbf{H}}_k\right) + \sigma_n^2 }\right),
\\
\nabla_{\mathbf{W}} R &= \frac{1}{\ln 2}\sum_{k=1}^{K} \left(\frac{\bar{\mathbf{H}}_k \mathbf{W}}{ \text{tr}\left(\mathbf{V} \bar{\mathbf{H}}_k\right) + \sigma_n^2 } -  \frac{\bar{\mathbf{H}}_k \mathbf{W}_k}{ \text{tr}\left(\mathbf{V}_{\bar{k}} \bar{\mathbf{H}}_k\right) + \sigma_n^2}\right),
\end{align}
where $\mathbf{Z}=\mathbf{F}\mathbf{V}\mathbf{F}^{\mathsf{H}}\in \mathbb{C}^{N\times N}$, $\mathbf{Z}_{\bar{k}}=\mathbf{F}\mathbf{V}_{\bar{k}}\mathbf{F}^{\mathsf{H}}\in \mathbb{C}^{N\times N}$, $\mathbf{V}=\mathbf{W}\mathbf{W}^{\mathsf{H}}\in \mathbb{C}^{L\times L}$, $\mathbf{V}_{\bar{k}}=\mathbf{W}_{\bar{k}}\mathbf{W}^{\mathsf{H}}_{\bar{k}}\in \mathbb{C}^{L\times L}$, $\tilde{\mathbf{H}}_k=\mathbf{h}_k\mathbf{h}^{\mathsf{H}}_k\in \mathbb{C}^{N\times N}$, $\bar{\mathbf{H}}_k=\mathbf{F}^{\mathsf{H}}\tilde{\mathbf{H}}_k\mathbf{F}\in \mathbb{C}^{L\times L}$. Here, $\mathbf{W}_{\bar{k}}$ is obtained by replacing the $k$-th column of $\mathbf{W}$ by an all-zero vector.

\subsubsection{Unrolled PGA-based Design}
The PGA algorithm is capable of optimizing the hybrid transmit beamforming matrices $\mathbf{F}$ and $\mathbf{W}$ within a predefined number of layers. However, the key challenge in PGA algorithm stems from determining the optimal hyperparameters (i.e., step sizes $\mu_{\text{F}}$ and $\mu_{\text{W}}$ in our context), which can significantly affect the performance and convergence of the algorithm \cite{Pesquet2024unrolled}.
To this end, we resort to algorithm unrolling technique \cite{Monga2021Unrolling} to tune these hyperparameters in an unsupervised learning manner. The main principle of this technique is to transform each iteration of PGA algorithm into a network layer and stack these layers together \cite{Monga2021Unrolling,Pesquet2024unrolled}, resulting a multi-layer neural network, as shown in Fig. \ref{UPGA_architecture}. In the unrolled network, the trainable parameters, i.e., the step sizes for $I$ iterations, are defined as $\boldsymbol{\Phi}=[\boldsymbol{\mu}_1,\ldots,\boldsymbol{\mu}_I]$, where $\boldsymbol{\mu}_i=[\mu^{(i)}_\text{F},\,\mu^{(i)}_\text{W}]^{\mathsf{T}}$ stands for the step size vector to be learned for the $i$-th layer. To train the unrolled network, we employ a loss function that takes into account the outputs across all intermediate layers, expressed as \cite{Lavi2023unrolling}
\begin{align}
\mathcal{L}(\boldsymbol{\Phi})=-\frac{1}{|\mathcal{D}|}\sum^{|\mathcal{D}|}_{d=1}\frac{1}{I}\sum^{I}_{i=1}\log{1+i}R(\mathbf{h}^{(d)}_k,\mathbf{F}^{(i)},\mathbf{W}^{(i)})
\end{align}
in which $\mathcal{D}$ is the data set containing communication channel realizations.
Then the hyperparameter matrix $\boldsymbol{\Phi}$ is tuned via
\begin{align}
\boldsymbol{\Phi}^{\star} = \mathrm{arg} \min_{\boldsymbol{\Phi}}\mathcal{L}(\boldsymbol{\Phi}).
\end{align}

\subsubsection{Simulation}
In this section, the performance of unrolled neural network tailored to hybrid beamforming design is tested. The system parameters are set as $N=16$, $L=6$, and $K=4$. The noise variance is set as $\sigma^2_n=1$. The SNR is defined as $\text{SNR}=10\mathrm{log}_{10} (P_t/\sigma^2_n)$. The channel $\{\mathbf{h}_k\}^{K}_{k=1}$ is modeled as Rayleigh fading channel. The sizes of data sets for training and testing are 1000 and 100. For training phase, stochastic gradient descent optimizer with a learning rate of 0.005 is used. The fixed step sizes for the PGA  and initial step sizes for unrolled PGA are both set as 0.05. In Figs. \ref{fig_SR_iter} and \ref{fig_SR_snr}, the hybrid beamforming  design using unrolled neural network is named as `unrolled PGA', and PGA-based hybrid beamforming  design as `PGA'. 

The achievable sum rate versus the number of iterations/layers for $\text{SNR}=10\,\text{dB}$ is shown in Fig. \ref{fig_SR_iter}. Both the PGA and unrolled PGA are observed to stabilize at distinct constant values, demonstrating the strong convergence of the proposed schemes. Notably, the unrolled PGA converges much faster than the benchmark PGA, highlighting the effectiveness of algorithm unrolling in optimizing the hybrid transmit beamformer. The reason for unrolled PGA to outperform the PGA is that the unrolled PGA algorithm is capable of learning the update rules from data, allowing it to fit the objective function. This capacity allows the unrolled PGA algorithm to escape local optima more effectively than the PGA \cite{Wang2025Robust}.  


\begin{figure}[tbp]

\centering 
\includegraphics[width=0.4\textwidth]{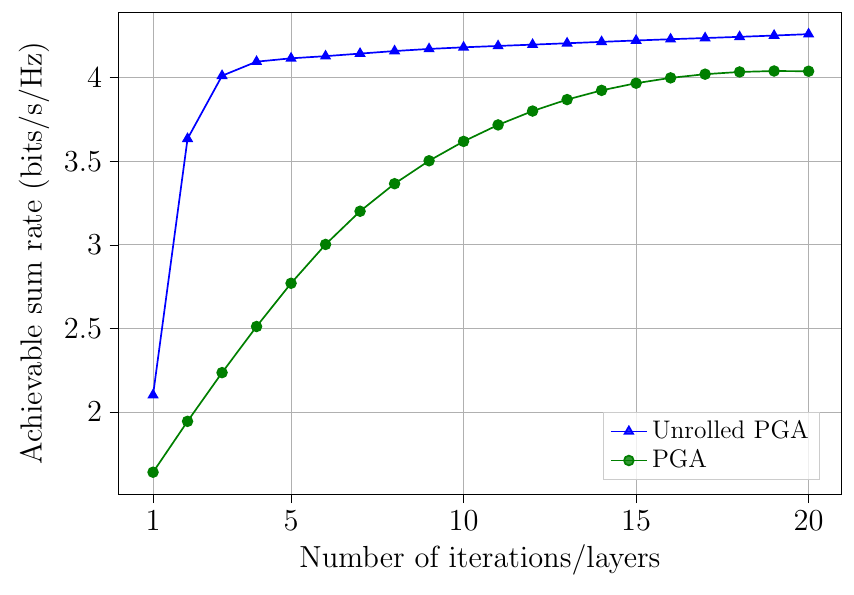}

	\caption{{Achievable sum rate versus number of iterations/layers with $\text{SNR}=10~\text{dB}$, $N=16$, $K=4$, and $L=6$.}}
	\label{fig_SR_iter}
\end{figure}

\begin{figure}[tbp]
\centering 
\includegraphics[width=0.4\textwidth]{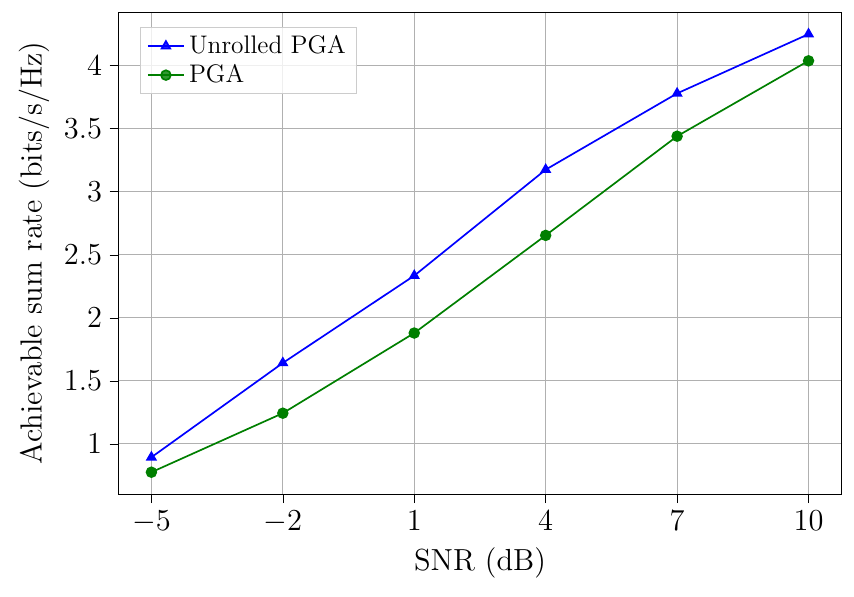}
	\caption{{ Achievable sum rate versus SNR with $N=16$, $L=6$, and $K=4$.}}
	\label{fig_SR_snr}

\end{figure}

%
%

Figure~\ref{fig_SR_snr} shows the achievable sum rate versus SNR varying from $-5$ to $10$\,dB. 
It is observed that the unrolled PGA consistently achieves a higher sum rate than the standard PGA across the entire SNR range. This implies that the unrolled PGA maintains superior performance over the benchmark PGA even at low SNR levels.


\subsection{Case Study III: Autoencoder-Based ISAC  Constellation Design} 
\label{sec:constellation}

A key enabler in realizing the promise of ISAC is the development of waveforms capable of supporting both information transmission and target sensing. In this context, the design of modulation constellations plays a pivotal role, as it directly influences the performance of both communication and sensing functionalities.

\begin{figure*}     
\centering
\scalebox{0.85}{
\begin{tikzpicture}[
    node distance=10pt and 15pt,
    block/.style={
        draw, 
        rectangle, 
        rounded corners, 
        minimum height=20pt,
        text width=2.6cm,
        align=center
    },
    lbl/.style={font=\footnotesize, text=magenta},
    arrow/.style={-Stealth, semithick},
    dashbox/.style={
        draw=black, 
        dashed, 
        thick, 
        rounded corners,
        inner sep=8pt
    }
]

\node (input) {$\mathbf{m} \in \{0,1\}^K$};
\node[block, right=of input] (encoder) {Encoder {\color{cyan}$\Psi_{E}$}};
\node[right= 1cm of encoder] (x) {$x \in \mathbb{C}$};

\node[block, above right=1cm and -.5cm of x] (comm) {$f_{Y|X}(y|x)$};
\node[block, right=1cm of comm] (decoder) {Decoder {\color{cyan}$\Psi_{\rm comm}$}};
\node[right=of decoder] (output) {$\hat{\mathbf{m}} \in \{0,1\}^K$};

\node[lbl, above=of encoder] (txlabel) {ISAC BS};
\node[lbl, below=of decoder] (rxlabel) {Comm. receiver (UE)};
\node[lbl, below=of comm] (commlabel) {Comm. channel};

\node[block, below right=1.4cm and 0.5cm of x] (radar) {$f_{Z|X}(z|x)$};
\node[block, left=2.6cm of radar] (detector) {Presence Detector \\ {\color{cyan}$\Psi_{\rm radar}$}};
\node[lbl, above=of radar] (radarlabel) {Radar channel};
\node[left= 1cm of detector] (output_R) {$P_D$};

\node[below left=-0.2cm and -1cm of output] (cellphone) {\includegraphics[width=1.4cm]{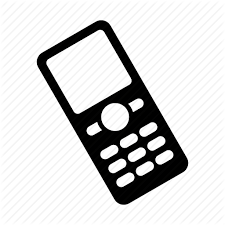}};
\node[right=0.2cm of radar] (target) {\includegraphics[width=2cm]{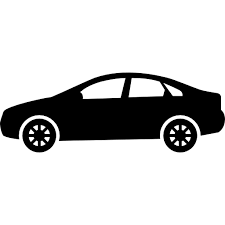}};
\node[lbl, below= -.7cm of target] (ratgetlabel) {Radar target};

\begin{scope}[arrow]
    \draw (input) -- (encoder);
    \draw (encoder) -- (x);
    \draw (x) -| (comm);
    \draw (x) -| (radar);
    \draw (comm) -- node[above] {$y$} (decoder);
    \draw (decoder) -- (output);
    \draw (detector) -- (output_R);
    \draw (radar) -- node[above] {$z$} (detector);
\end{scope}

\node[dashbox, fit=(encoder) (detector)] {};

\end{tikzpicture}
}
  \caption{{Block diagram of an ISAC system consisting of an ISAC base station (transmitter and radar), a UE communication receiver, and a car as the radar target.
 The objective is to transmit a message $\mathbf{m}$ to UE and detect the presence of the target using a single transmitted signal $x$. The received signals after passing through the communication and radar channels are denoted by $y$ and $z$, respectively, where $y \sim f_{Y|X}(y|x)$ and $z \sim f_{Z|X}(z|x)$. The neural networks $\Psi_{E}$, $\Psi_{\rm comm}$, and $\Psi_{\rm radar}$ represent the encoder, communication decoder, and radar detector.}}
\label{fig:ISAC_diagram}

\end{figure*}

A fundamental challenge in ISAC arises from the inherent trade-off between the randomness that benefits communication and the deterministic structure preferred for accurate sensing. Traditional constellations typically offer either high communication efficiency or high sensing accuracy—but not both. Specifically, QAM is optimized for communication, while PSK is more suitable for sensing \cite{yang2023random,du2024pcs,Keshavarz2025const}. Thus, constellation shaping has emerged crucial for balancing sensing and communication requirements in ISAC systems, and forming a unified waveform.

Waveform design in ISAC is typically formulated as an optimization problem that balances communication and sensing objectives, such as maximizing data rate and minimizing BER for communication, while improving range and velocity resolution for sensing, through a joint cost function. Specifically, optimization problem \eqref{eq:weightedsum} can incorporate meaningful sensing and communication metrics for waveform design. In general, the metrics listed in Table~\ref{table:comm:metric} and Table~\ref{table:radar:metric} may be used. It is often desirable to design a waveform that simultaneously minimizes the BER for communication and maximizes  \( P_D \)  for radar.

In this case study, we present an example of how such an optimization problem can be solved using machine learning.  
Before presenting the design details, we first describe the problem and the system model.

 

Consider the ISAC system illustrated in Fig.~\ref{fig:ISAC_diagram}, where, at each channel use, the ISAC base station aims to transmit a binary message vector \( \boldsymbol{m} \in \{0,1\}^K \) and simultaneously sense a target using a single transmitted signal.
Since each input \( \boldsymbol{m} \) is a binary vector of length \( K \), there are \( M = 2^K \) possible distinct message vectors. Traditional approaches may use an \( M \)-ary constellation, such as \( M \)-PSK, to map each message to a constellation point \( x(\boldsymbol{m}) \in \mathbb{C} \) and transmit it.

 The transmitted signal \( x(\boldsymbol{m}) \in \mathbb{C} \) passes through both the communication channel \( f_{Y|X}(y | x) \) and the sensing channel \( f_{Z|X}(z | x) \). Assuming that both channels are AWGN channels, the communication and sensing signals are respectively given by 
\begin{subequations}
\begin{align}
y &= x(\boldsymbol{m}) + n_{\rm comm},  \label{eq:comm-channel}\\
z &= x(\boldsymbol{m}) + n_{\rm radar}, \label{eq:radar-channel}
\end{align}
\end{subequations}
where \( n_{\rm comm} \sim \mathcal{CN}(0, \sigma^2_c) \) and \( n_{\rm radar} \sim \mathcal{CN}(0, \sigma^2_{\rm radar}) \) are the complex Gaussian noises associated with each channel, respectively.

 ISAC research \cite{yang2023random,du2024pcs,Keshavarz2025const}, shows that constellation structure plays a key role in balancing performance trade-offs between these two functions. While  QAM-like constellations favor communication and constant modulus constellations like PSK benefits sensing. Hence, intermediate designs like multi-ring constellations offer a promising balance, and thus are more suited for joint sensing and communication.

In the following, we illustrate how a combination of supervised and unsupervised   learning can be used to design effective constellations for ISAC.

\subsubsection{Communication}
The goal of communication is to send a binary message vector \( \boldsymbol{m} \) such that it can be reconstructed at the UE with minimal errors.
To enable this, we can design an end-to-end communication system based on an autoencoder architecture. In Fig.~\ref{fig:ISAC_diagram}, the encoder and decoder neural networks are denoted by \( \Psi_{\rm E} \) and \( \Psi_{\rm comm} \), respectively.

To minimize the BER, the autoencoder must learn to generate \( M = 2^K \) well-separated clusters in the latent space. Thus, the autoencoder effectively designs an \( M \)-ary constellation. For an AWGN channel, a  well-trained autoencoder  should maximize the minimum distance between constellation points to improve robustness against noise. At each channel use, one of these constellation points is then transmitted.

The \textit{autoencoder-based end-to-end design} framework has been successfully applied to various communication scenarios (e.g., \cite{o2017introduction,zhang2021svd,zhang2023wcnc}). The objective is to train a DNN that accurately estimates the transmitted bits, i.e., to minimize the number of errors. To this end, the \textit{cross-entropy loss function} is commonly used to guide the training process.

Let \( \boldsymbol{m}_j \) denote the \( j \)th message bit, and let \( \hat{\boldsymbol{m}}_j \) represent the corresponding likelihood estimate produced by the neural network  \( \Psi_{\rm comm} \). The cross-entropy loss is then defined as  
\begin{align} \label{eq:comm_loss}
\mathcal{L}_{\rm comm} = -\mathbb{E}\Big[\sum_{j=1}^{K} \boldsymbol{m}_j \log{\hat{\boldsymbol{m}}_j}\Big],
\end{align}
where \( \mathbb{E} \) denotes the expectation operator.

\subsubsection{Sensing}

For sensing, we assume a simple task where the BS acts as a radar and wants to detect a target, as shown in Fig.~\ref{fig:ISAC_diagram}. Assume that that the \textit{target} be present ($T=1$) or absent ($T=0$) with equal probability, i.e., \( p(T=1) = 0.5 \).

For radar, we use a supervised approach to learn the model \( \Psi_{\rm radar} \) by defining a loss function for target detection.
 One may also define loss functions for other  metrics, such as angle estimation. Our loss function for target detection is the \emph{binary cross-entropy} loss function, given by
\begin{align} \label{eq:radar_loss}
\mathcal{L}_{\rm sens} = -\mathbb{E}\left[ T \log{\hat{T}} + (1 - T) \log {1 - \hat{T}} \right],
\end{align}
in which, \(\hat{T}\) represents the DNN estimation, and \(T\) denotes the actual presence of the target.

\subsubsection{ISAC}

In a setting where separate signals are sent for communication and sensing, the above optimizations are straightforward and usually result in QAM-like constellations for communication and constant-modulus modulation, such as PSK, for sensing.  
However, in an ISAC system where the same signal is used for both sensing and communication—as in our case in Fig.~\ref{fig:ISAC_diagram}—finding an appropriate constellation becomes more challenging, as it must take the competing requirements of both tasks into account.

With the goal of minimizing the BER and maximizing \( P_D \) using a single signal \( x \) in the ISAC system, the overall loss function is formulated as the weighted sum of the communication and radar loss functions, given by
\begin{align}
\mathcal{L}_{\rm ISAC} = \eta\mathcal{L}_{\rm radar} +  \bar \eta \mathcal{L}_{\rm comm},
\end{align}
where $0 \le \eta\leq 1 $ is a hyper-parameter that defines the relative importance of communication versus radar. 
 The values of \(\eta\) close to 1 correspond to radar-centric scenarios, while values close to 0 correspond to communication-centric scenarios. The loss functions $\mathcal{L}_{\rm comm}$ and $\mathcal{L}_{\rm sens}$ are defined in   \eqref{eq:comm_loss} and \eqref{eq:radar_loss}.

After defining the network structure, we simultaneously learn the neural network models \( \Psi_{\rm E} \), \( \Psi_{\rm comm} \), and \( \Psi_{\rm radar} \) through appropriate training. These correspond to the encoder, communication decoder, and radar detector, respectively. Table~\ref{tab:NN_architecture} summarizes the number of neurons in the hidden layers and the activation functions used in each neural network. It can be seen that all three models share the same structure for their hidden layers, but their input and output layers differ. We use the ReLU activation function between hidden layers.

\begin{table}[t]
\caption{Neural networks architecture}
\label{tab:NN_architecture}
\centering
\scriptsize 
\begin{tabular}{|c|c|c|c|}
\hline
\textbf{Model} & \textbf{Input} & \textbf{Hidden Layers} & \textbf{Output} \\
\hline
\( \Psi_{\rm E} \) & $K$ & $(16, 32, 16)$ & $2$ (linear) \\
\hline
\( \Psi_{\rm comm} \) & $2$ & $(16, 32, 16)$ & $K$ (softmax) \\
\hline
\( \Psi_{\rm radar} \) & $2$ & $(16, 32, 16)$ & $1$ (sigmoid) \\
\hline
\end{tabular}
\end{table}

Figure~\ref{fig:const} shows examples of learning-based constellations. 
To obtain these, we set the message length to \( K = 5 \), resulting in a constellation size of \( M = 32 \). The model is trained using the Adam optimizer with a learning rate of 0.001 for 500 epochs, with \( 10^6 \) samples per epoch and a batch size of \( 10^5 \).

As illustrated in Fig.~\ref{fig:const}(a), the NN-based algorithm generates novel constellations by varying the trade-off parameter \( \eta \), which balances sensing and communication performance. Specifically, the figure shows constellations corresponding to \( \eta = 0.9 \) (sensing-centric), \( \eta = 0.7 \) (balanced), and \( \eta = 0.05 \) (communication-centric). It is seen that, PSK-like structures are effective in sensing-centric scenarios, QAM-like structures emerge in communication-centric cases, and multi-ring constellations are optimal when balancing both objectives. 
In Table~\ref{tab:eta_metrics}, the SER, \( P_D \), and \( P_{\rm FA} \)  of the proposed designs are provided. 
Threshold chosen such that $P_{fa}$ is as small as about $ 10^{-3}$. {For comparison with traditional constellations, we have also included the performance of 32-QAM and 32-PSK in this table. }

    \begin{figure*}[t]
        \centering
\includegraphics[width=.795\linewidth]{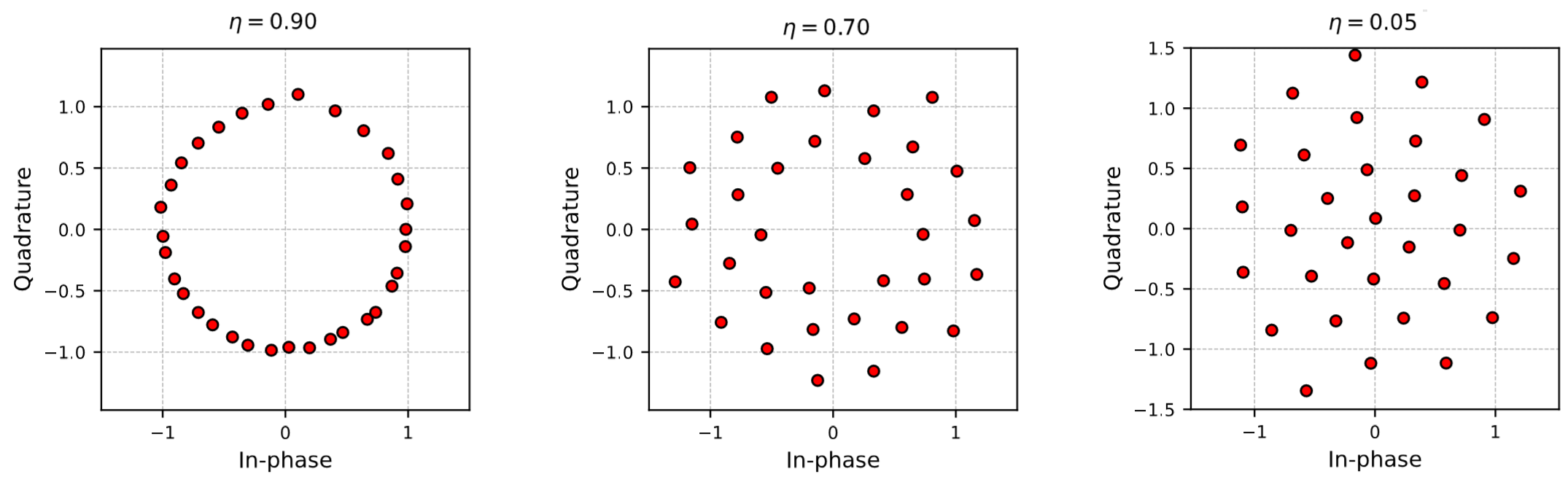} 
    
\caption{Comparison of various 32-point constellations designed using neural network for different ISAC scenarios. The value of the trade-off parameter \( \eta \) is indicated in each subfigure.} \vspace{-4mm}
    
    \label{fig:const}
    \end{figure*}

\begin{table}[th!]
\centering 
{
\caption{Performance metrics for learning-based constellations with different \( \eta \) and traditional constellations}
\label{tab:eta_metrics}
\begin{tabular}{c c c c c}
\hline
Constellation Type &  & SER & $P_d$ & $P_{fa}$ \\
\hline \hline
\multirow{3}{*}{Learning-based} & $\eta = 0.05$ & $10^{-1.38}$ & 0.157 & 0.045 \\
 & $\eta = 0.70$ & $10^{-1.19}$ & 0.570 & 0.0067 \\
 & $\eta = 0.90$ & $10^{-0.46}$ & 0.935 & 0.0088 \\
\hline
\multirow{2}{*}{Traditional} & 32-QAM & $10^{-1.39}$ & 0.170 & 0.0240 \\
 & 32-PSK & $10^{-0.49}$ & 0.935 & 0.0085 \\
\hline
\end{tabular}
}
\end{table}

{Other performance metrics may also be used to compare these constellations. In particular, the MMSE of each constellation can be evaluated using \cite[Eq. 24]{lozano2006optimum}.
Furthermore, the CRLB performance of the proposed ML-based constellation represents another important evaluation criterion. The evaluation of the CRLB for phase and frequency estimation in classical modulation schemes has been studied in \cite{Moeneclaey1998,Andrea1994,Steendam2001_LowSNR,Tavares2001}. Since evaluating the exact CRLB is mathematically involved, the modified CRLB (MCRB) was proposed to provide a more tractable bound \cite{Andrea1994}. The MCRBs of phase ($\phi$) and frequency ($\nu$) for PSK and QAM modulated signals are given by \cite{Andrea1994,Tavares2001,Steendam2001_LowSNR} $\mathrm{MCRB}_{\phi} = \frac{N_0}{2E_s} \cdot \frac{1}{L}$ and $\mathrm{MCRB}_{\nu} = \frac{N_0}{2E_s} \cdot \frac{3}{\pi^2 L (L^2 - 1)}$
where $L$ is the number of samples used for estimation, and $\frac{E_s}{N_0}$ is the symbol energy-to-noise ratio. The MCRB provides a looser but simpler bound that is accurate at high SNRs and identical for PSK and QAM constellations.

At low SNRs, closed-form CRLBs have been derived only for certain  constellations such as BPSK and QPSK \cite{Steendam2001_LowSNR}. The MCRB becomes less accurate in this regime, and the exact CRLB strongly depends on the modulation type. Since QAM and PSK share similar MCRBs, it is reasonable to expect comparable results for learning-based constellations. However, due to their irregular and asymmetric structures, evaluating low-SNR CRLBs for ML-based constellations remains a challenging, yet interesting open problem.

Finally, range accuracy, can also depend on the constellation, because it is tied to how well the received signal allows estimation of parameters such as delay or phase under noise.
This relationship is captured by the CRLB for time-delay (or phase) estimation, given by 
$\mathrm{CRB}_{\tau} = \frac{1}{8\pi^{2} \beta^{2} \, \mathrm{SNR}}$ \cite[Eq. 62]{LiuAn2022}
where $\beta$ is the effective root-mean-square bandwidth of the signal. 
The effective $\beta$  can change slightly depending on the symbol energy distribution, the instantaneous power envelope, and the phase structure of the constellation, and is mostly influenced by pulse shaping.

}

\section{Lessons Learned, Limitations \& Future Research Directions}\label{sec:introduction}

{
The range of ISAC problems that have shown promise in benefiting from AI is wide. In this section, we first summarize key lessons learned and then discuss  directions for future research.

\subsection{Lessons Learned}

A key takeaway from this tutorial is that ISAC problems are typically trade-off problems—meaning they involve the simultaneous optimization of multiple, often conflicting, objectives. Consequently, regardless of the solution approach—whether traditional multi-objective optimization or emerging learning-based methods—the results often lead to Pareto-optimal solutions. 
An example of such solutions is one that balances communication throughput and sensing accuracy, where improving one objective typically comes at the cost of the other.  

 AI-empowered solutions have both advantages and limitations when applied to ISAC systems. We discuss these below, noting that there is currently little guidance on when each approach is most suitable.

\subsubsection{When to Use Each AI Model?}\label{sec:when}

In this paper, we have discussed various AI models. Below is a brief summary of when each model is most appropriate for ISAC applications.
 
\begin{itemize}
   \item 
    \textit{Single-Task Learning:} Designed to optimize performance for one specific task, without leveraging shared features from related tasks. For example, training a model solely for target detection in ISAC.

\item \textit{Multi-Task Learning:} Suited for two or more related tasks that can share features, e.g., joint channel estimation and symbol detection, or detecting pedestrians and
predicting the road’s curvature in a  self-driving car model.

  \item   \textit{Multi-Objective Learning:} Optimizes multiple, and often competing, objectives in a single task, e.g., a car model trained solely
to detect pedestrians (one task) but has two objectives:
 to maximize  accuracy and to minimize inference time. 

    \item 
 \textit{Model-Based Learning:} Best for problems with known physical models but practical uncertainties, e.g., hybrid beamforming with impairments or waveform design combining analytical and learned components.
    
    \item \textit{Deep Learning:} Fits high-dimensional, non-linear problems lacking tractable models, e.g., end-to-end waveform design or complex resource allocation.
    \item 
    \textit{End-to-End Learning:} Trains a model to directly map raw inputs to final outputs, jointly optimizing all intermediate processing stages, e.g., learning ISAC waveforms and detection algorithms in one unified framework without separately optimizing each block.

\end{itemize}

}

 {

\subsubsection{Benefits of AI-Empowered ISAC Systems}\label{sec:limitations}
We have discussed several advantages of AI-based ISAC over classical optimization approaches throughout the paper. These advantages can be classified as:

\begin{itemize}
    \item \textit{Reducing Computational Complexity:} Many AI-based ISAC designs belong to this category. For example, as demonstrated in Case Study~I, lightweight unsupervised AI models can largely reduce the computational complexity of classical optimization techniques while maintaining a negligible performance or accuracy gap. This gap can be further minimized by increasing dataset size and fine-tuning the model architecture, complexity, and generalization capability through techniques such as regularization and cross-validation.

    
\item \textit{Solving Non-Convex  Problems:}
From Case Study II, we learned that  algorithm unfolding is particularly effective for solving non-convex problems in communications and ISAC, as the trained unfolded model can better avoid poor local minima and often achieves improved suboptimal solutions. While the per-layer computations mirror those of the original algorithm, efficiency gains arise from requiring significantly fewer layers than the number of algorithm iterations.

\item \textit{Improving Accuracy:}
From Case Study~III, we saw that autoencoder-based constellation design can flexibly balance the sensing–communication trade-off and produce novel, previously unseen solutions that outperform both traditional constellations (e.g., QAM) and probabilistic-shaped constellations in terms of sensing accuracy and communication SER.

\item \textit{Simplifying System Design:}
Some DL models, such as the end-to-end learning, offer a powerful and streamlined alternative to traditional block-by-block system design. By optimizing multiple components, e.g., channel coding, modulation, and beamforming, in a single framework, end-to-end learning reduces system design complexity compared to traditional approaches. This approach could advance the development of AI-native air interfaces, a long-term objective in AI-driven communication systems.

\item \textit{Enhancing Robustness:}
  Learning-based approaches can enhance robustness in ISAC systems, particularly when traditional model-based methods struggle with model mismatches or hardware impairments. Such mismatches may arise from imperfect or quantized CSI, rapidly changing channel conditions, hardware limitations, or the inherent complexity of ISAC operations. By leveraging real-time data, AI models can adapt to these uncertainties and design robust waveforms and beamformers compared to convex optimization \cite{nguyen2024joint}, and  demonstrates superior error-rate performance under hardware impairments \cite{mateos2022end}.

\end{itemize}

\subsubsection{Limitations of AI-Based ISAC Systems}\label{sec:limitations}

Despite its potential benefits, AI-based ISAC system designs   have several limitations. These include 
}

{
\begin{itemize}

    \item 
In supervised ISAC, collecting and labeling large datasets for ISAC systems—especially in dynamic wireless environments—can be challenging and costly. Fortunately, as noted earlier, real-world datasets are becoming increasingly available \cite{alkhateeb2023deepsense}.

\item  AI-aided ISAC models trained on specific datasets may not generalize well to new signal characteristics, dynamic propagation conditions, or unseen noise and interference. This limits their robustness in real environments, where signals are constantly affected by multipath fading and co-channel interference.


\item Deep learning models  can be hard to interpret due to their complexity and lack of transparency.
Such  AI models can inherit biases of their training data, leading to unfair or discriminatory outcomes of the end-to-end ISAC systems. 

{
\item Learning-based ISAC waveforms and beamformers   may also suffer from generalization limitations. Both sensing and communication performance metrics degrade when the CSI, which is used to train the model,  becomes aged/outdated or when the  system topology changes. In such cases, the models must be retrained by considering the latest CSI and updating the system topologies.} 

\item 
AI-based ISAC may raise privacy and security concerns when processing sensitive data, such as user location, and are vulnerable to attacks such as adversarial examples that can manipulate inputs and disrupt system behavior.

\end{itemize}
 
 It is worth noting, however, that the last three limitations are currently paid increasingly much attention in several research communities 
 and there are good promisees to overcome them, or  develop guarantees on performance, generalizability, and privacy. These are, however, outside of the scope of this tutorial. 
}

\subsection{Future Research Directions}

 Here, we identify several directions for future research in the area of AI-empowered ISAC. This list is not intended to be complete, but it focuses more on ISAC open problems,
 particularly in  waveform and beamforming design.

\subsubsection{End-to-End ISAC}

While AI-based constellation design for interference mitigation in both communications and radar has been studied recently \cite{waldschmidt2021automotive,nartasilpa2018communications,zhang2024deep}, there is currently no work for ISAC. 
Future directions in end-to-end learning-based ISAC can focus on addressing this problem, { e.g., by extending Case Study III to a multi-cell scenario.} In addition, addressing practical challenges, such as minimizing symbol error rate  in the presences of interference in various forms, including self-interference, inter-cell interference, and inter-user interference. Further, extending these methods to multi-user and multi-target scenarios, while enhancing spectral efficiency and ensuring scalability, will be critical to realizing the full potential of ISAC systems in practical deployments.


While current approaches \cite{mateos2022end,mateos2024semi,zheng2024end} have demonstrated promising path in jointly optimizing sensing and communication tasks, their performance, particularly in realistic scenarios with interference, remains unknown. These works also ignore interference in its various forms including self-interference \cite{li2020self,balatsoukas2021joint}, inter-cell interference\cite{zhang2023zicicc}, and inter-user interference \cite{vaezi2025interference}, among others. Future research should explore advanced techniques to integrate interference mitigation into end-to-end learning frameworks for both single-carrier and multi-carrier methods, including OFDM and OTFS. Problem of interference reduction involves using end-to-end learning to develop new, flexible constellations that are specifically designed to be aware of self-interference. To  this end, the research could leverage autoencoder-based techniques for self-interference cancellation, which is a promising  since interference can be identified at the transmitter.
{In addition, a deeper analysis of the low-SNR performance of ML-based constellations, including their CRLB for phase, frequency, and range estimation, is needed to clarify their advantages and limitations. Further, ML-based constellations could potentially be designed to achieve a desired CRLB bound.}

\subsubsection{Transformer Neural Networks for ISAC}

Intelligent vehicular networks are a key use case of ISAC in 6G \cite{Tang2020a}. Classical waveform and beamformer designs, such as beam codebooks and analytical optimizations, may fall short in high-speed vehicular environments. To address this, AI-driven ISAC designs can leverage multimodal sensory data—such as radar, vision, and LiDAR—to enhance waveform and beamforming performance under challenging propagation conditions \cite{cheng2023_survey}.
For example, the high-resolution images from cameras connected to roadside BSs, point clouds from LiDAR, reflected electromagnetic signals from transmitted waves, localization information from GPS-enabled sensors can be collected and used to enhance the design of ISAC waveforms.
In \cite{Cui2024Transformer},   a transformer-based beamforming framework is proposed that leverages 3D ResNet-18 to extract and fuse multi-modal features (vision, radar, LiDAR, and position) for sensing-assisted communication in dynamic and complex road conditions.

AI techniques can extract 3D features from multimodal inputs (vision, radar, LiDAR, position). Transformer networks \cite{Vaswani2017},
the foundation of large language models like ChatGPT and Gemini, can fuse these features to learn ISAC waveforms and beamformers. 
{
Unique features of  transformers such as long-range dependency modeling,   self-attention, parallel processing, and positional encoding   can be leveraged for enhancing ISAC designs over the existing deep learning models.

Transformers offer strong potential for ISAC systems by jointly learning  sensing and communication tasks through multi-head self-attention. The global self-attention mechanisms offered by generative transformers can be leveraged to dynamically aggregate information across distributed access points (APs), users, and targets without relying on any predefined graphical structures. They enable each network entity (APs, user, and target) to attend to every other entity, thereby capturing full-order dependencies and modeling global cooperation across the distributed ISAC-enabled entities. Their ability to model global dependencies across multimodal data enables efficient unified waveform and beamformer design that can outperform traditional approaches based on  convolutional neural networks and  graph neural networks (GNN),   in dynamic propagation environments envisioned for 6G. Unlike models relying on local or predefined structures, transformers dynamically assign attention weights to all features, allowing task-oriented focus for channel estimation, target tracking, and signal processing. Masked multi-head attention and customized embeddings \cite{Kocharlakota2024,Kocharlakota2024_paper} further integrate sensing and communication side information, while hybrid transformer–GNN models \cite{velivckovic2017graph,zhang2025pilot} enhance multi-user coordination. Despite challenges such as high computational cost, long-sequence handling, and interpretability, transformers enable faster, parallel optimization of ISAC waveforms and beamforming with superior adaptability in complex environments.

}

\subsubsection{Digital Twin-Aided ISAC Systems}
Digital twins are expected to play a key role in enabling 6G wireless technologies, offering proactive analytics and efficient training of AI models to replicate real-world propagation environments \cite{Khan2022}. A digital twin comprises pre-trained models that support real-time decision-making for multi-functional wireless services.

In the context of ISAC, digital-twin-aided architectures are particularly promising for dynamic and distributed scenarios. By collecting spatial and temporal information from distributed APs and sensors, digital twins can construct 3D models of the surrounding environment, capturing positions, orientations, and shapes of objects such as scatterers, targets, and users. These digital replicas enable simulation of communication and sensing channels, which can be used to design ISAC transmission strategies and train AI models for unified waveform and beamformer design.

Advanced AI tools in data analysis, pattern recognition, and real-time reasoning further enhance digital twin capabilities for ISAC tasks—ranging from waveform and beamformer optimization to spectral-efficient constellations, beam pattern generation, interference mitigation, and resource allocation.
A key challenge is making digital twins self-configurable with online learning capabilities. However, realizing practical digital twins for ISAC demands substantial engineering effort to accurately replicate complex physical environments using fast and efficient AI models.

{ \subsubsection{Learning-Based ISAC in FR3} 
Most existing ISAC designs focus on sub-6 GHz or mmWave frequencies. However, the upper mid-band spectrum spanning 7–24 GHz, also known as frequency range 3 (FR3) \cite{bazzi2025upper,baduge2025frequency}, is emerging as a key enabler for 6G ISAC. It offers the potential to bridge the wide-area coverage advantages of sub-6 GHz (FR1)  with the high-capacity benefits of mmWave (FR2). However, 
directly applying learning-based algorithms developed for FR1 or FR2 to FR3 ISAC may not be feasible due to the distinct propagation characteristics, antenna configurations, and temporal dynamics of the FR3 band \cite{baduge2025frequency}. FR3 channels exhibit moderate sparsity, higher angular resolution, and mid-range Doppler variations, which require tailored feature representations and model architectures. Moreover, as large-scale arrays in FR3 approach the near-field regime, traditional far-field assumptions—central to many FR1 and FR2 algorithms—break down, necessitating range–angle joint modeling and beam focusing rather than simple steering. To effectively operate in FR3, learning algorithms must integrate frequency-aware normalization, domain adaptation, and physics-informed priors, ensuring that they can leverage FR3’s unique sensing–communication trade-offs while maintaining robustness to moderate blockage, beam misalignment, and frequency-dependent scaling effects.}

 \section{Conclusions}\label{sec:conclusion}

{
This tutorial has provided a comprehensive overview of AI-driven ISAC systems, emphasizing their significance in 6G networks. We have explored fundamental ISAC concepts, classical multi-objective optimization–based design approaches, learning-based optimization techniques, and AI-based unified waveform, beamformer, and constellation designs. We have elaborated on the necessity of adopting low-complexity learning-based alternatives to computationally intensive classical analytical optimization techniques for designing ISAC waveforms and beamformers.

Through three case studies, covering learning-based designs of ISAC waveforms, beamformers, and constellations, we have demonstrated AI’s potential to improve ISAC algorithms while highlighting design trade-offs and computational efficiency. 
Lastly, we have outlined key lessons learned to summarize our findings and proposed future research directions to guide researchers in advancing learning-based ISAC technology.
Based on the lessons learned, we recommend adopting learning-based signal processing algorithms over classical optimization-based methods to simplify design and enhance performance, particularly in the following cases:
(i) when traditional models are incomplete or inadequate to capture key operating characteristics; (ii) when no known algorithms exist for sophisticated multi-objective optimization problem formulations; and (iii) when known optimal algorithms are prohibitively complex for real-time implementation in practical settings. 
}

 \section*{Acknowledgments}

 The authors would like to thank Janith Dassanayake, Lei Wang, and Amirhossein Keshavarz for their help in simulation of Case Study I, Case Study II and Case Study III in Sections~VI.A, VI.B and VI.C, respectively.


  \linespread{1.0}



\begin{IEEEbiography}
	[{\includegraphics[width=1in,height=1.25in,clip,keepaspectratio]{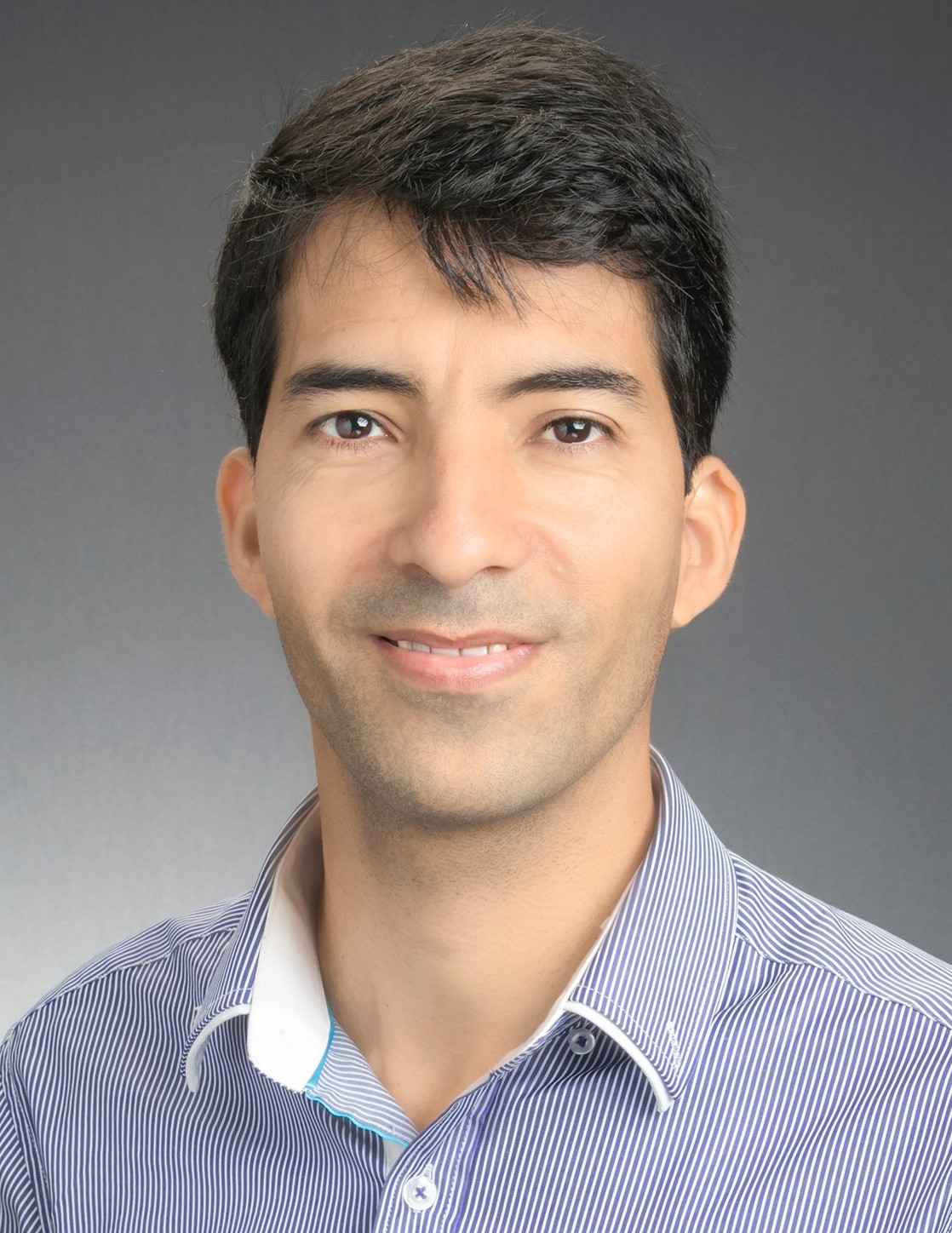}}]{Mojtaba Vaezi} (S’09–M’14–SM’18) received his B.Sc. and M.Sc. degrees from Amirkabir University of Technology (Tehran Polytechnic) and his Ph.D. from McGill University, all in Electrical Engineering. From 2015 to 2018, he was a Postdoctoral Research Fellow and Associate Research Scholar at Princeton University. Currently, he serves as an Associate Professor in the Department of Electrical and Computer Engineering at Villanova University. Prior to his tenure at Princeton, he worked as a researcher at Ericsson Research in Montreal, Canada. His research interests encompass signal processing and machine learning for wireless communications, focusing on physical layer security and sixth-generation (6G) radio access technologies. Among his  publications is the book \textit{Multiple Access Techniques for 5G Wireless Networks and Beyond} (Springer, 2019).

Dr. Vaezi has held editorial positions at several IEEE journals and currently serves as an Editor for \textsc{IEEE Transactions on Wireless Communications} and a Senior Editor for \textsc{IEEE Communications Letters}. Previously, he served as a Senior Area Editor for \textsc{IEEE Signal Processing Letters}, an Editor for \textsc{IEEE Transactions on Communications}, and an Associate Editor for \textsc{IEEE Communications Magazine}. He has received numerous academic, leadership, and research awards, including the McGill Engineering Doctoral Award, the IEEE Larry K. Wilson Regional Student Activities Award in 2013, the Natural Sciences and Engineering Research Council of Canada (NSERC) Postdoctoral Fellowship in 2014, the Ministry of Science and ICT of Korea’s Best Paper Award in 2017, the IEEE Communications Letters Exemplary Editor Award in 2018, the 2020 IEEE Communications Society Fred W. Ellersick Prize, the 2021 IEEE Philadelphia Section Delaware Valley Engineer of the Year Award, and the National Science Foundation (NSF) CAREER Award in 2023.
\end{IEEEbiography}	

\begin{IEEEbiography}
	[{\includegraphics[width=1in,height=1.25in,clip,keepaspectratio]{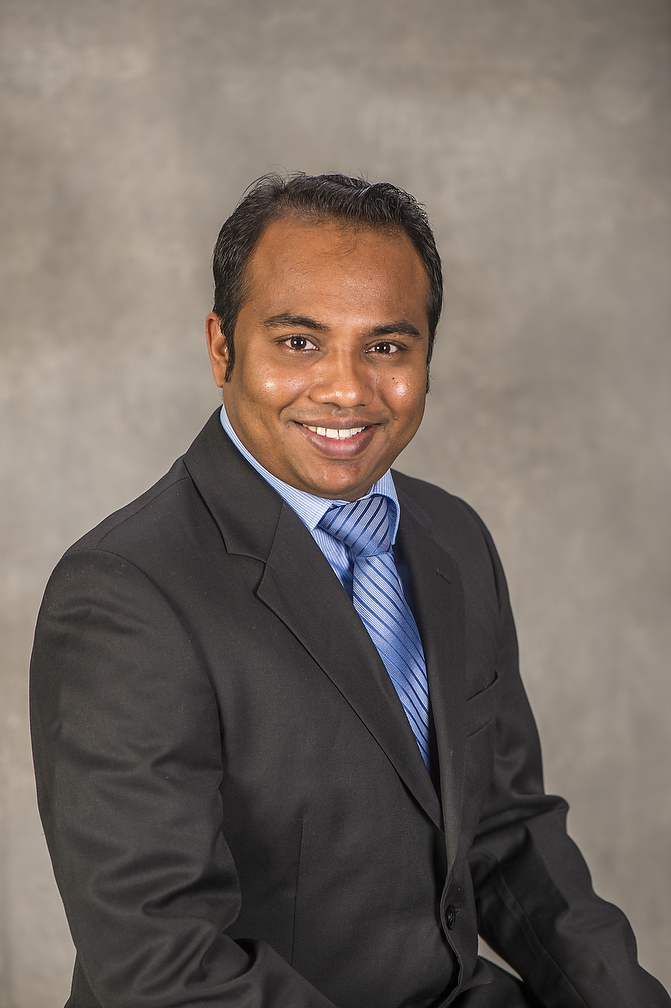}}]{Gayan Amarasuriya Aruma Baduge} (S'09, M'13, SM'19) received the B.Sc. degree in  Engineering (with first class 	Hons.)  from the Department of Electronics and	Telecommunications Engineering, University of
	Moratuwa, Moratuwa, Sri Lanka, in 2006, and the	Ph.D. degree in Electrical Engineering from the	Department of Electrical and Computer Engineering,
	University of Alberta, Edmonton, AB, Canada, in	2013.  He was   a Postdoctoral Research Fellow	with the Department of Electrical Engineering,	Princeton University, Princeton, NJ, USA from 2014 to 2016.	Currently, he is a professor   in     the School of Electrical,  Computer, and Biomedical   Engineering  in Southern Illinois University, IL, USA.	
    
    Dr. Aruma Baduge has held editorial positions at various IEEE journals, and currently  he serves as an Editor of \textsc{IEEE Transactions on Wireless Communications} and  \textsc{IEEE Wireless Communications Letters}. Previously, he was an  Editor for \textsc{IEEE Transactions on  Communications}, \textsc{IEEE Communications  Letters}, and  \textsc{IEEE Open Journal of the Communications Society}.

\end{IEEEbiography} 

\begin{IEEEbiography}[{\includegraphics[width=1in,height=1.25in,clip,keepaspectratio]{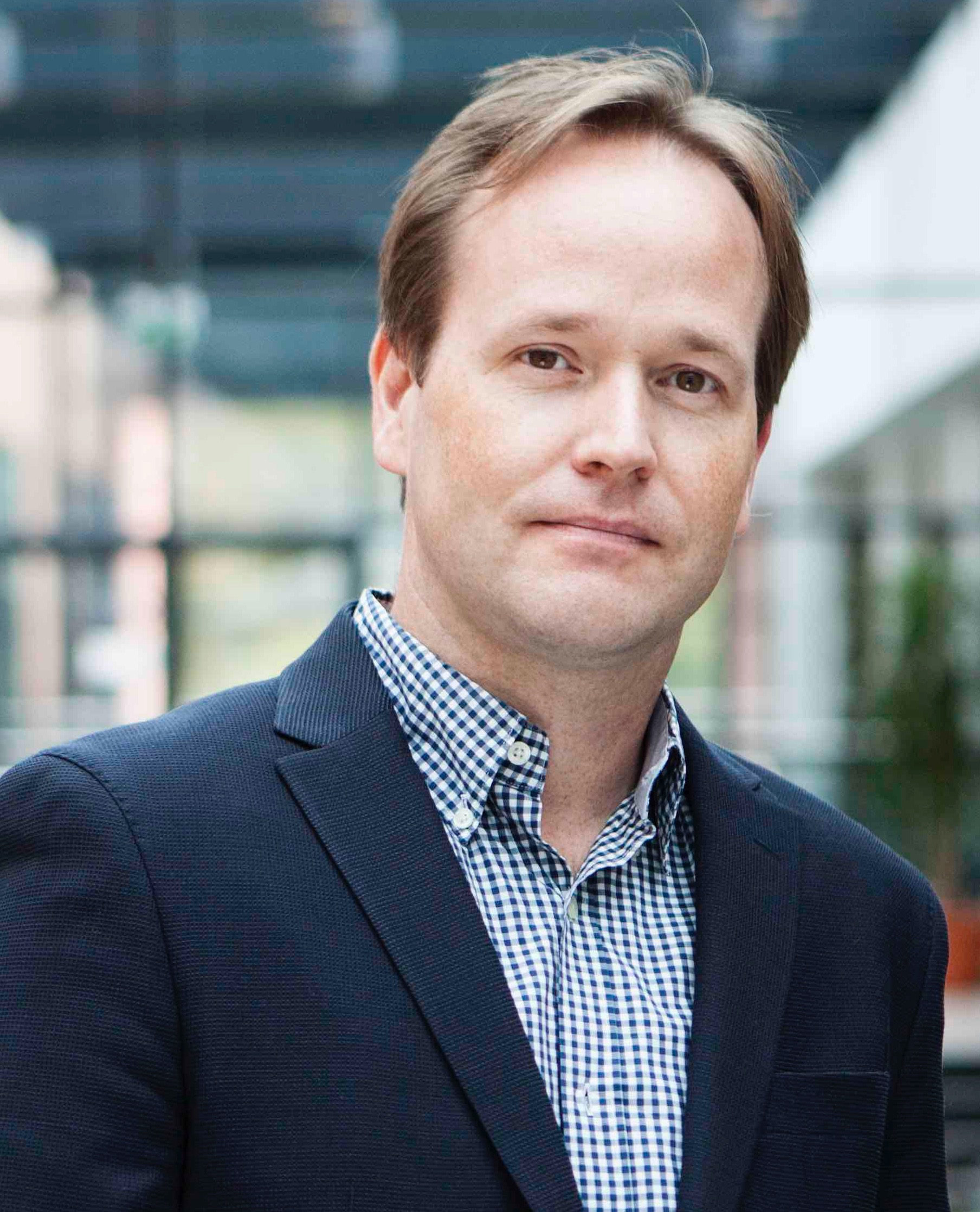}}]{Esa Ollila} (M'06,SM'21) received the M.Sc. degree in mathematics from the University of Oulu,  Finland, in 1998, the Ph.D. degree (Hons.) in statistics from the University of Jyv\"askyl\"a,  Finland, in 2002, and the D.Sc.(Tech) degree (Hons.) in signal processing from Aalto University, Finland, in 2010.  From 2004 to 2007, he was a Post-Doctoral Fellow and from August 2010 to May 2015, he was an Academy Research Fellow of the Research Council  of Finland. From 2010 to 2011, he was a Visiting Post-Doctoral Research Associate with the Department of Electrical Engineering, Princeton University,  NJ, USA.  Currently, he is a professor in the School of Electrical Engineering at Aalto University.  He has coauthored a  textbook "Robust Statistics for Signal Processing" published by Cambridge University Press in 2018.  He is on the  Board of Directors of European  Association for Signal Processing (EURASIP) and an elected member of IEEE SPS Signal Processing Theory and Methods (SPTM) Technical Committee (2022-2027) where he serves as  the chair of the Awards subcommittee.  He was the General Co-Chair for EUSIPCO-2023 and  served as  an Associate Editor for the  Scandinavian Journal of Statistics during 2021-2025. 
\end{IEEEbiography}

\begin{IEEEbiography}[{\includegraphics[width=1in,height=1.25in,clip,keepaspectratio]{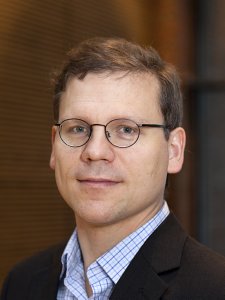}}]{Sergiy A. Vorobyov} (M’02--SM’05--F’18) 
	is a Professor with the Department of Information and Communications Engineering, Aalto University, Finland. Since his graduation, he held various faculty and research positions with the University of Alberta, Canada; the Joint Research Institute between Heriot-Watt University and Edinburgh University, U.K.; Darmstadt University of Technology and Duisburg-Essen University, Germany; McMaster University, Canada; the Institute of Physical and Chemical Research, Japan; and the National University of Radio Electronics, Ukraine. His research interests include optimization and multi-linear algebra methods in signal processing and data analysis; statistical, array, and graph signal processing; estimation, detection and learning theory and methods; computational imaging; multi-antenna, very large, cooperative, and cognitive systems, and integrated sensing and communications.
	
	Dr. Vorobyov is the recipient of the 2004 IEEE Signal Processing Society Best Paper Award, the 2007 Alberta Ingenuity New Faculty Award, the 2011 Carl Zeiss Award (Germany), the 2012 NSERC Discovery Accelerator Award, and IEEE ICASSP 2023 Top 3\% paper recognition, and other awards. He was a Senior Area Editor for the IEEE Signal Processing Letters in 2016—2020, an Associate Editor for the IEEE Trans. Signal Processing in 2006–2010 and the IEEE Signal Processing Letters in 2007–2009. He was a member of the Sensor Array and Multi-Channel Signal Processing and Signal Processing for Communications and Networking Technical Committees of the IEEE Signal Processing Society in 2007–2012 and 2010–2016, respectively. He was the Track Chair for Asilomar 2011, Pacific Grove, CA, USA; the Technical Co-Chair for IEEE CAMSAP 2011, Puerto Rico; the Tutorial Chair for ISWCS 2013, Ilmenau, Germany; the Technical Co-Chair for IEEE SAM 2018, Sheffield, U.K.; the Technical Co-Chair for IEEE CAMSAP 2023, Costa Rica; and the General Co-Chair for EUSIPCO 2023, Helsinki, Finland.
\end{IEEEbiography}

\end{document}